\documentclass[aps,showpacs,twocolumn,prd,superscriptaddress,nofootinbib]{revtex4}

\usepackage{amsmath}
\usepackage{latexsym}
\usepackage{graphicx}
\usepackage{color}

\newcommand{\Rext}{R_\mathrm{ext}}
\newcommand{\Mf}{M_{\rm f}}
\newcommand{\MAH}{M_{\rm AH}}
\newcommand{\MADM}{M_{\rm ADM}}
\newcommand{\Mrest}{M_{\rm rest}}
\newcommand{\Mchirp}{\mathcal{M}}
\newcommand{\Oi}{\Omega_{\rm i}}
\newcommand{\Of}{\Omega_{\rm f}}
\newcommand{\beq}{\begin{equation}}
\newcommand{\eeq}{\end{equation}}
\newcommand{\bea}{\begin{eqnarray}}
\newcommand{\eea}{\end{eqnarray}}
\newcommand{\ba}{\begin{array}}
\newcommand{\ea}{\end{array}}
\newcommand{\omQ}{\omega_{\rm QNM}}
\newcommand{\cval}{\dot{\Omega}_0}
\newcommand\degrees[1]{\ensuremath{#1^\circ}}

\begin{document}

\title{Mergers of nonspinning black-hole binaries: Gravitational radiation characteristics}

\author{John G. Baker}
\affiliation{Gravitational Astrophysics Laboratory, NASA Goddard Space Flight Center, 8800 Greenbelt Rd., Greenbelt, MD 20771, USA}
\author{William D. Boggs}
\affiliation{University of Maryland, Department of Physics, College Park, MD 20742, USA}
\author{Joan Centrella}
\affiliation{Gravitational Astrophysics Laboratory, NASA Goddard Space Flight Center, 8800 Greenbelt Rd., Greenbelt, MD 20771, USA}
\author{Bernard J. Kelly}
\affiliation{Gravitational Astrophysics Laboratory, NASA Goddard Space Flight Center, 8800 Greenbelt Rd., Greenbelt, MD 20771, USA}
\author{Sean T. McWilliams}
\affiliation{Gravitational Astrophysics Laboratory, NASA Goddard Space Flight Center, 8800 Greenbelt Rd., Greenbelt, MD 20771, USA}
\affiliation{University of Maryland, Department of Physics, College Park, MD 20742, USA}
\author{James R. van Meter}
\affiliation{Gravitational Astrophysics Laboratory, NASA Goddard Space Flight Center, 8800 Greenbelt Rd., Greenbelt, MD 20771, USA}

\date{\today}

\begin{abstract}
We present a detailed descriptive analysis of the gravitational
radiation from black-hole binary mergers of nonspinning black holes,
based on numerical simulations of systems varying from equal-mass to a
6:1 mass ratio.  Our primary goal is to present relatively complete
information about the waveforms, including all the leading multipolar
components, to interested researchers. In our analysis, we pursue the
simplest physical description of the dominant features in the
radiation, providing an interpretation of the waveforms in terms of an
{\em implicit rotating source}.  This interpretation applies uniformly
to the full wave train, from inspiral through ringdown.  We emphasize
strong relationships among the $\ell=m$ modes that persist through the
full wave train.  Exploring the structure of the waveforms in more
detail, we conduct detailed analytic fitting of the late-time
frequency evolution, identifying a key quantitative feature shared by
the $\ell=m$ modes among all mass ratios.  We identify relationships,
with a simple interpretation in terms of the implicit rotating source,
among the evolution of frequency and amplitude, which hold for the
late-time radiation.  These detailed relationships provide sufficient
information about the late-time radiation to yield a predictive model
for the late-time waveforms, an alternative to the common practice of
modeling by a sum of quasinormal mode overtones. We demonstrate an
application of this in a new effective-one-body-based analytic
waveform model.
\end{abstract}


\maketitle

\section{Introduction}

The final merger of two black holes (BHs) having comparable masses will
produce an intense burst of gravitational radiation, and is expected
to be one of the strongest sources in the gravitational-wave sky.
Mergers of stellar black holes are key targets for ground-based
detectors such as LIGO, VIRGO, and GEO600, and knowledge of the merger
waveforms is an important component of improving the detectability of
such systems.  The space-based LISA detector will observe mergers of
massive black holes at high signal-to-noise ratios, allowing tests
of general relativity in the strong-field, dynamical regime.

Today, numerical relativity (NR) studies are beginning to progress toward a
full description of black-hole binary merger systems.  For
noneccentric inspirals, this space is spanned by seven parameters:
the symmetric mass-ratio $\eta=m_1 m_2/(m_1+m_2)^2$, and the six
combined components of the black holes' spin vectors.  Considerable
study has been focused on the fiducial center point of this parameter
space, the case of equal-mass nonspinning black-hole mergers.  After
the series of breakthroughs that ushered in an era of rapid progress
in the field
\cite{Bruegmann:2003aw,Pretorius:2005gq,Campanelli:2005dd,Baker:2006yw},
several investigations assessing the accuracy of the available equal
mass waveforms and applying them to data analysis were conducted
\cite{Baker:2007fb,Buonanno:2006ui,Baker:2006kr,Baker:2006ha,Boyle:2007ft}.

In this paper, we undertake a descriptive study of the waveforms
generated in the late inspiral and merger of black-hole binaries for
the subspace of nonspinning black holes, parametrized only by
$\eta$.  Our study is based on a series of numerical simulations,
discussed in Sec.~\ref{sec:simulations}, covering at least the last
$\gtrsim 4$ orbits of nonspinning black-hole binary mergers with mass
ratios extending to 6:1 ($\eta\approx0.122$).  Several of the
simulations presented here have already been applied in a recent
paper, focusing on the development of a faithful analytic waveform
model \cite{Buonanno:2007pf}.  Here we provide details of these and
additional simulations, together with considerable analysis, focused
on providing a qualitative and quantitative picture of how the
waveforms from nonspinning black-hole mergers depend on $\eta$.
Nonspinning black-hole binary merger waveforms were previously
examined in Ref.~\cite{Berti:2007fi}, but our analysis is novel and
complementary to that work.  Our descriptive presentation puts
emphasis on the relationships between waveforms from the different
mass-ratio cases and different harmonic modes, with references to
Ref.~\cite{Berti:2007fi} where related observations have been
made. Our approach to describing the inspiral-merger-ringdown
transition is particularly distinct, founded in a uniform approach
that describes all stages of this process in similar terms, and
ultimately suggesting a complementary physical picture.

Black-hole-binary merger waveforms have been noted for their
``simplicity.''  For the nonspinning systems the simple physics of the
coalescence is exposed by a spherical harmonic decomposition of the
waveforms.  In Sec.~\ref{sec:description} we walk readers through the
basic features of the radiation, characterizing amplitude and phase
evolution of the multipolar components, and discussing relationships
among the simulations representing different mass ratios, and among
the multipolar components of each simulation. As we analyze the
waveforms we develop a conceptual interpretation of the basic waveform
features. In this interpretation we consider the structure of an {\em
implicit rotating source}, which could have generated the measured
radiation through its rotational motion.  This allows a uniform
interpretation that applies throughout the coalescence process:
inspiral, merger and ringdown.

In Sec.~\ref{sec:description2}, we examine the strong final burst of
radiation beginning $\sim20M$ before the formation of a common
horizon.  We quantitatively describe the phasing in terms of an
analytic model, based on a continuous, monotonically increasing
frequency.  We find, in particular, that the peak rate of change in
frequency, appropriately scaled, is the same across all $\ell=m$ modes
and mass ratios.  We also identify relationships among the mode
amplitudes and phases, which are connected to an approximately linear
relationship between angular momentum and frequency:
$d^2J/d\omega^2\sim0$.  We interpret these relationships in terms of
the implicit source.

Finally, In Sec.~\ref{sec:newEOB}, we demonstrate the utility of what
we have learned in our waveform characterization by applying some of
the quantitative features we have uncovered in a new variation on the
analytic waveform model in \cite{Buonanno:2007pf}, which was based on
the Effective-One-Body (EOB) resummation of the Post-Newtonian(PN)
approximation to inspiral dynamics \cite{Buonanno:1998gg}.  In
particular, we provide a distinct late-time waveform model,
alternative to the common ``spectroscopic'' model
\cite{Dreyer:2003bv,Berti:2005ys} based on sums of quasinormal mode
overtones.

\section{Overview}
\label{sec:concept}

We begin with some examples of gravitational strain waveforms as they
might be observed by gravitational-wave instruments.  In observational
work, and PN analysis, it is customary to describe the radiation in
terms of gravitational-wave strain, $h$.  In representing the strain,
it is convenient to combine the two real waveform polarization
components, $h_+$ and $h_\times$, into one complex strain waveform,
\begin{equation}
h = h_+ + i h_\times.
\end{equation}
We decompose the strain waveforms measured on a sphere of radius
$\Rext$, into spin-weighted spherical harmonic components, $h_{\ell
m}$.  The details of the decomposition, and how the waveform
information is extracted from the numerical simulations, are given in
Appendix~\ref{appendix:Radiation}.

The waveforms in this section are aligned in time and phase so that
the maximum strain amplitude occurs at $t=0$.  The remaining figures
of this paper will be aligned in a similar way, but with $t=0$ marking
the time of peak (2,2) mode energy flux, $\dot{E}_{22}$ (unless stated
otherwise).

\begin{figure*}
  \includegraphics*[width=7in]{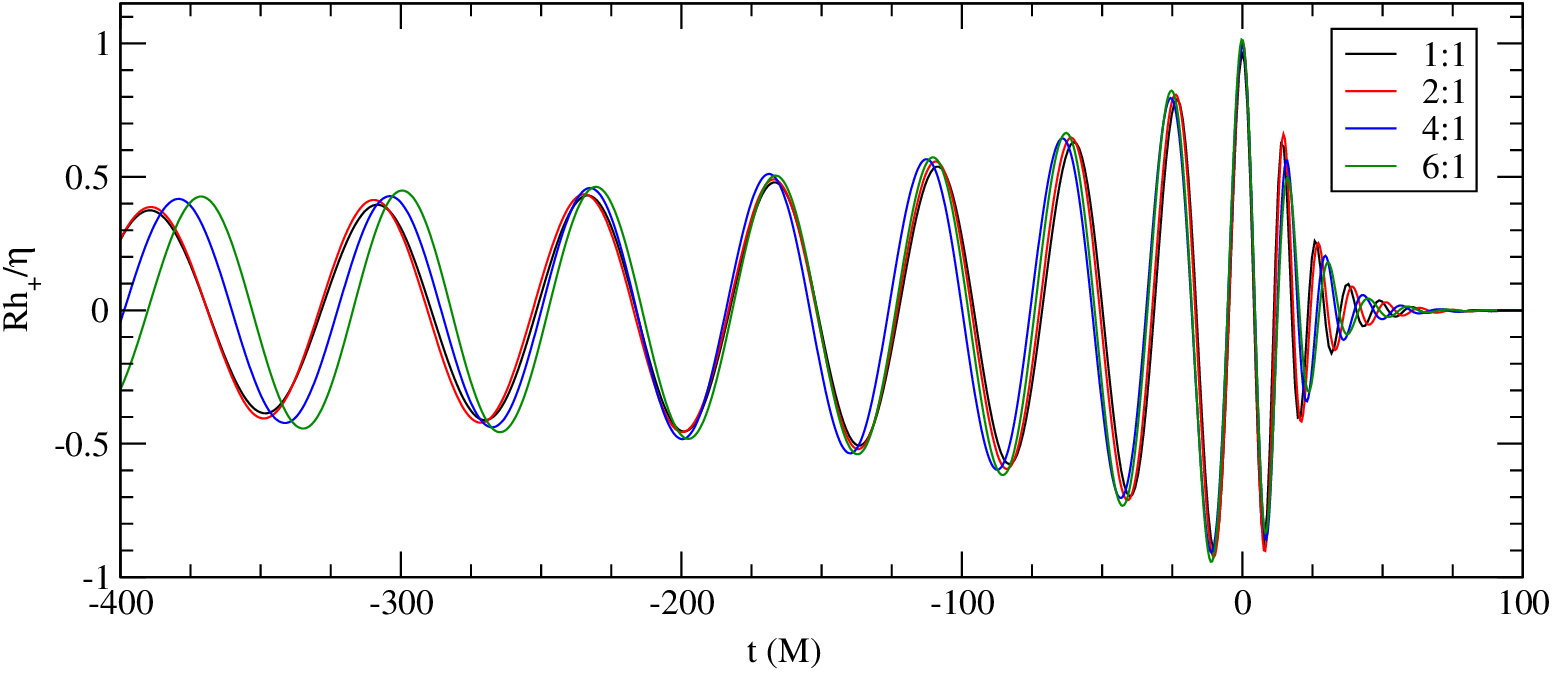} \caption{The plus
  component of the strain, $h_+$, is shown for different mass ratios using
  the $\ell=2$ and $\ell=3$ modes.  The observer is located at
  distance $R$ along the axis of the system, at
  inclination (footnote \ref{inclination}) $\theta=\degrees{0}$ and azimuthal
  angle $\phi=\degrees{0}$. The strains are scaled by symmetric mass
  ratio and aligned such that, for each mass ratio, the peak of
  $h_{22}$ occurs at $t=0$. The phases are rotated such that the
  phases are \degrees{0} at $t=0$.}  \label{fig:waves}
\end{figure*}

Fig.~\ref{fig:waves} shows waveforms from mergers of nonspinning black
holes for various mass ratios, as observed at distance $R$ on the
rotational/orbital axis of the system.  The figure shows $h_+$ for
each of the four mass ratios 1:1, 2:1, 4:1, and 6:1. For these
observers the observed waveforms will be circularly polarized, so that
$h_{\times}$ is \degrees{90} out of phase with $h_{+}$.  We use units
in which $G = 1$ and $c = 1$ and express both time and spatial
distances in terms of the total mass $M$, where $M \sim 5 \times
10^{-6} (M/M_{\odot}) {\rm sec} \sim 1.5 (M/M_{\odot}) {\rm km}$.

\begin{figure*}
\includegraphics*[width=3.5in]{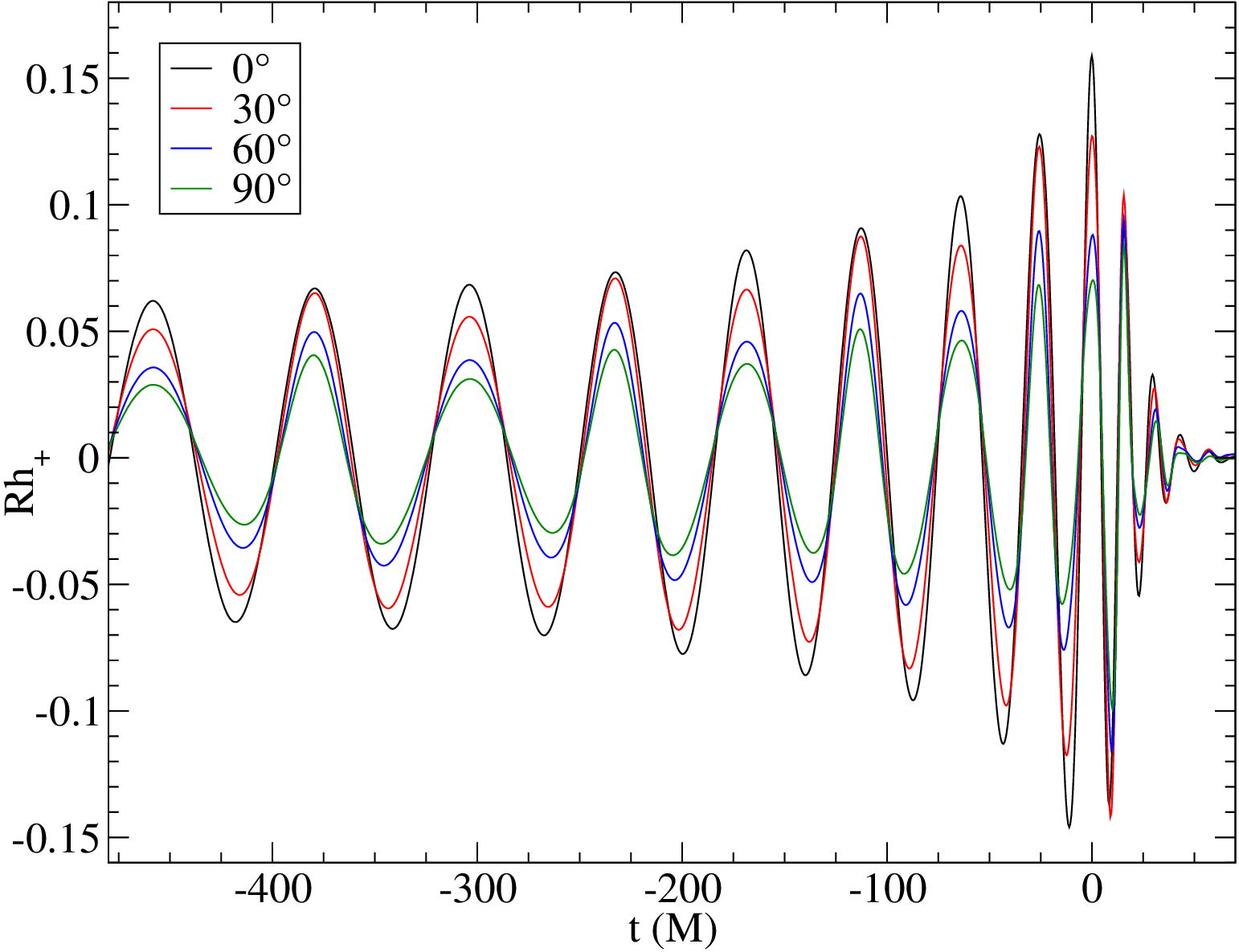}
\includegraphics*[width=3.5in]{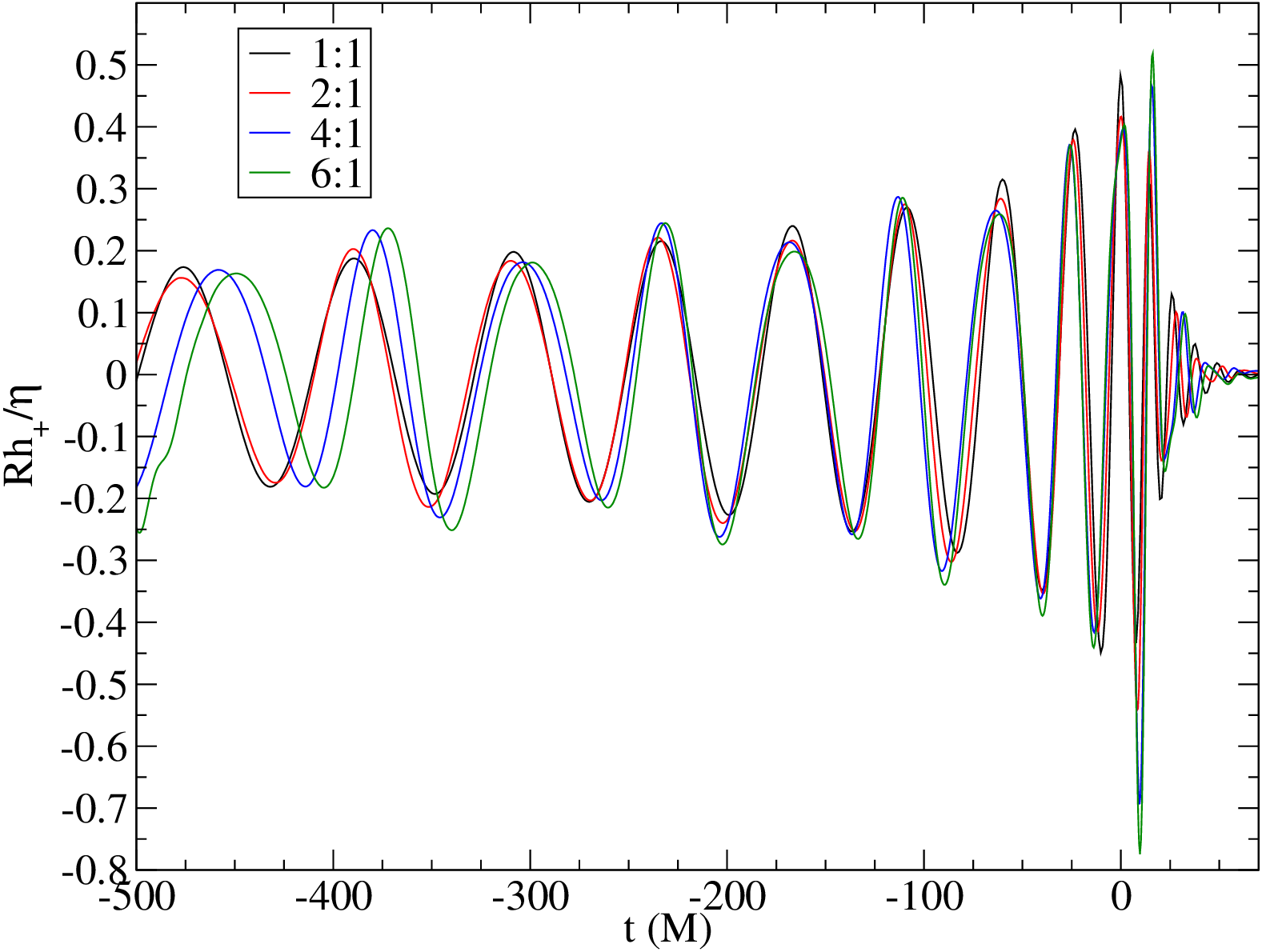}
\caption{Waveform variation
 with inclination.  The left panel shows how the wave shape changes
 with inclination $\theta$ for a 4:1 mass ratio. $h_+$ is plotted,
 using the sum of multipolar modes up to $\ell=5$.  The curves are
 labeled by the inclination in degrees of the observer, and the time
 axis is labeled so that the peak of $h_{22}$ occurs at $t=0$. The
 right panel shows how the wave shape changes with mass ratio. The
 waveforms $h_+$ are computed at an inclination of \degrees{90} from
 the $\ell=2$ and $\ell=3$ modes. The different mass ratios are
 aligned in time so that the peak of $h_{22}$ occurs at $t = 0$. The
 rotational phases are matched to \degrees{0} at $t=0$. In both
 panels, the azimuthal angle is
\degrees{0}.}
\label{fig:waveform_lm}
\end{figure*}

More typically, the observer will not be located on the system's
orbital axis.  The left panel of Fig.~\ref{fig:waveform_lm} shows
$h_+$ for the 4:1 case.  The strain is measured at an azimuthal angle
of \degrees{0} and various
inclinations\footnote{\label{inclination}The inclination angle is
defined here as the angle between the line-of-sight with respect to
the detector and the orbital axis of the binary.  This is the same
angle referred to as ``inclination'' in the PN/NR literature, and most
equations are constructed using that definition.  However, the
astronomical literature has often defined inclination to be the angle
between the line of sight and the orbital plane of the binary,
resulting in a \degrees{90} inconsistency.}. The detailed shapes of
the waveforms change as the system is reoriented so that the observer
moves off the system's rotational axis.  For larger inclinations
(closer to being viewed edge-on) there are notable modulations at half
the base gravitational-wave frequency.

The right panel of Fig.~\ref{fig:waveform_lm} shows $h_+$ for
different mass ratios oriented at an inclination of \degrees{90} and
an azimuthal angle of \degrees{0}.  For this orientation, $h_+$
constitutes the full strain waveform.  For larger mass ratios, the
lower frequency modulation increases.
For gravitational-wave observations of sufficiently strong binary
black-hole sources, the types of differences shown in
Fig.~\ref{fig:waveform_lm} could be exploited to estimate the
inclination and mass-ratio of the source system.

For observational purposes, the combined waveform information encoding
all possible source orientations can be conveniently represented in
terms of spin-weighted spherical harmonic components [see
Eqs.(\ref{eqn:psi4lm_def})-(\ref{eqn:hlm_def})], providing a neat
description of the leading waveform features, The multipolar
decomposition is even more valuable as a tool for exposing the
hallmark simplicity of the merger radiation.  The readily apparent
simplicity in the waves viewed from the system's orbital axis in
Fig.~\ref{fig:waves} extends to each of the spherical harmonic
components. Viewed off axis, these components linearly combine to
yield the more complex appearance of the waveforms in
Fig.~\ref{fig:waveform_lm}.

This characterization of the gravitational radiation from a merging
black-hole binary in terms of circular polarization was first
recognized in the Lazarus project studies \cite{Baker:2002qf}.  In
this picture, the radiation can be represented by a slowly varying
amplitude and a polarization angle; see Eq.~(\ref{eqn:strain_modes})
below.  This description relies on how the radiation appears to
distant observers located on the rotational axis of the system.  Other
observers will typically see elliptically polarized waves, having a
generally simple pattern that conforms to the rotational nature of the
source.  In the equatorial plane, the radiation reduces to the plus
polarization, corresponding to the observer seeing no circulation in
the source.  Looking along the negative rotation axis, the observer
sees circular polarization with the opposite helicity.

Each of the spherical harmonic waveform components exhibits circular
polarization with steadily varying phase and amplitude, providing a
natural framework for developing a practical and intuitive
understanding of binary black-hole merger radiation.  Our basic
waveform description, in Sec.~\ref{sec:description}, and the more
detailed analysis that follows, is based on this spherical harmonic
decomposition.

As we describe the waveforms, we will also suggest a heuristic
interpretation of what the radiation tells us about the motion and
structure of the binary black-hole source. In the weak-field
description of radiation from a rotating object, the multipolar
waveform components of the gravitational radiation can be associated
with dynamics of multipolar moments of the radiating source
\cite{Thorne:1980ru}. It is useful, in conceptualizing the full
coalescence radiation from inspiral through merger and ringdown, to
think of the multipolar radiation description as providing information
about the motion of a changing source object, described as a sum of
several multipolar mass moments. This source object is what we will
interpret as an effective rigid rotator radiation source, with a
slowly changing structure.  We refer to this as the \emph{implicit
rotating source} (IRS).

In the process of coalescence, the source begins as a separated
black-hole binary system and ends as a single distorted black hole.
For nonspinning binary mergers, numerical and PN results consistently
indicate that the radiation is circularly polarized, in the sense
first recognized in the Lazarus project studies \cite{Baker:2002qf},
not only in the inspiral, but uniformly through the merger and
ringdown.  In our conceptual source description, this pattern of
circular polarization is consistent with radiation generated by
rotational motion of each source multipolar moment, where the
polarization phase is tied to the instantaneous orientation of the
source. Similarly, we think of the amplitude of the radiation
multipole as related, through some generalization of the quadrupole
formula, to the magnitude of the implicit source multipole.

We can write each multipolar component in a specific polar form
natural for circularly polarized radiation:
\begin{equation}
\label{eqn:strain_modes}
r h_{\ell m}(t) =
\begin{cases}
H_{\ell m}e^{i m \Phi^{(h)}_{\ell m} (t)} & \left(m>0\right), \\
H_{\ell m}e^{i m \Phi^{(h)}_{\ell m} (t)} \, (-1)^{\ell} & \left(m<0\right).
\end{cases}
\end{equation}
Each amplitude $H_{\ell m}$ is expected to be a slowly varying
function of time and can be conceptually considered as a function only
of the magnitude of the source multipolar moment and its rotational
frequency. The additional sign for some $m<0$ cases allows a
consistent interpretation for the component phases and the component
amplitudes.  The waveform phase is given here by $m \Phi^{(h)}_{\ell
m}$, but in most of our analysis we refer to $\Phi^{(h)}_{\ell m}$,
which we call the \emph{rotational phase}.  In our terms of our
implicit source heuristic, the rotational phase for some particular
$(\ell,m)$ mode can be thought of as the azimuthal orientation of the
specified multipolar source component. In the inspiral, where the
source can be considered as a separated binary, $\Phi^{(h)}_{\ell m}$
should coincide with the orbital phase, independent of $(\ell,m)$.  In
the post-Newtonian expansion \cite{Blanchet06}, all the
$\Phi^{(h)}_{\ell m}$ defined here agree to at least 2PN order, while
the amplitudes $H_{\ell m}$ remain real and non-negative.  The
equatorial-plane symmetry of these mergers ensures that $H_{\ell
m}=H_{\ell{(-m)}}$ and that $\Phi^{(h)}_{\ell
m}=\Phi^{(h)}_{\ell(-m)}$, so that we need consider only $m>0$ modes
for this analysis.

The expansion (\ref{eqn:strain_modes}) is not appropriate for the
$m=0$ modes. This points to an important caveat to our implicit
rotating source interpretation of the radiation, that it applies to
the degree that the radiation is circularly polarized. While not
strictly vanishing, the $m=0$ waveform components, and other
deviations from circular polarization are generally extremely small,
and largely unmeasurable at the resolutions of the simulations we
study.  For the most part, we will not address deviations from
circular polarization, and the consequent limitations of our
implicit-rotating-source interpretation in this paper, focusing for now
on the dominant features of the radiation.

\section{Simulations}
\label{sec:simulations}

Our analysis is based on four simulations, representing mass ratios
1:1, 2:1, 4:1 and 6:1.  Results from the 1:1 and 4:1 simulations have
appeared in previous publications (\cite{Baker:2006kr} and
\cite{Buonanno:2007pf}, respectively). More recently,
higher accuracy simulations have been presented by other groups for
the 1:1 case \cite{Husa:2007hp,Boyle:2007ft}. Our older waveform is
sufficiently accurate for our present purpose, to examine the general
features of the nonspinning merger waveforms.

Our numerical simulations are carried out with the \textsc{hahndol}
evolution code \cite{Imbiriba:2004tp}, which uses finite-differencing
methods to solve a 3+1 formulation of Einstein's equations on a
Cartesian grid.  For initial data we solve the elliptic equation given
by Brandt and Br\"ugmann for conformally flat data in which the black
holes are represented by punctures \cite{Brandt97b}.  This is
performed numerically using the multigrid solver \textsc{amrmg}
\cite{Brown:2004ma}, which is second-order-accurate but tuned to give
truncation errors typically much smaller than those produced by the
evolution code.  The momentum parameters are chosen according to the
2PN-accurate quasicircular approximation given by Kidder
\cite{Kidder95a}, which has been found to result in low eccentricity.
We evolve these data using the moving puncture method
\cite{Campanelli:2005dd,Baker:2005vv} with a modified version of the
Baumgarte-Shapiro-Shibata-Nakamura equations \cite{Shibata95,Baumgarte99}.  Specifically, as
suggested in \cite{vanMeter:2006g2n}, we replaced the conformal factor
variable $\phi$ with $\exp(-2\phi)$, which vanishes at the punctures.
Further, we added the constraint-damping terms suggested in
\cite{Duez:2004uh}, and the dissipation terms suggested in
\cite{Kreiss73,Huebner99}.  For the gauge we use the specific 1+log lapse and
Gamma-freezing shift conditions recommended for moving punctures in
\cite{vanMeter:2006vi}.

Accurate simulations require adequate spatial resolution near the
black holes (length scales $\sim M$) as well as in the wave zone where
the gravitational waves are extracted (length scales $\sim (10-
100)M$).  To this end, the grid has multiple refinement levels,
determined adaptively near the black holes, but fixed in regions
farther away (typically, $|x|>30M$) where the waves are extracted; all
grid refinement is handled within the framework of the software
package \textsc{paramesh} \cite{Paramesh}.  The adaptive mesh
refinement criterion near the black holes is designed to keep the
scale of the square root of an invariantly defined curvature
component, known as the Coulomb scalar
\cite{Beetle:2004wu,Burko:2005fa}, roughly constant with respect to
the grid spacing.  Interpolation in guard-cells between refinement
regions is fifth-order-accurate, coupling with differencing stencils
to yield at least fourth-order accuracy in the bulk.

Spatial derivatives are taken by sixth-order-accurate differencing
stencils, with the exception of advection derivatives, which are
handled by fifth-order-accurate mesh-adapted differencing for greater
stability \cite{Baker:2005xe}\footnote{Sixth-order center-differenced
advection is unstable, and sixth-order lopsided advection is too
costly in terms of \textsc{paramesh} guardcells, which motivated our
particular modification.}. Time integration is performed with a
fourth-order Runge-Kutta algorithm.

The initial configurations of the simulations we analyze are given in
Table~\ref{table:RunData}.  In each case, the initial separation was
chosen to be large enough to result in at least five orbits.  The
finest resolution, $h_f$, ranged from $M/32$ to $3M/224$, as required
to adequately resolve the black hole with the smaller mass in each
case.  The outer boundary was typically at $|x|>1000M$, far enough
away to prevent reflections from reaching the wave-extraction region
during the simulation.
\begin{center}
\begin{table*}
  \caption{\label{table:RunData} Physical and numerical parameters of
  the initial data for all the runs presented. $m_{1,p}$ and $m_{2,p}$
  are the puncture masses of the two pre-merger holes. $r_0$ and $P_0$
  are the initial coordinate separation and (transverse) linear
  momentum, respectively, giving rise to a total initial orbital
  angular momentum $J_0$. $h_f$ is the spatial resolution of the
  highest refinement level for each run.  $\MADM$ is the total energy
  of the initial data.  The total, infinite-separation mass $M$ of the
  system is measured in two ways -- $\MAH$, the sum of the initial
  (apparent) horizon masses of the two holes, and $\Mrest$, the sum of
  the ADM energy and the binding energy from effective-one-body theory
  \cite{Buonanno:1998gg}. Finally, $\eta$ is the resulting symmetric
  mass ratio, as determined from the two holes' horizon masses.  }
\begin{tabular}{c c||rrrrr||cccc}
\hline \hline
Mass ratio& $h_f$ & $m_{1,p}$ & $m_{2,p}$  & $r_0$ & $P_0$ & $J_0^2$ & $\MADM$ & $\MAH$& $\Mrest$&$\eta$     \\
\hline
1:1  & $M/32$   & 0.4872 & 0.4872 & 10.800 & 0.09118 & 0.9847 & 0.9907 &  $\cdots$ & 1.0005 & 0.2500    \\
\hline
2:1 & $3M/160$ & 0.3202 & 0.6504 & 8.865 & 0.09330 & 0.8271 & 0.9889 &
0.9989 & 0.9990 & 0.2228 \\
\hline
4:1  & $3M/224$ & 0.1890 & 0.7900 &  8.470 & 0.06957 & 0.5893 & 0.9929 &  1.0003 & 1.0004 & 0.1601     \\
     &  $M/64$  & 0.1890 & 0.7900 &  8.470 & 0.06957 & 0.5893 & 0.9929 &  1.0003 & 1.0004 & 0.1601     \\
     & $3M/160$ & 0.1890 & 0.7900 &  8.470 & 0.06957 & 0.5893 & 0.9930 &  1.0003 & 1.0005 & 0.1601     \\
\hline
6:1  & $M/64$   & 0.1338 & 0.8490 &  8.003 & 0.05559 & 0.4449 & 0.9942 &  1.0000 & 1.0001 & 0.1226    \\
\hline \hline
\end{tabular}
\end{table*}
\end{center}

We have measured the individual black-hole masses using the
\emph{apparent-horizon mass} $m_i$, a quantity calculated from the
area of each hole's horizon, which we locate using the
\textsc{AHFinderDirect} code \cite{Thornburg:2003sf}. From these
horizon masses, we calculate the symmetric mass ratio $\eta \equiv m_1
m_2/(m_1+m_2)^2$.  This gives the most precise specification of the
actual mass ratio attained in our simulations.  In the text we will
refer to the simulations by the mass ratio (\emph{e.g.} 4:1).

We define the total, infinite-separation, mass $M$ of the system as an
analogue for the total rest mass parameter used in PN studies. We
measure $M$ in two ways:
\begin{equation}
\MAH \equiv m_1 + m_2,
\end{equation}
the sum of the individual BH horizon masses, and
\begin{equation}
\Mrest = \MADM - E_b,
\end{equation}
defined as the difference between $\MADM$, the total energy of the
initial data, and the (negative) binding energy of the binary.  The
binding energy $E_b$ is estimated from an effective-one-body PN
treatment \cite{Buonanno:1998gg}, given the initial angular momentum
$J_0$. The result shows a very close correspondence between $\MAH$ and
$\Mrest$, with differences at the level $10^{-4}$.  For the rest of
this paper, we use $M = \MAH$, except for the 1:1 simulation data,
where $\MAH$ was not available for technical reasons.

In interpreting the late-time radiation, it is valuable to know the
mass and spin of the final Kerr black hole formed by the merger.  We
discuss the state of the final black hole, determined consistently by
several means, in Appendix \ref{appendix:EndState}.

For the 4:1 mass ratio case, we have carried out runs at three
different resolutions in order to assess the quality of the
simulations.  The convergence of the constraints and waveforms is
discussed in Appendix~\ref{appendix:Convergence}.

\begin{figure}
\includegraphics*[width=3.5in]{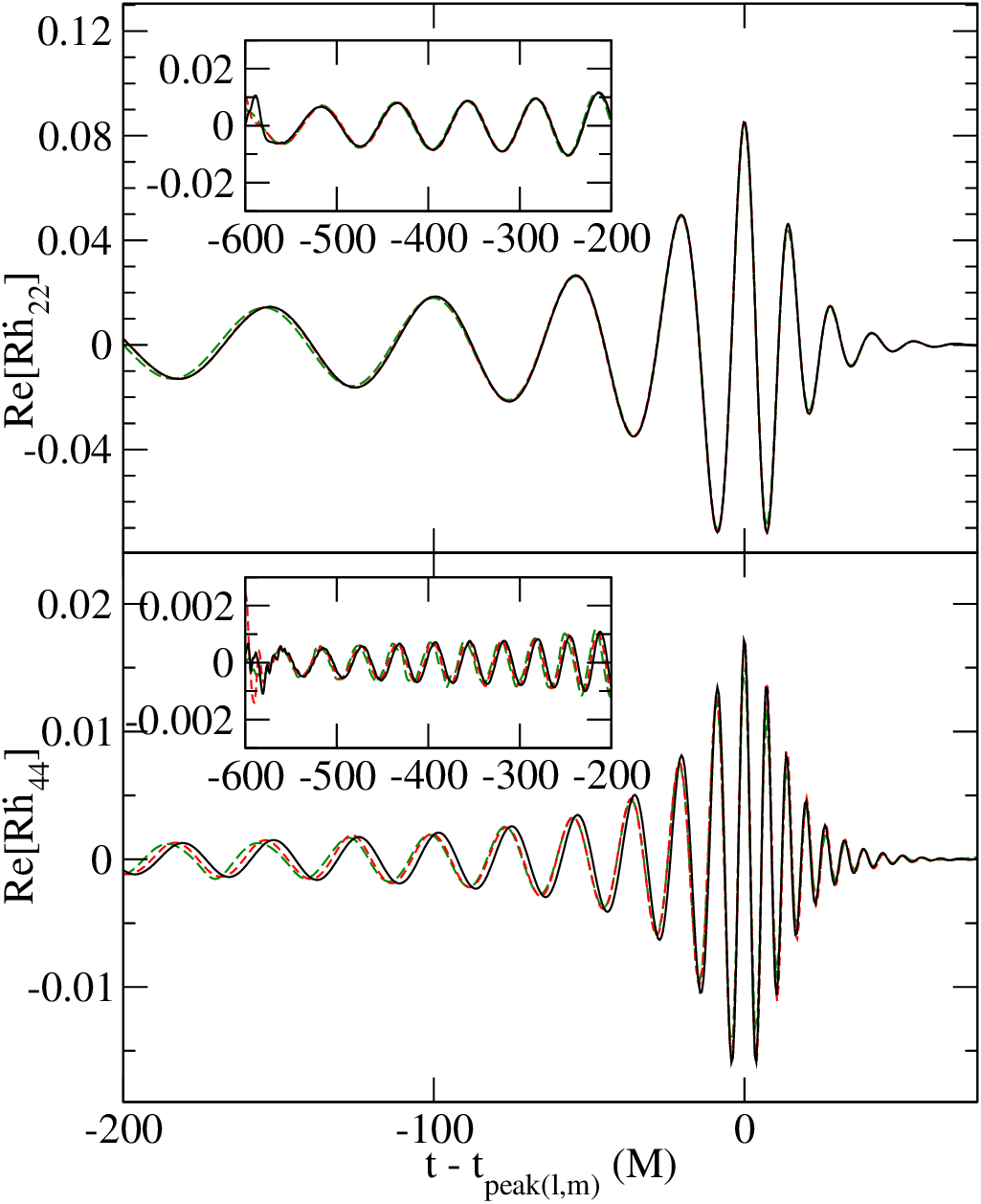}
\caption{The (2,2) (top panel) and (4,4) (bottom panel) strain-rate
waveform modes from the 4:1 mass-ratio case.  Three resolutions are
shown, time- and phase-shifted to match at their respective peak
amplitudes. Excellent agreement is seen to persist throughout most of
the simulation.}
\label{fig:hdot22_44}
\end{figure}

\begin{figure}
\includegraphics*[width=3.5in]{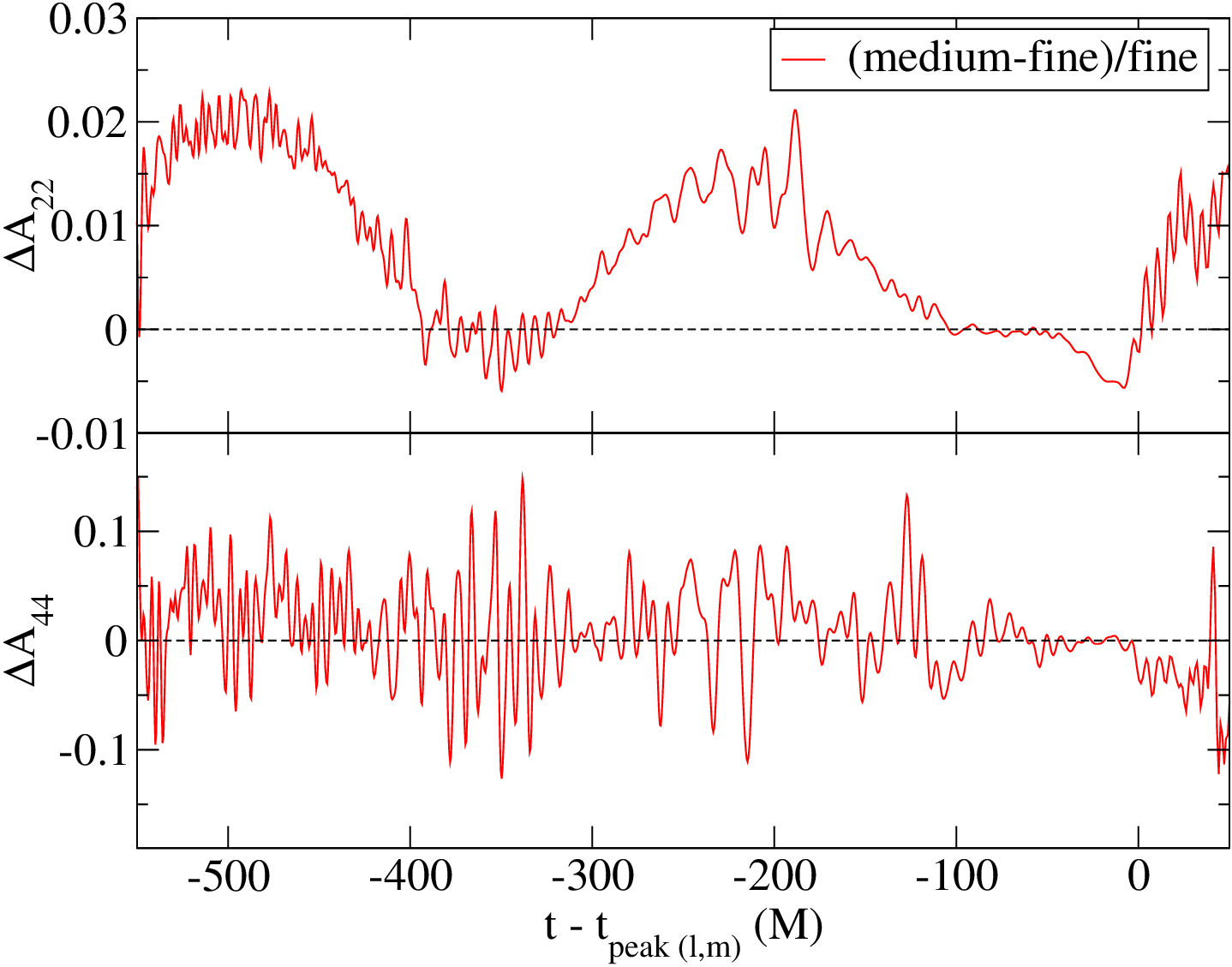}
\caption{The relative amplitude error $(A_{\rm medium}
- A_{\rm fine})/A_{\rm fine}$ for the (2,2) (top panel) and (4,4)
(bottom panel) strain-rate waveform from the 4:1 mass-ratio case.}
\label{fig:amperrorMAQ}
\end{figure}

\begin{figure}
\includegraphics*[width=3.5in]{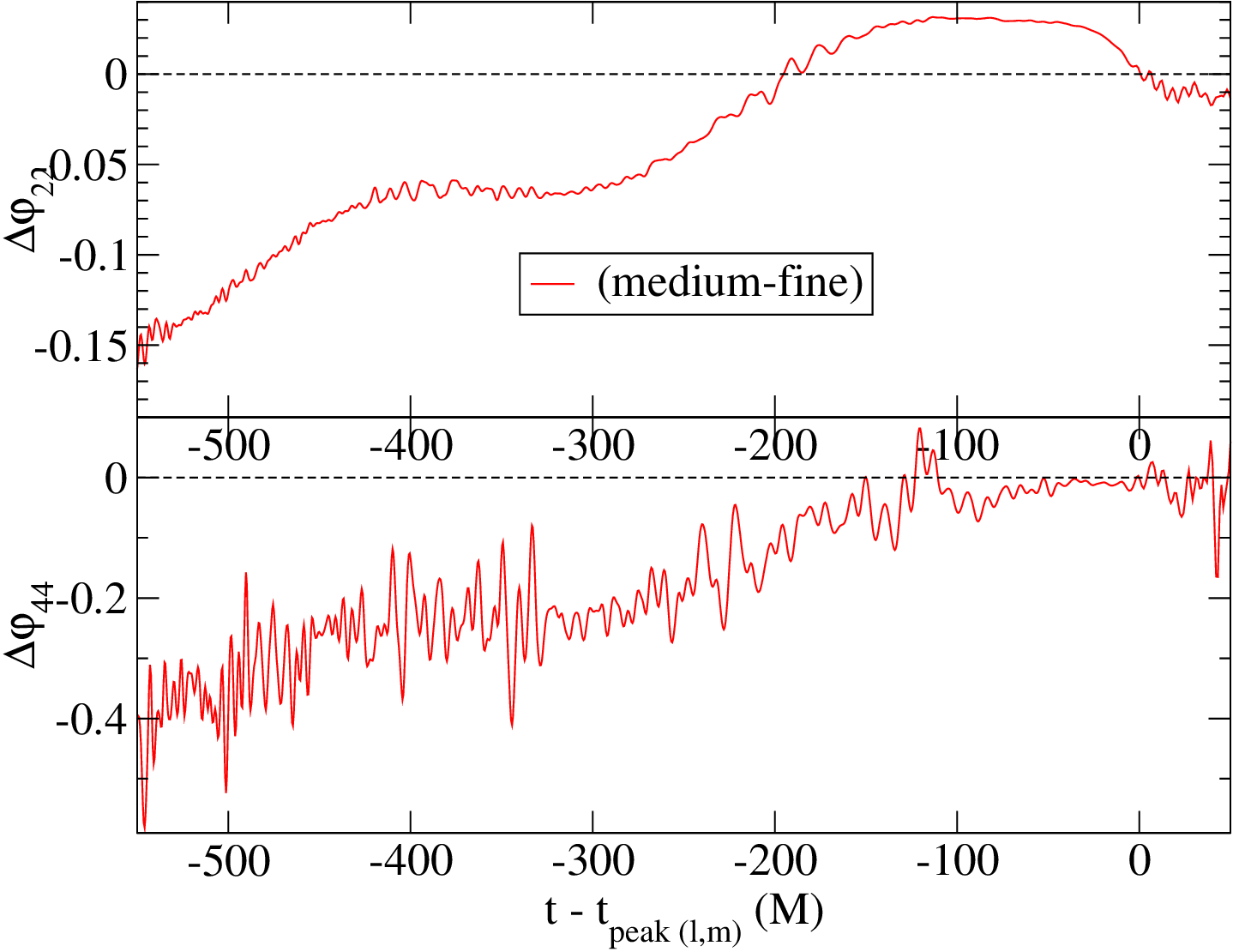}
\caption{The absolute phase error $(\varphi_{\rm medium} -
\varphi_{\rm fine})$ (in radians) for the (2,2) (top panel) and
(4,4) (bottom panel) strain-rate waveform from the 4:1 mass-ratio
case.}
\label{fig:phaseerrorMAQ}
\end{figure}

The most important products of our simulations are the gravitational
radiation waveforms, which we extract from the evolved simulation data
as explained in Appendix~\ref{appendix:Radiation}.  Strain-rate
waveforms for the 4:1 case at various resolutions are shown in
Fig.~\ref{fig:hdot22_44}, where the times and phases have been shifted
to agree at the moment of peak energy flux, as is generally done in
our analysis below.  We can get some measure of the error in the
waveforms by comparing the difference between the high and medium
resolution simulations. Fig.~\ref{fig:amperrorMAQ} shows the relative
differences in amplitudes, scaled by the high-resolution result.
Ignoring the high frequency noise, the (2,2)-mode differences (upper
panel) indicate a combination of a secular amplitude difference and a
sinusoidal effect, which results from the combination of the
eccentricity in the orbital dynamics and the difference in peak time
due to limited resolution.  These combine to give differences
generally at the 3\% level, somewhat smaller at late times. The
eccentricity plays less of a role in the (4,4) differences (lower
panel), as the relative secular error is much ($\sim 5$ times)
larger. Sinusoidal eccentricity effects are also visible in the
phasing error [see Fig.~\ref{fig:phaseerrorMAQ}]. Overall, we find
waveform amplitude and phase errors to be consistent with between
fourth- and fifth-order convergence [see Appendix~\ref{appendix:Convergence}].

Assuming fourth-order convergence, and using Richardson extrapolation,
our nominal expectation for these simulations leads to an error
estimate for the high-resolution simulation applied in our analysis of
$\sim 1.2 \times$ the difference shown in Fig.~\ref{fig:amperrorMAQ}.
To be conservative, we could instead assume second-order convergence,
which would lead to an error estimate of $\sim 2.8 \times$ the
difference shown in Fig.~\ref{fig:amperrorMAQ}.

The errors for the 2:1 case should be comparable to the 4:1 case.  The
resolution for the 6:1, scaled by the smaller black hole's mass is
about 15\% lower than lowest resolution 4:1 simulation, suggesting
errors eight times larger, if we conservatively assume fourth-order
convergence and that the errors around the smaller black hole
dominate.  The errors for the 1:1 mass ratio case are discussed in
Ref.~\cite{Baker:2006kr}.

\section{Descriptive analysis of waveforms}
\label{sec:description}

In this section we provide a descriptive analysis of the waveforms
from our simulations.  We try to serve two purposes in analyzing the
radiation.  In the first place, we are hoping to provide material for
gravitational-wave observers, and others outside the field of
numerical relativity, which makes clear some of the general
characteristics of the radiation from these mergers.  Beyond that, we
also push the analysis in more detail, hoping to generate deeper
insight into the physics which generates the radiation.  Through this
analysis we explore the similarities and differences for the various
mass-ratio simulations, and among the different multipole components
of the 4:1 case case.  In this way, we examine the waveform amplitudes
and energy, and the waveform phasing.  As we proceed, we will
interpret the results in terms of our implicit-rotating-source model,
building up a heuristic description that applies through the inspiral,
merger and ringdown of the binary.

Following a brief discussion of strain rate in \ref{ssec:strainrate},
we study the waveform amplitudes and the associated energetics of the
merger in \ref{ssec:amplitude}. In \ref{ssec:phasing} we address the
polarization phase of several waveform modes, relating them to a
common implicit source phase.  Next, in Sec.~\ref{sec:description2},
we will examine the late-time frequency evolution and the relation to
amplitude in more quantitative detail through the
inspiral-merger-ringdown transition.

\subsection{Strain-rate}
\label{ssec:strainrate}

In the Introduction, we motivated the spherical harmonic phase and
amplitude waveform decomposition with a discussion in terms of strain
$h$.  In analyzing our numerical simulation results, however, we can
work more directly with the \emph{strain rate} $\dot h(t)=d h(t)/dt$
[see Appendix~\ref{appendix:Radiation} for a more detailed
discussion].  As with the strain decomposition
(\ref{eqn:strain_modes}), we will expand the strain rate as
\begin{equation}
\label{eqn:strainrate_modes}
R\dot{h}_{\ell m}(t) =
\begin{cases}
i\,A_{\ell m}e^{i m \Phi_{\ell m} (t)} & \left(m>0\right) \\ -
i\,A_{\ell m}e^{i m \Phi_{\ell m} (t)} \, (-1)^{\ell} &
\left(m<0\right),
\end{cases}
\end{equation}
with $A_{\ell m}$ real and non-negative.\footnote{A similar
expression, differing from (\ref{eqn:strain_modes}) by an overall
sign, would be equally appropriate for direct interpretation of
numerically derived $\psi_4$ waveforms.}  Direct differentiation of
(\ref{eqn:strain_modes}) reveals the relationships between the phases
and amplitudes defined in (\ref{eqn:strain_modes}) and
(\ref{eqn:strainrate_modes}), with $A_{\ell m}\sim
|m|\dot{\Phi}^{(h)}_{\ell m}H_{\ell m}+{\cal O}(5PN)$ while
$\Phi^{(h)}_{\ell m}$ differs from $\Phi_{\ell m}$ only at 2.5PN
order.  Note that the differentiation produces a phase shift of
$\pi/2$, so that strain-rate phases should be defined in the slightly
unconventional form (\ref{eqn:strainrate_modes}) if we wish to
preserve the property that all $\Phi_{\ell m}$ are equal to orbital
phase in the limit of well-separated binaries. This representation
allows a more meaningful comparison of the phases of different
multipolar modes, ensuring that instantaneous phase corresponds to the
orientation of the binary system during the inspiral.

The strain-rate amplitude is directly related to the radiation power
for the $(\ell,m)$ mode by Eq. (\ref{eq:energy-amplitude}),
$\dot{E}_{\ell m}=(A_{\ell m})^2/16\pi$.  Henceforth we shall use the
strain-rate-derived rotational phase $\Phi_{\ell m}$, rather than
$\Phi^{(h)}_{\ell m}$.  We also limit our presentation to the $m>0$
modes, as equatorial symmetry implies $\Phi_{\ell m}=\Phi_{\ell (-m)}$
and $A_{\ell m}=A_{\ell (-m)}$.

Unless otherwise indicated, in the remainder of the paper, the time
axis of each plot will be shifted so that the peak of $\dot E_{2 2}$
(and hence of the strain-rate amplitude $A_{22}$) occurs at $t =
0$. As the (2,2) mode is strongly dominant, this will closely
approximate the peak time of the total $\dot E$.

\subsection{Amplitude and energetics}
\label{ssec:amplitude}

We first study wave amplitudes across modes and mass ratios.  Since we
are examining strain rate waveforms, the modal energy flux is
effectively equivalent to the square of the mode amplitude, as in
Eq. (\ref{eq:energy-amplitude}). Preferring the most physical language
we will express our modal amplitude comparisons as energy flux
comparisons.  In terms of the implicit rotating source model, we can
think of the energy carried by the radiation as energy lost by the
source.

\begin{figure}
\includegraphics*[width=3.5in]{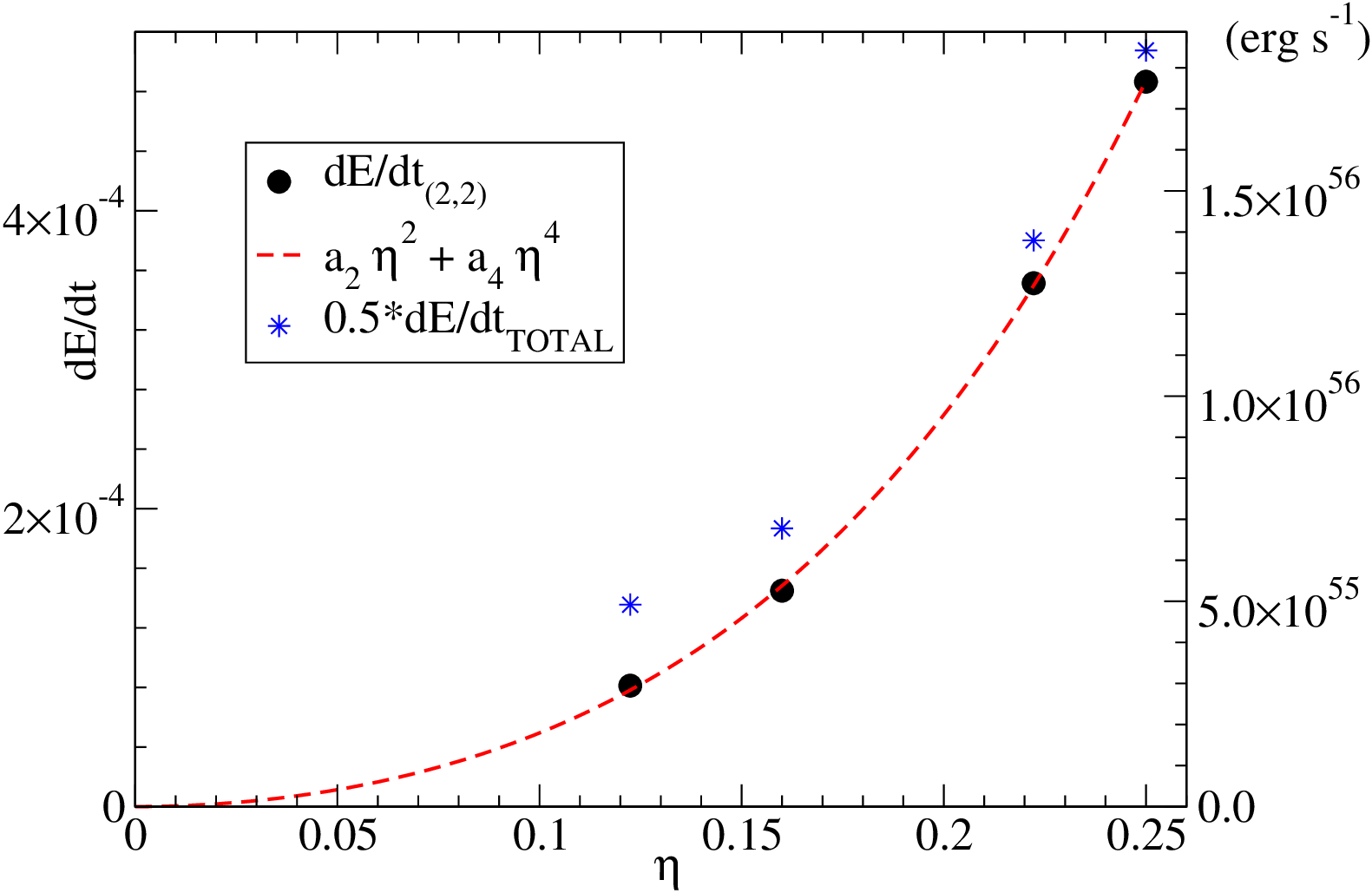}
\caption{The (2,2) mode of the peak power versus mass-ratio $\eta$
(solid circles), with a numerical fit to the quadratic-quartic formula
(\ref{eqn:Edot22_quartic_fit}). Also shown is the full peak power,
divided by two (asterisks). The growing difference between the two as
$\eta \rightarrow 0$ indicates the increasing importance of the
subdominant modes in unequal-mass cases. The right-hand scale gives
the peak power in c.g.s units.}
\label{fig:peak_power}
\end{figure}

In Fig. \ref{fig:peak_power}, we plot the actual peak values of the
dominant (2,2) energy flux contribution from
Eq. (\ref{eq:energy-amplitude}) as a function of symmetric mass ratio
$\eta$. As this mode contribution is proportional to
$|\dot{h}_{22}|^2$, and we expect $\lim_{\eta\rightarrow0}
|\dot{h}_{22}| \rightarrow 0$ (the test-particle limit), we fit it to
a quadratic-quartic form, obtaining the fit\footnote{A quadratic-cubic
form is equally plausible, but fit the numerical data worse in this
case.}:
\begin{eqnarray}
\label{eqn:Edot22_quartic_fit}
\dot{E}_{22} (\eta) &=& (4.40 \pm 0.17) \times 10^{-3} \eta^2 \nonumber \\
 && + (5.43 \pm 0.31) \times 10^{-2} \eta^4.
\end{eqnarray}
We also plot the peak of the total energy flux for each mass ratio,
scaled by one-half, since the (2,2) and (2,-2) modes contribute
equally to $\dot{E}$. The difference between $\dot{E}_{\rm TOTAL}/2$
and $\dot{E}_{22}$ increases as $\eta \rightarrow 0$, reflecting the
increased importance of other modes for unequal masses.

\begin{figure}
\includegraphics*[width=3.5in]{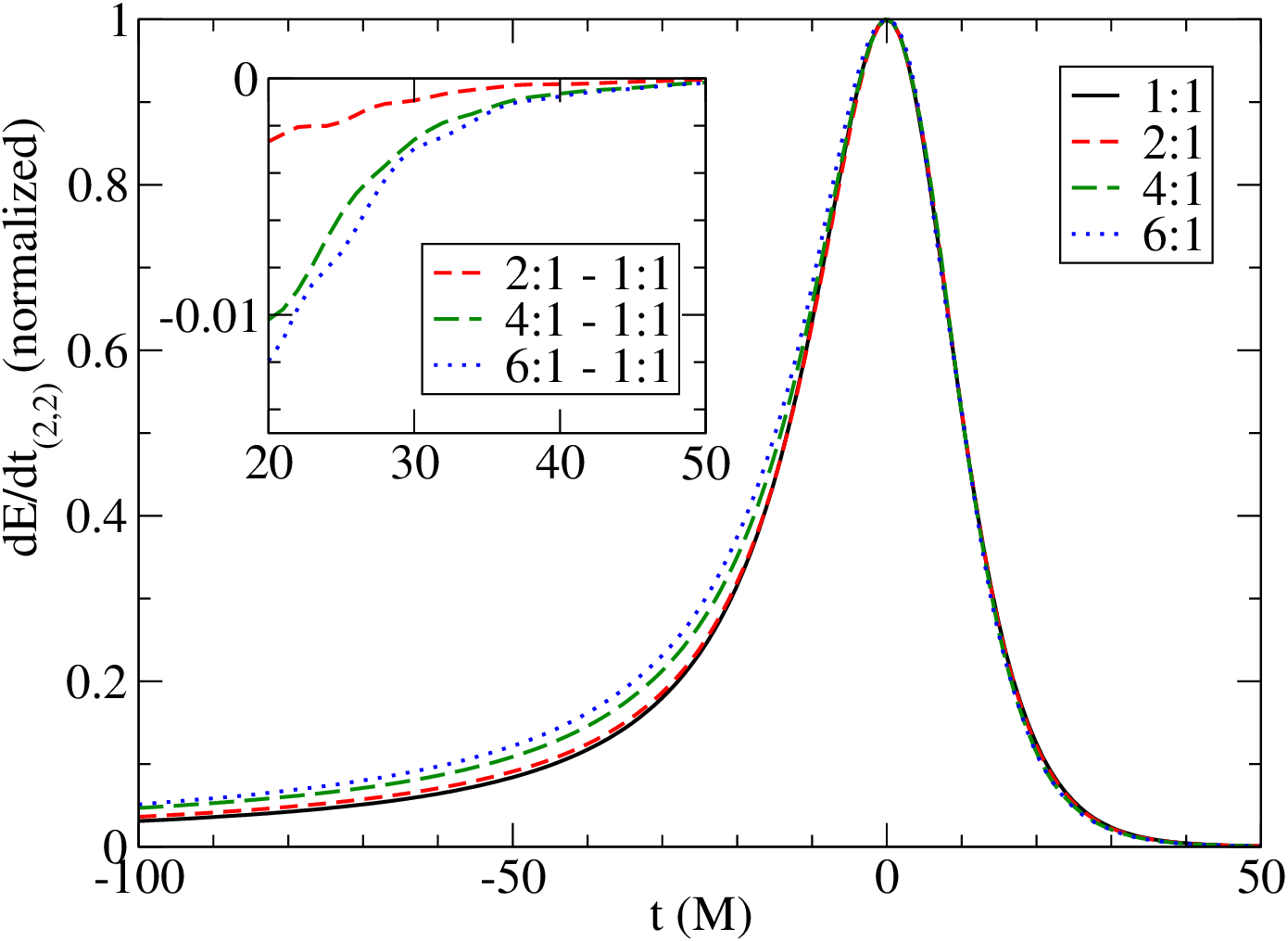}
\caption{The shape of the (2,2) contribution to $\dot{E}(t)$ for all
  mass ratios, scaled to unity at the peaks. We have shifted the time
  axis so that the peak in $\dot{E}_{(2,2)}$ occurs at $t=0$. The
  inset focuses on the ringdown portion of the curves.}
\label{fig:peak_shape_eta}
\end{figure}

Aside from the value of the radiation maxima, it is interesting to see
how the radiation power evolves in time near the peak.  In
Fig. \ref{fig:peak_shape_eta}, we show shapes of the dominant (2,2)
contributions to the peak energy fluxes (\ref{eq:energy-amplitude})
for each mass ratio.  These are scaled to the same peak height to
allow shape comparison, and shifted in time so that the peaks of
$\dot{E}_{22}$ are aligned.  We note the striking similarity of the
peak shape and duration across all mass ratios.  During the
late-inspiral phase, the more extreme mass ratios appear to radiate
more energy; however, since we have normalized each curve by peak
height, this only means that the equal-mass binary experiences a
steeper climb to its peak power rate.  Nevertheless, the different
mass ratios follow similarly shaped tracks approaching merger, and
differences have been effaced by $\sim 10 M$ before the peak
power. The post-peak portion of the curve is determined by the
dominant quasinormal mode (QNM) damping time, which varies only
slightly with underlying Kerr spin for moderate spins [see, for
instance, tables in \cite{Berti:2005ys}]. The lower spins on the black
holes formed by smaller-$\eta$ mergers should cause power to fall off
faster in these cases.  The inset in Fig. \ref{fig:peak_shape_eta}
shows the difference in fall-off rate relative to the equal-mass case
from $20M$ after the peak.

\begin{figure}
\includegraphics*[width=3.5in]{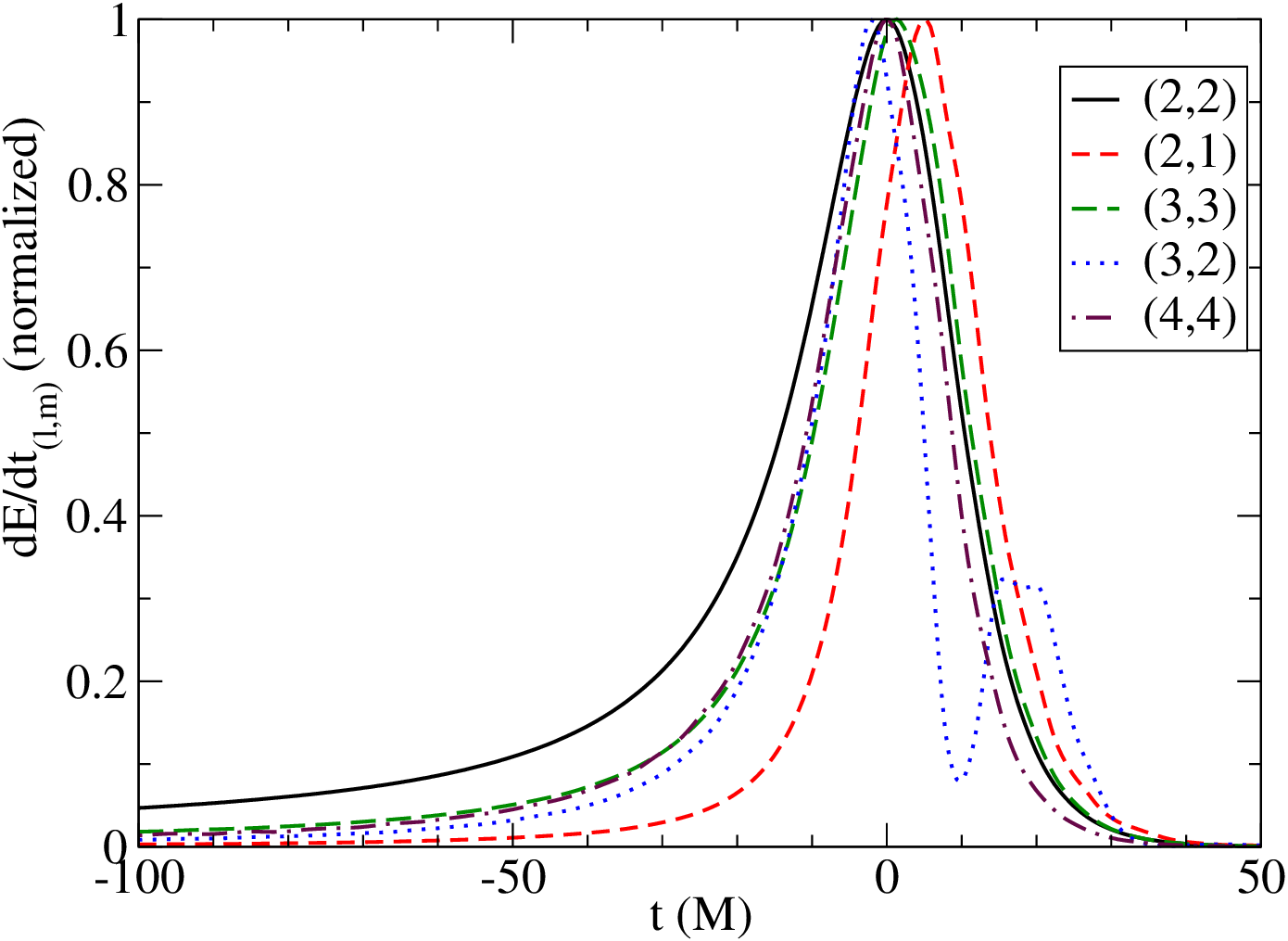}
\caption{The shape of the different multipolar contributions to
  $\dot{E}(t)$ for the 4:1 case, scaled to unity at the peaks. We have
  shifted the time axis so that the peak in $\dot{E}_{(2,2)}$ occurs
  at $t=0$.}
\label{fig:peak_shape_lm}
\end{figure}

In Fig. \ref{fig:peak_shape_lm}, we concentrate on the 4:1 mass ratio,
plotting the strongest modal contributions to the flux, again scaled
to the same peak height. Here we have applied only an overall time
shift, so that the total flux peak is at $t=0$.  The energy profiles
of the radiation burst in modes with $\ell=m$ are similar, all peaking
at approximately the same time.  The subdominant modes are relatively
stronger in the burst than in the inspiral, so that they show up here
as weaker in the approach to the peak even after scaling to the peak
radiation.  We note that the subdominant modes with $\ell=m$, $(3,3)$
and $(4,4)$, are particularly similar in this regard.  Similarity in
$\ell=m$ modes, and distinction in the other modes is a general
feature of the bursts in several ways.

\begin{figure}
\includegraphics*[width=3.5in]{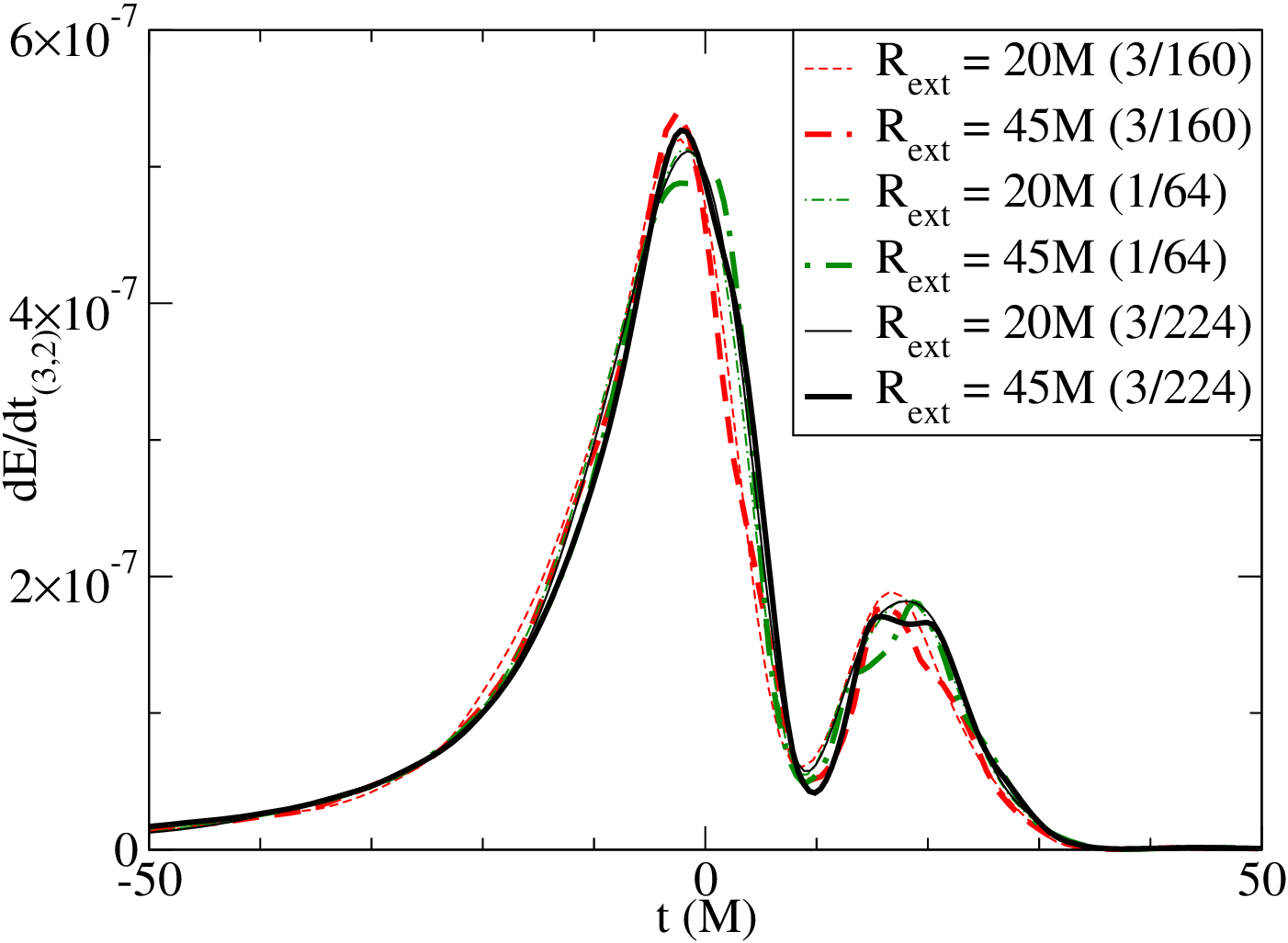}
\caption{The (3,2)-mode ``double-bump'' in $\dot{E}(t)$ for the 4:1
  case.  We show the inner two extraction radii ($20 M$ and $45 M$)
  for each of our three resolutions, shifting the time axis so that
  the peak in $\dot{E}_{(2,2)}$ occurs at $t = 0$. Though there is
  variation in the detailed shape, the overall double-bump envelope
  seems to be robust.}
\label{fig:double_bump_X4}
\end{figure}

\begin{figure}
\includegraphics*[width=3.5in]{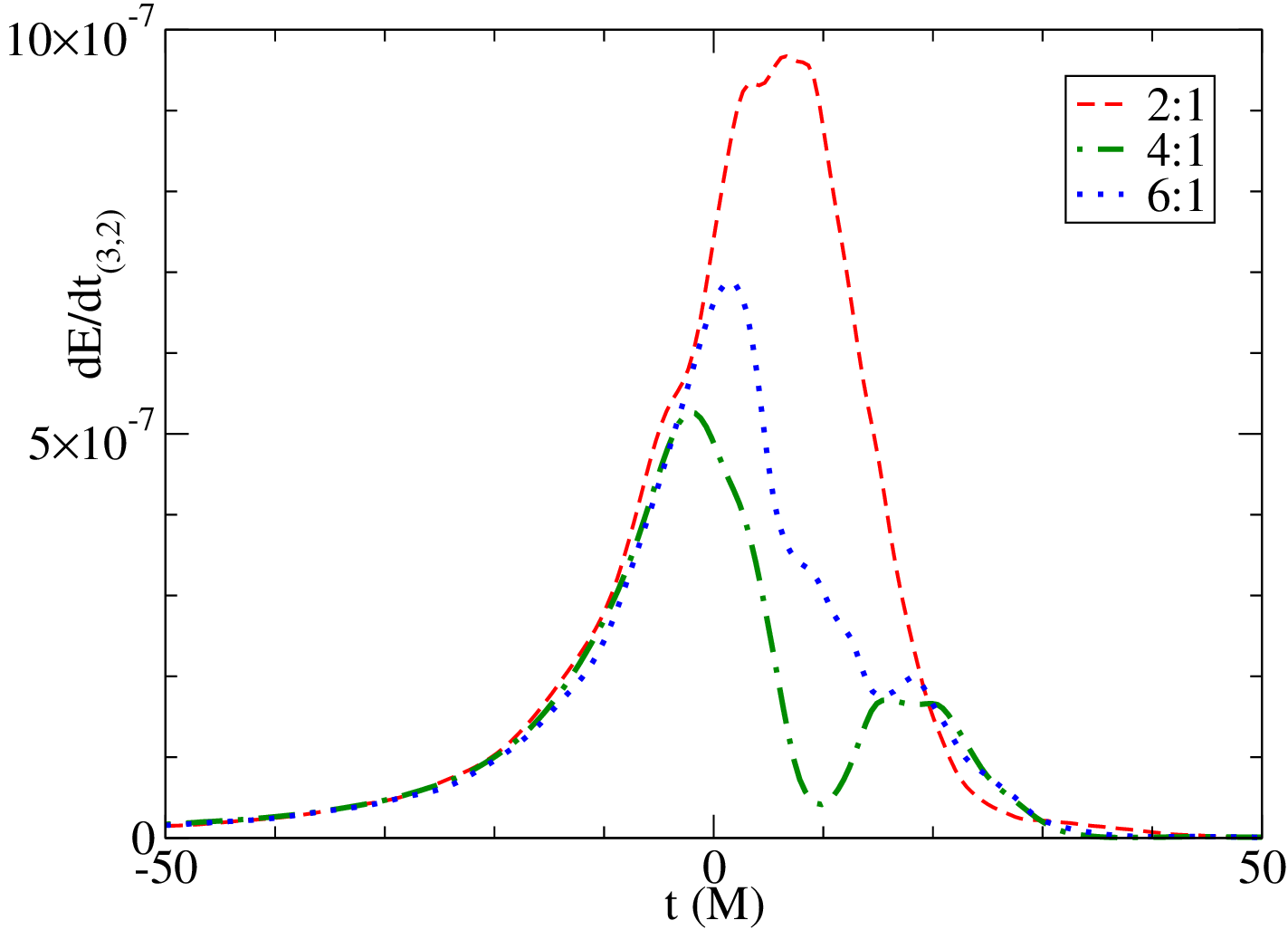}
\caption{The (3,2)-mode in $\dot{E}(t)$ for the 2:1 (red/dashed), 4:1
  (green/dot-dashed), and 6:1 (blue/dotted) cases, extracted at $\Rext
  = 45M$.  We have shifted the time axis of each data set so that the
  peak in $\dot{E}_{(2,2)}$ occurs at $t = 0$. The amplitudes are
  unscaled. It is clear that the 4:1 case has the most pronounced
  deviation from a single well-defined peak.}
\label{fig:double_bump_Xall}
\end{figure}

The shape of the peaks with $\ell\neq m$ are particularly distinct.
The (2,1) mode peaks particularly late, and the burst is much stronger
than the inspiral.  We note that the (3,2) mode shows a double-bump in
its contribution to the energy flux. From
Fig. \ref{fig:double_bump_X4}, this appears to be robust in its gross
shape over resolution and extraction radius.  From
Fig. \ref{fig:double_bump_Xall} we note, however, that the extent of
this double-bump effect is very dependent on mass ratio; it is less in
the 6:1 case, and not evident at all in the 2:1 case. Later we will
also note irregularities in the late-time frequency evolution of this
mode, apparently indicating a deviation from circular polarization in
this case.

\begin{figure}
\includegraphics*[width=3.5in]{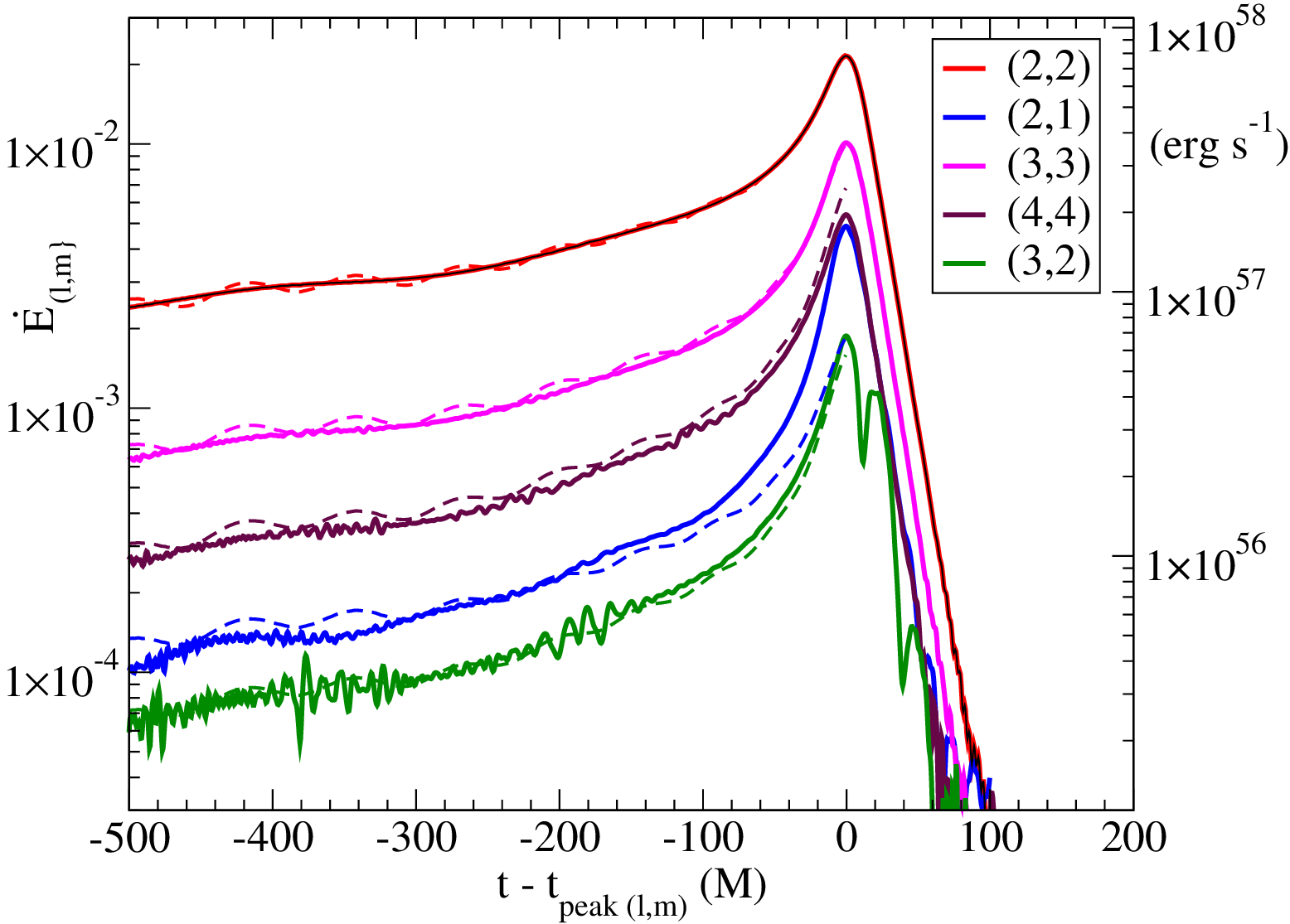}
\caption{Relative power $\dot{E}_{\ell,m}$ for different multipolar
modes of the 4:1 case, on a logarithmic scale (solid lines).  For
comparison, we also plot (dashed lines) the fractional power expected
from each of these modes based on Eq. (\ref{eqn:Edot_modes}), and the
restricted PN waveforms of Eqs. (30)-(36) of
\cite{Buonanno:2007pf}. The different modes have been aligned to peak
at $t = 0$. The right-hand scale gives the power in c.g.s units.}
\label{fig:relative_power}
\end{figure}

In both Fig. \ref{fig:peak_shape_eta} and Fig. \ref{fig:peak_shape_lm}
we have normalized the mode-flux peaks for the purposes of shape
comparison. It is also important to understand the relative strengths
of each mode.  We show in Fig. \ref{fig:relative_power} the relative
mode contributions to $\dot{E}(t)$ for several dominant modes over the
final inspiral and merger of the 4:1 case.  In the merger-ringdown
peaks, as in the inspirals, the $\ell=m$ modes dominate the energy
flux, followed by the $\ell=|m|+1$ modes.  More discussion of the
relative mode strengths for general nonspinning mergers is given in
Ref.\cite{Berti:2007fi}.

In addition to the direct mode contributions, we plot the PN-derived
energy ``partitions'' -- the power emitted in each mode as a fraction
of the total, according to the leading-order ``restricted'' PN
expressions found in Eqs. (30)-(36) of \cite{Buonanno:2007pf}, where
the underlying orbital frequency was derived from the (2,2) mode.
These are shown in Fig. \ref{fig:relative_power} using dashed
lines. Until near the peak time, where we discontinue the PN curves,
we see that the partitioning tracks the numerics very well for all
modes except (2,1), which grows visibly faster than the restricted
amplitude prediction after $\sim 100 M$ before merger.  A similar
study \cite{Buonanno:2007pf} showed that the restricted
(leading-order) approximation for the amplitudes consistently
overestimates the strength of the radiation.  This shows that the
leading-order partitioning of energy can provide a simple, but more
accurate, approximation of the mode amplitudes.  We will take
advantage of this in Sec.~\ref{sec:newEOB} to provide more accurate
amplitudes in a variation on the analytic EOB-based waveform model
studied in \cite{Buonanno:2007pf}.

\subsection{Waveform phasing}
\label{ssec:phasing}

In gravitational-wave observations, the waveform phase provides most
of the time variation in the signals, and consequently is critically
important in encoding observable information about the source. Here we
consider the phasing of the leading spherical harmonic waveform
components. Following the discussion above, we conceptually interpret
each waveform phase as describing the orientation of a particular
$(\ell,m)$ multipole of an implicit rotating source of the
gravitational waves.

Direct comparative analysis of phases provides a stronger probe of the
phase relations among the multipolar modes than the comparative
analysis of frequencies conducted in a number of previous studies of
numerical simulations.  Here we will discuss waveform phasing in terms
of $\Phi_{\ell m}$ as it appears in (\ref{eqn:strainrate_modes}),
which we will compute from each strain-rate mode $\dot{h}_{\ell
m}(t)$. As noted in Sec. \ref{sec:concept}, we expect all phases to
agree in the large-separation limit.

We first compute the strain-rate waveform phase $\varphi_{\ell m}$
using the conventional decomposition:
\begin{equation}
\label{eqn:hdotlm_polar}
R \dot{h}_{\ell m}(t) = V_{\ell m}(t) e^{i \varphi_{\ell m} (t)},
\end{equation}
with $V_{\ell m}$ real and non-negative.  Then, setting this equal to
(\ref{eqn:strainrate_modes}), and solving for the rotational phase
$\Phi_{\ell m}(t)$, we find
\begin{equation}
\label{eqn:phase_modes}
\Phi_{\ell m}(t)=
\begin{cases}
\frac{1}{m}\left(\varphi_{\ell m}(t)-\frac{\pi}{2}+
    2\pi n_{\ell m}\right) & \left(m>0\right) \\
\frac{1}{m}\left(\varphi_{\ell m}(t)+\frac{\pi}{2}+2\pi n_{\ell m}+
    \ell\pi\right) & \left(m<0\right).
\end{cases}
\end{equation}
The $\frac{\pi}{2}$ term results from the factor $i$ in
(\ref{eqn:strainrate_modes}), while the $(-1)^\ell$ factor there
produces the $\ell\pi$ term for $m<0$. The $n_{\ell m}$ terms express
the $2\pi$ ambiguity in defining waveform phase.  Considering any
$(\ell,m)$ mode in isolation leads to an $m$-fold degeneracy in the
associated rotational phase.  We resolve this degeneracy by choosing
the pair $\{n_{22},n_{33}\} \in \{(0,1)\times(0,1,2)\}$ that yields
the closest consistency between $\Phi_{22}$ and $\Phi_{33}$ at early
times (near $t=-400 M$).  We then determine the remaining $n_{\ell m}$
for closest early consistency with $\Phi_{22}$.  This gives us a phase
for each mode that can be interpreted as the rotational phase of the
implicit rotating source that produced that component of the
radiation.

\begin{figure*}
\includegraphics*[width=2.6in]{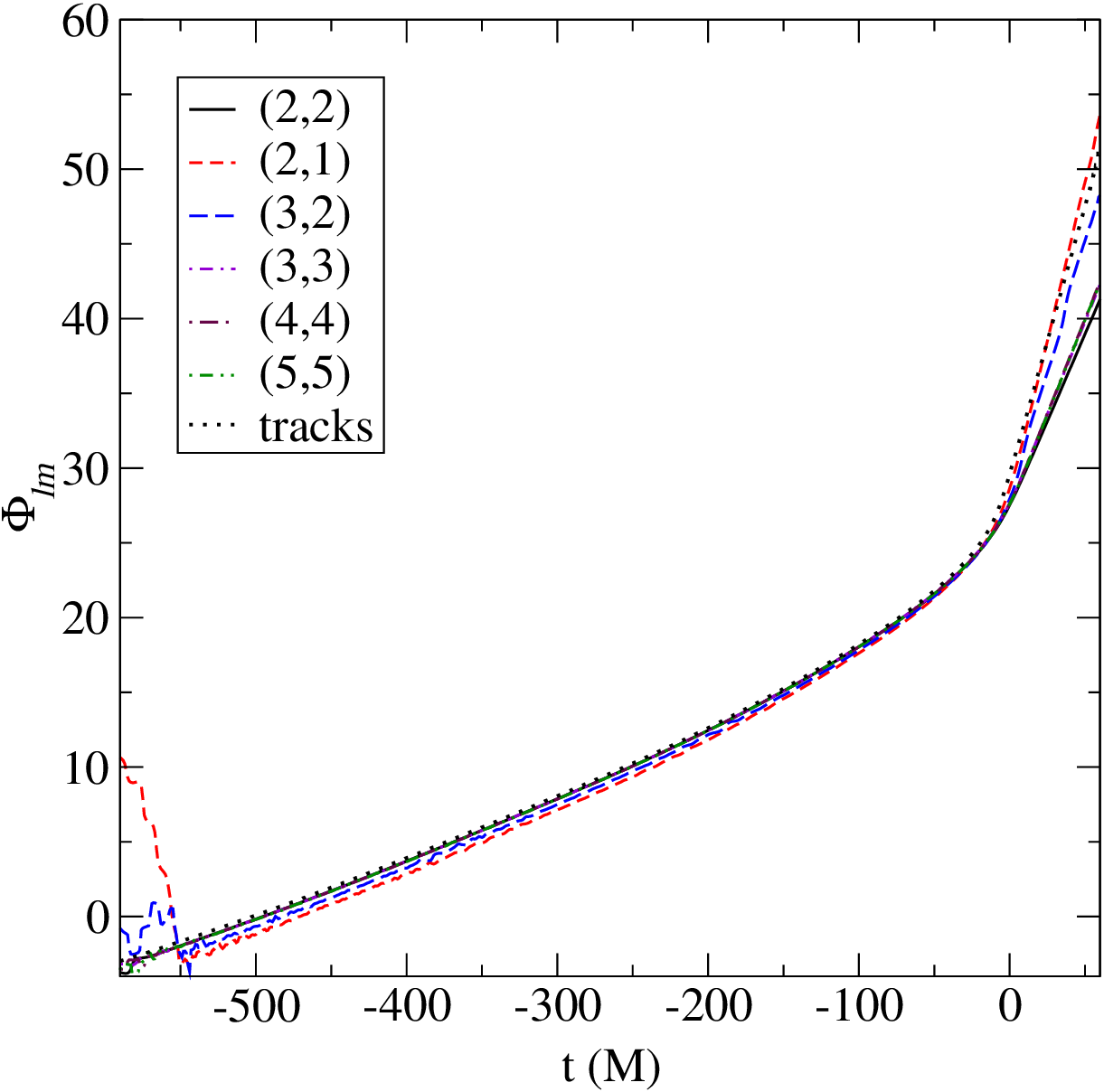}
\includegraphics*[width=4.0in]{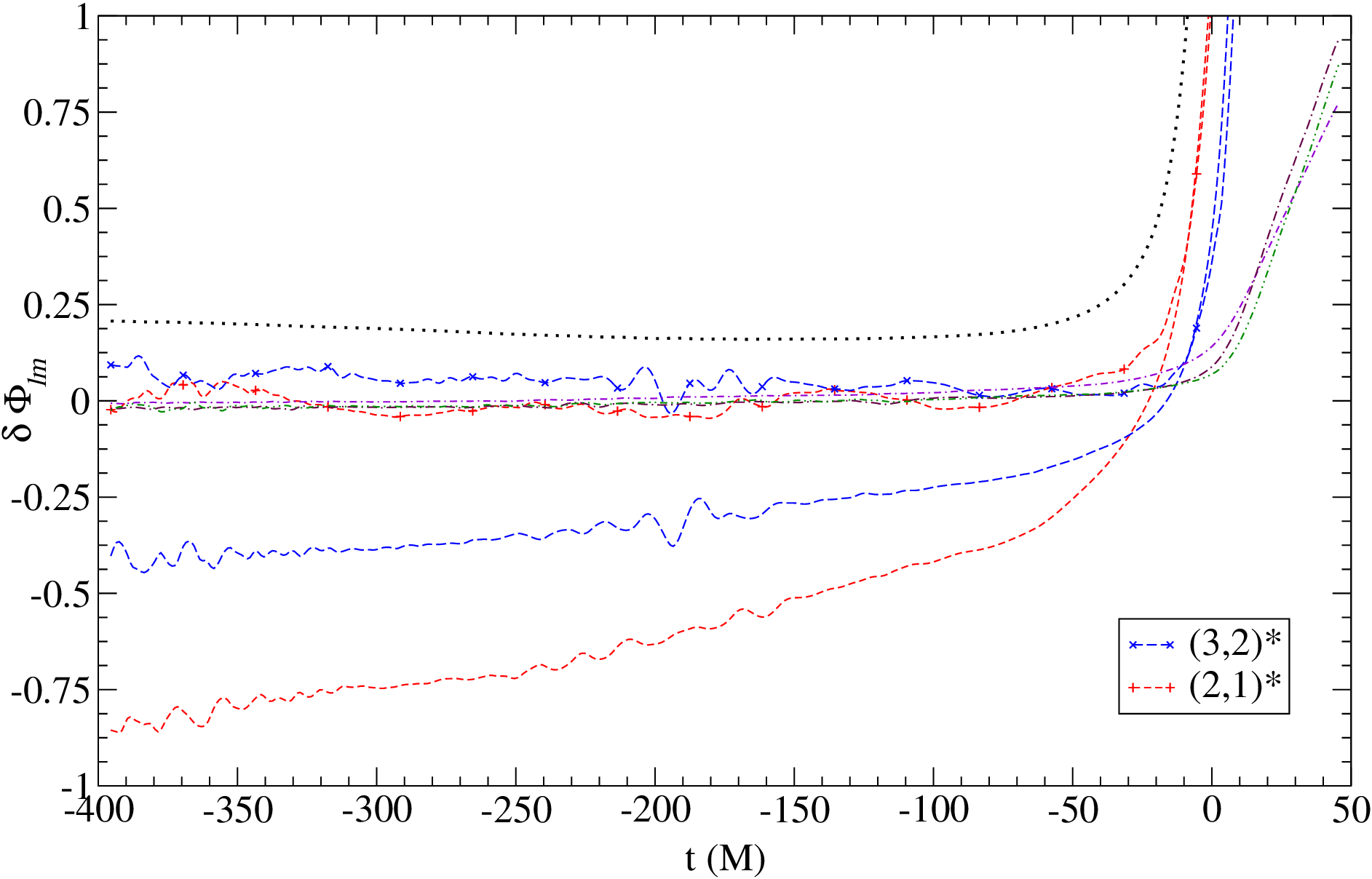}
\caption{Rotational phase $\Phi_{\ell m}$ calculated from puncture tracks and
different multipolar strain-rate modes extracted at 45 $M$ for a 4:1
mass ratio.  The rotational phase is computed from each mode using
(\ref{eqn:phase_modes}) and from the angle of the vector connecting
the two punctures. (The puncture tracks remain in the $x$-$y$ plane.) 
The phases are aligned so that the peak of $\dot E$ occurs at $t=0$
for all phases. The left panel compares the different calculations of
$\Phi_{\ell m}$, while the right panel compares the differences
between the phases and the rotational phase from the (2,2) mode . In
the right panel, we add Richardson extrapolations of the (2,1) and
(3,2) modes based on the 45 $M$ and 90 $M$ extractions of these modes
and assuming an $R^{-2}$ error. The Richardson extrapolations are
distinguished by asterisks in the figure. The differences are smoothed
to better show the trends.}
\label{fig:phase_lm}
\end{figure*}

\begin{figure*}
\includegraphics*[width=3.0in]{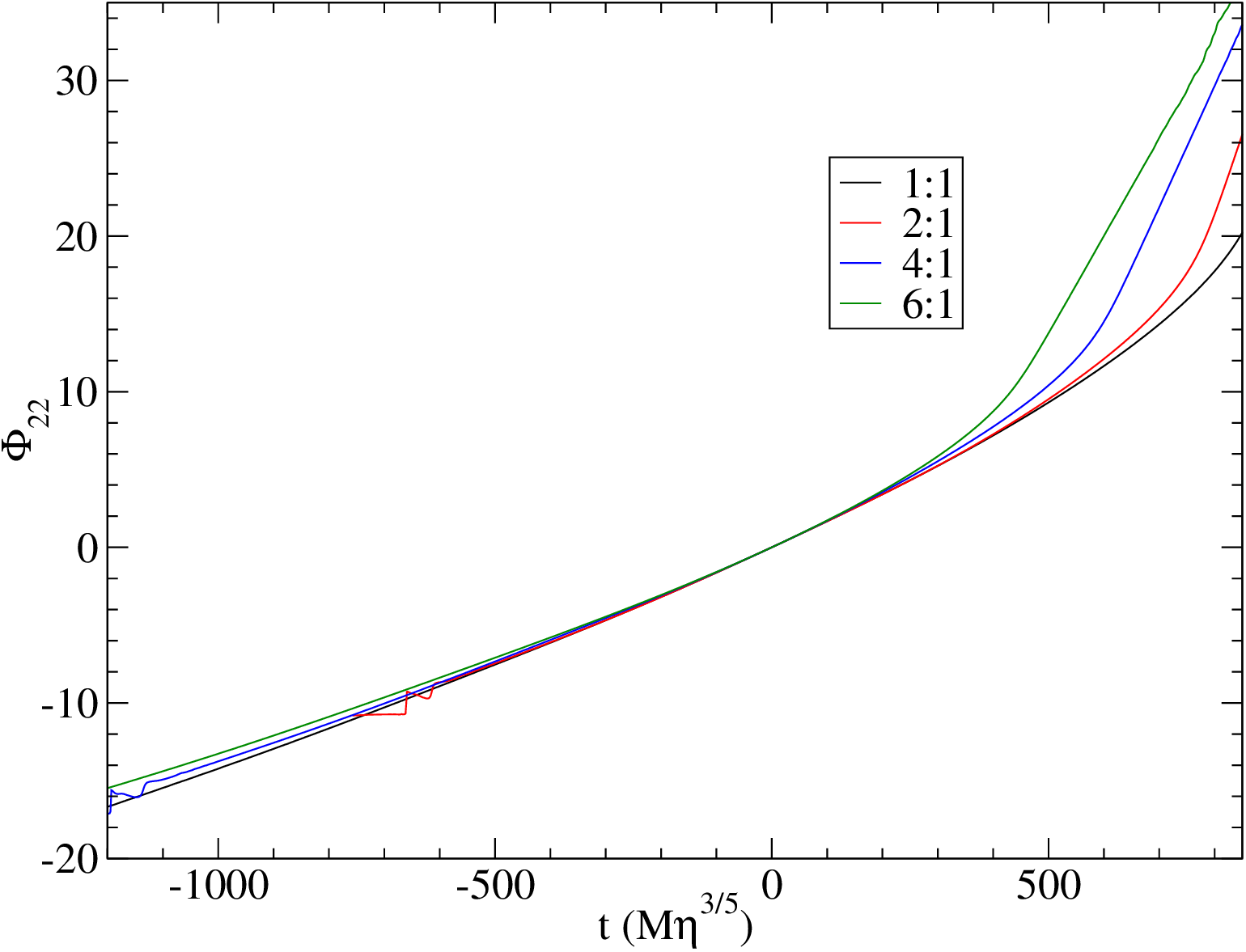}
\includegraphics*[width=3.0in]{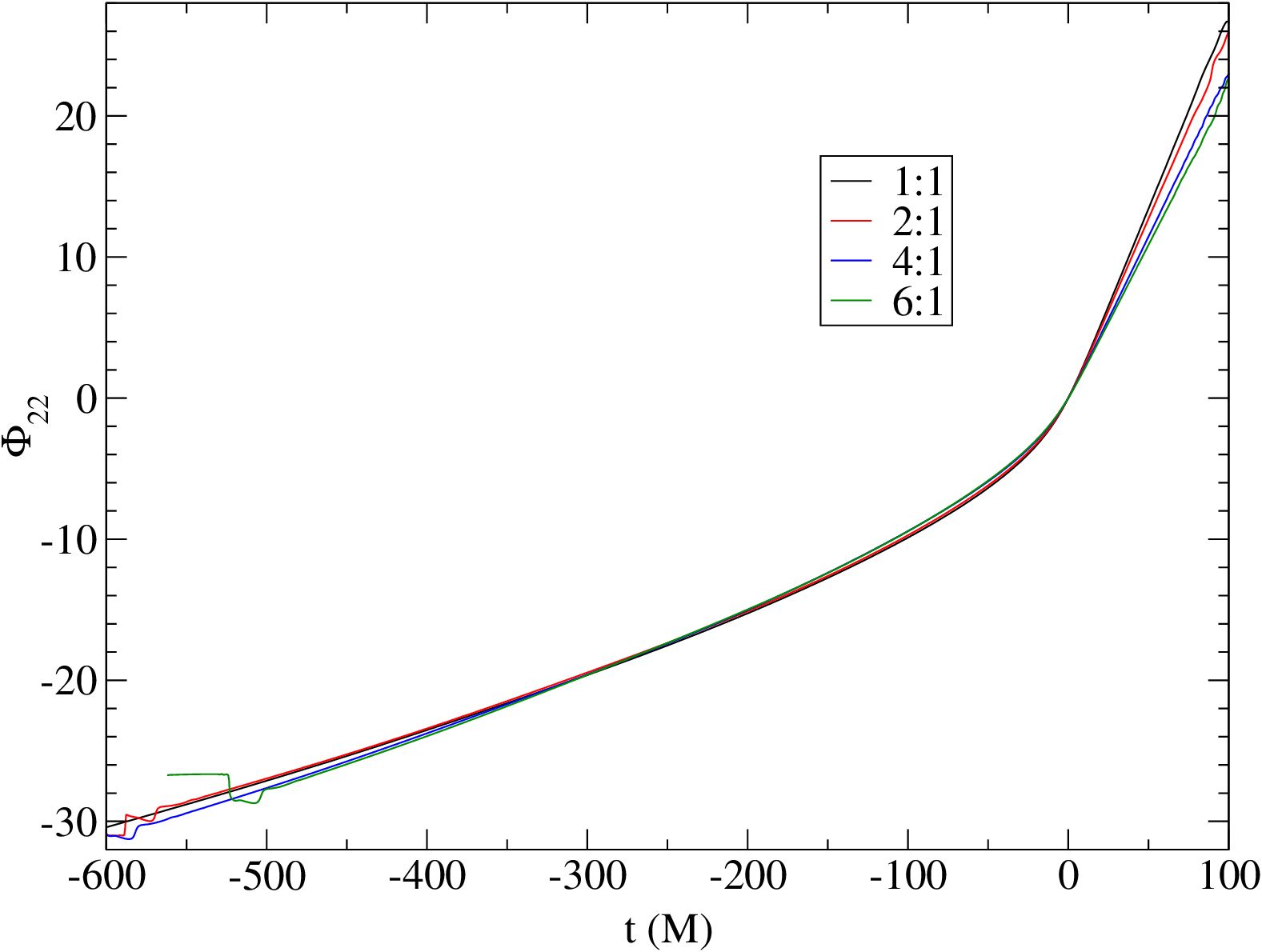}
\caption{Rotational phase $\Phi_{22}$ for different mass ratios, as computed from $\dot h_{22}$.
In the left panel, we time shifted the curves such that the chirp
frequencies $\Omega_{22} \Mchirp = \dot{\Phi}_{22} \Mchirp$ are equal
at $t=0$. In the right panel, we aligned the curves such that the
$\dot h_{22}$ peaks occur at $t=0$, and the phases are 0 at this
time.}
\label{fig:phase_eta}
\end{figure*}

Fig.~\ref{fig:phase_lm} shows the rotational phase $\Phi_{\ell m}$
from several modes of the strain-rate waveforms from the
highest-resolution ($3M/224$) 4:1 run, together with the rotational
phase calculated from the tracks of the punctures.  The left panel
shows that for the inspiral portion of the evolution, all waveform
phases, extracted at $\Rext = 45M$, agree extremely well, except for
the $(2,1)$ and $(3,2)$ modes, which differ by a significant part of a
radian.  The relative difference of each mode from the $(2,2)$ mode is
shown in the right panel.  The differences between the $\ell=m$ modes
are $\lesssim \pi/60$, much smaller than the $2\pi/m$ ambiguity in
defining rotational phase from the waveforms.

Note that the early part of the $(2,1)$ waveforms contain, by far, the
longest wavelength radiation present, suggesting a greater potential
for problems caused by extracting the waveforms too close to the
source, not yet in the wave zone.  For the relatively deviant $(2,1)$
and $(3,2)$ modes, we performed Richardson extrapolation with respect
to extraction radius $\Rext$, using the values extracted at $45 M$ and
$90 M$ and assuming an $\Rext^{-2}$ error for each mode.  These
Richardson-extrapolated phases are also shown in the right panel of
Fig.~\ref{fig:phase_lm}, subtracted from the (2,2) rotational
phase. Richardson extrapolation evidently reduces the early phase
deviations in these modes considerably.

The rotational phases calculated from the $\ell=m$ modes agree to
within 0.025 radians during the inspiral for several hundred $M$
before the peak $\dot E$, effectively identical within the
uncertainties of the numerical approach. This phase agreement is
consistent with expectations based on the PN analysis, for which all
$(\ell,m)$ phases should agree up to 2PN order.  These rotational
phases also agree to within about $0.2$ radians with the
coordinate-dependent rotational phases measured from the puncture
tracks, after shifting the puncture track phase by an overall factor
of $\pi/2$.  Heuristically, we can think of each $(\ell,m)$ radiation
waveform mode as having been generated by the rotation of its own
implicit source component.  The phase agreement would then be
interpreted as indicating that these implicit source components remain
aligned through the inspiral. This is to be expected for a system
which can be effectively described as an orbiting pair of point
particles.

It is, perhaps, more remarkable that a very tight agreement among the
$\ell=m$ mode persists throughout the merger and even into the
ringdown, remaining within about 1 radian until $\sim 50M$ after the
merger, when the amplitude has already diminished
significantly. According to our interpretation, this phase agreement
suggests that a significant portion of the implicit rotating radiation
source maintains some structural integrity throughout the coalescence.
That is, the implicit source we have considered appears to exhibit
considerable ``rigidity'' through merger.  This is only possible
because of the close relationship among the fundamental $\ell=m$
quasinormal ringdown frequencies [see Sec.~\ref{ssec:TransFreq}],
mimicking the harmonic frequency relationship that holds during the
inspiral.  For the $\ell\neq m$ modes this quasinormal frequency
relation does not hold and the phases must separate in the merger.  In
terms of our implicit source picture, these $\ell\neq m$ components of
the source seem to shear away from the main source structure to rotate
at a faster rate.

Heuristically, the puncture motion is strongly tied to the rotation of
the implicit source for most of the evolution.  During this period, it
is natural to think of the implicit source as an inspiralling pair of
pointlike objects moving on timelike world lines.  At late times the
orientation phase angle of the puncture track disassociates from the
waveform rotational phase.  The punctures veer away from the implicit
source at a late times as they fall into the final black hole. At this
point, though we can continue to consider an implicit rotating
radiation source, it no longer makes sense to think of that source as
a pair of pointlike objects.

Having compared the phases of different multipolar modes for the 4:1
case, we now consider how the phase evolution depends on mass-ratio.
There are various reasonable approaches to comparing the phases among
simulations of the different mass-ratio cases.  Having established
above the rotational phase consistency for the different $(\ell,m)$
modes, we will compare only the dominant $(2,2)$ phases.  An obvious
approach is to compare phases directly against time, scaled by the
total PN mass $M$.  In the early-time well-separated
limit, however, the leading-order PN analysis indicates that phases
for different mass ratios should evolve at similar rates when time is
scaled by the chirp mass $\Mchirp \equiv M \eta^{3/5}$
 
Fig.~\ref{fig:phase_eta} shows the rotational phase computed from the
(2,2) mode of strain rate for different mass ratios. In the left
panel, we align the rotational phases at an early time and scale time
by the chirp mass $\Mchirp$. For this plot, we shift the rotational
phases in (chirp) time so that at $t=0$, the chirp frequency, which is
rotational frequency multiplied by chirp mass, is 0.033 and the
rotational phase is 0.  Following this approach, we would expect good
phase agreement at times sufficiently early that only the
leading-order PN effects are significant.  However, for the late
portion of coalescence that we have simulated, we find that the
different mass ratios remain roughly in phase for several hundred
$\Mchirp$ before and after $t=0$, peeling away in order at late times.

In the right panel we compare phases in a manner common for numerical
relativity waveform comparisons, we shift the curves in time so that
each peak energy flux occurs at $t=0$, and we rotate the phases so
that the phases are 0 at this time. We scale the time by $M$. For the
equal-mass case $\Mchirp = 0.435275$, and for the other mass ratios it
is smaller, so all of the curves in the left panel are stretched by at
least a factor of $1/\Mchirp \approx 2.3$ in time relative to the
curves in the right panel.  In the $M$-scaled right panel, the
different mass ratios again remain approximately in phase for several
hundred $M$ before and after $t=0$.  At sufficiently late times, and
particularly for small $\eta$, we might expect this manner of
consistent phasing as the evolution of the system eventually [after
the innermost stable circular orbit] may become dominated by
the course of unstable geodesic trajectories around the larger black
hole [or an effective black hole in the effective-one-body (EOB)
framework]. In that case the frequency evolves independently of the
more strongly $\eta$-dependent rate of energy or angular momentum
loss.

\section{Detailed late-time analysis}
\label{sec:description2}

In Sec.~\ref{sec:description} we have presented general information
about the phasing and amplitudes of the radiation components. Our
analysis has stayed close to the standard numerical relativity
waveform analysis, though we have emphasized an interpretation in
terms of an implicit rotating source model. In this section we go
beyond the standard waveform presentation, exploring the radiation
with the hope of developing a deeper understanding of the simple
characteristics of the radiation as described above. Those features
and our heuristic interpretation suggest a new approach to examining
the structure of the late-time phasing and apparent relationships
between frequency and amplitude evolution.

In Sec.~\ref{ssec:TransFreq}, we examine the phasing again, seeking a
quantitative understanding of the late-time evolution of the
polarization frequency. We introduce a practical model that captures
the merger-ringdown transition without the need for multiple
quasinormal mode overtones. We investigate the implications of this
model in relating the frequency and amplitude close to merger in Sec.
\ref{ssec:FreqAmp}.

This section is more technical than Sec.~\ref{sec:description}, with
some subtle discussion of late-time radiation characteristics.  For
readers who may wish to jump ahead to Sec.~\ref{sec:newEOB}, we note
two results that we will carry forward: (1) a simple quantification of
the peak chirp rate $\dot\omega$ for $\ell=m$ modes, and (2) the idea
that $dJ/d\omega$ becomes approximately constant at late times, which
may serve as summary of relationships between frequency and amplitude
near the radiation peak.

\subsection{Waveform frequency evolution}
\label{ssec:TransFreq}

\begin{figure}
\includegraphics*[width=3.5in]{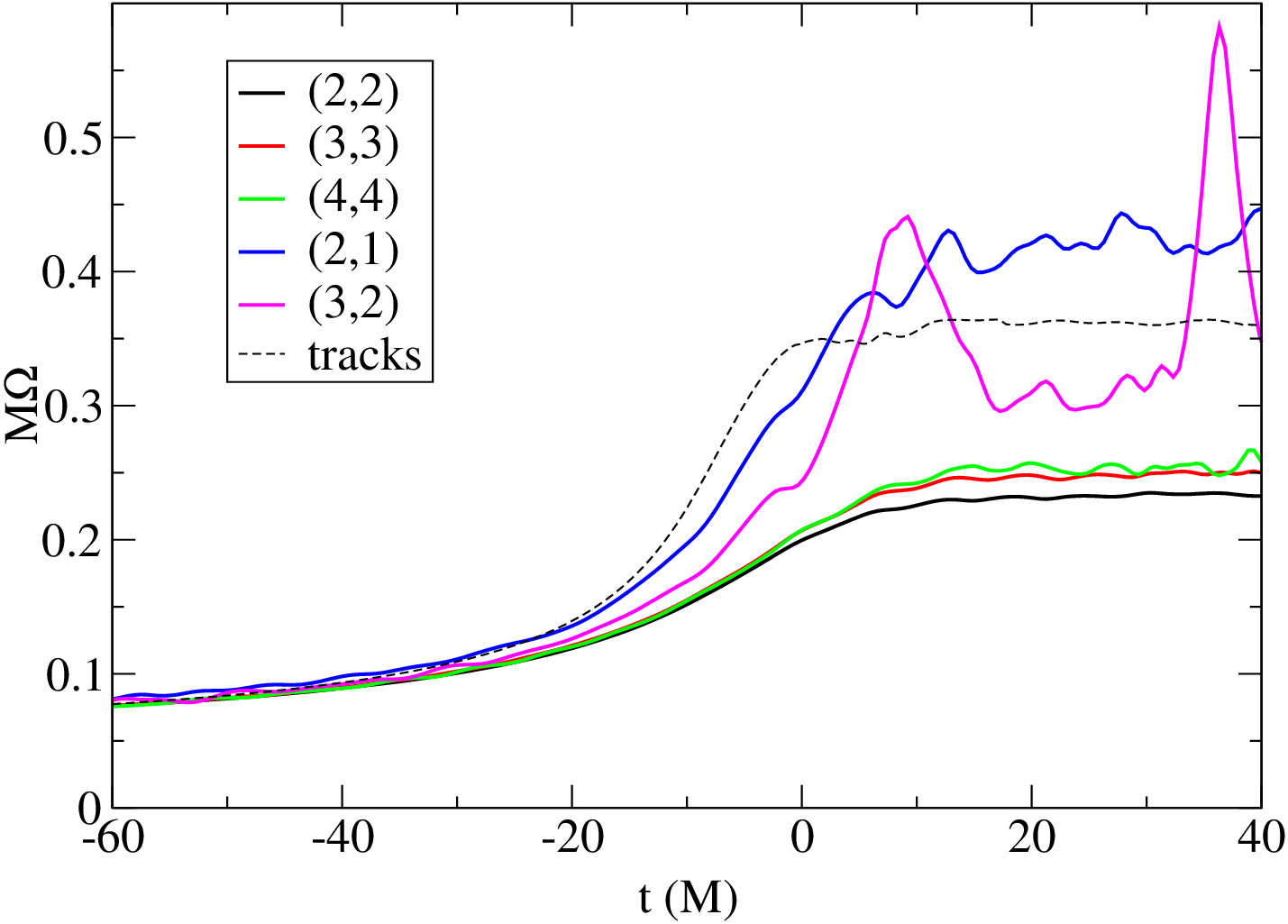}
\caption{Rotational frequency for several $(\ell,m)$ modes of the 
 4:1 mass ratio simulation.  These are time derivatives of the curves
 in Fig.~\ref{fig:phase_lm}.  For $\ell=m$ the rotational frequencies
 remain similar from inspiral through ringdown.  These are similar to
 the puncture track frequency until $t\sim-40M$, while the $(2,1)$
 mode begins to deviate somewhat earlier.}
\label{fig:trans_freq_X4_lm_nofit}
\end{figure}

First we will study the phasing in the merger and ringdown in more
quantitative detail by comparing the polarization frequency evolution
for each mode with a simple empirical model.  Based on our heuristic
model, we interpret the polarization frequency as corresponding to the
rotational frequency $(\ell,m)$ component of the implicit source.

Fig.~\ref{fig:trans_freq_X4_lm_nofit} shows the evolution of the
rotational frequency for several $(\ell,m)$ modes of the 4:1
mass-ratio case.  Similar frequency evolution curves are a common
feature in papers on numerical relativity waveforms [see, for example,
Refs. \cite{Baker:2006yw,Buonanno:2006ui,Berti:2007fi}].  These curves
are time derivatives of the phase evolution curves shown in
Fig.~\ref{fig:phase_lm}, zoomed in on the late-time behavior, near the
elbows in the phase curves.  The striking similarity in phasing for
the various $\ell=m$ modes implies similar frequency evolution, which
has been noted in previous studies \cite{Schnittman:2007ij}.

At late times, this similarity in frequency is made possible because
of a special approximate relationship among the fundamental $\ell=m$
quasinormal ringing frequencies, that they are nearly equal after
dividing by the azimuthal mode number $m$ to get what we call the
rotational frequency.  This has been considered in \cite{Mashoon85},
which pointed to a connection between the quasinormal ringing
frequencies and the frequencies of stable null orbits of a black hole
at the ``light ring'' with frequency $\Omega_{\rm
LR}=1/(a+M(r_+/M)^{(3/2)})$.  The association extends to charged
Kerr-Newman black holes and has been compared with recent precise
quasinormal ringing frequency calculations in \cite{Berti:2005ys} and
\cite{Berti:2005eb}.  Conceptually, this allows us to think of the
rotational frequency of the $\ell=m$ modes at late times as
corresponding to the rotational rate of gravitational perturbations
orbiting at the ``light ring''. This suggests a heuristic description
of our implicit rotating source at late times as a gravitational
distortion of the forming final black hole which predominantly
revolves around the black hole on null orbits at the light ring.

Returning to Fig.~\ref{fig:trans_freq_X4_lm_nofit}, we note that the
$(3,2)$ mode is different from all the others, showing two spikes,
near $t=9 M$ and $37M$.  Comparisons of waveforms extracted at
different radii, and from simulations of different resolutions,
suggest some sensitivity to extraction radius, but do not suggest that
the features will vanish in more accurate simulations or with more
distant wave extraction.  These anomalies may be related to the
unusual shape in the amplitude peaks noted in
Fig.~\ref{fig:double_bump_X4} above.  We will discuss this mode's
behavior further in Sec.~\ref{sec:discussion}.

\begin{figure*}
\includegraphics*[width=3.5in]{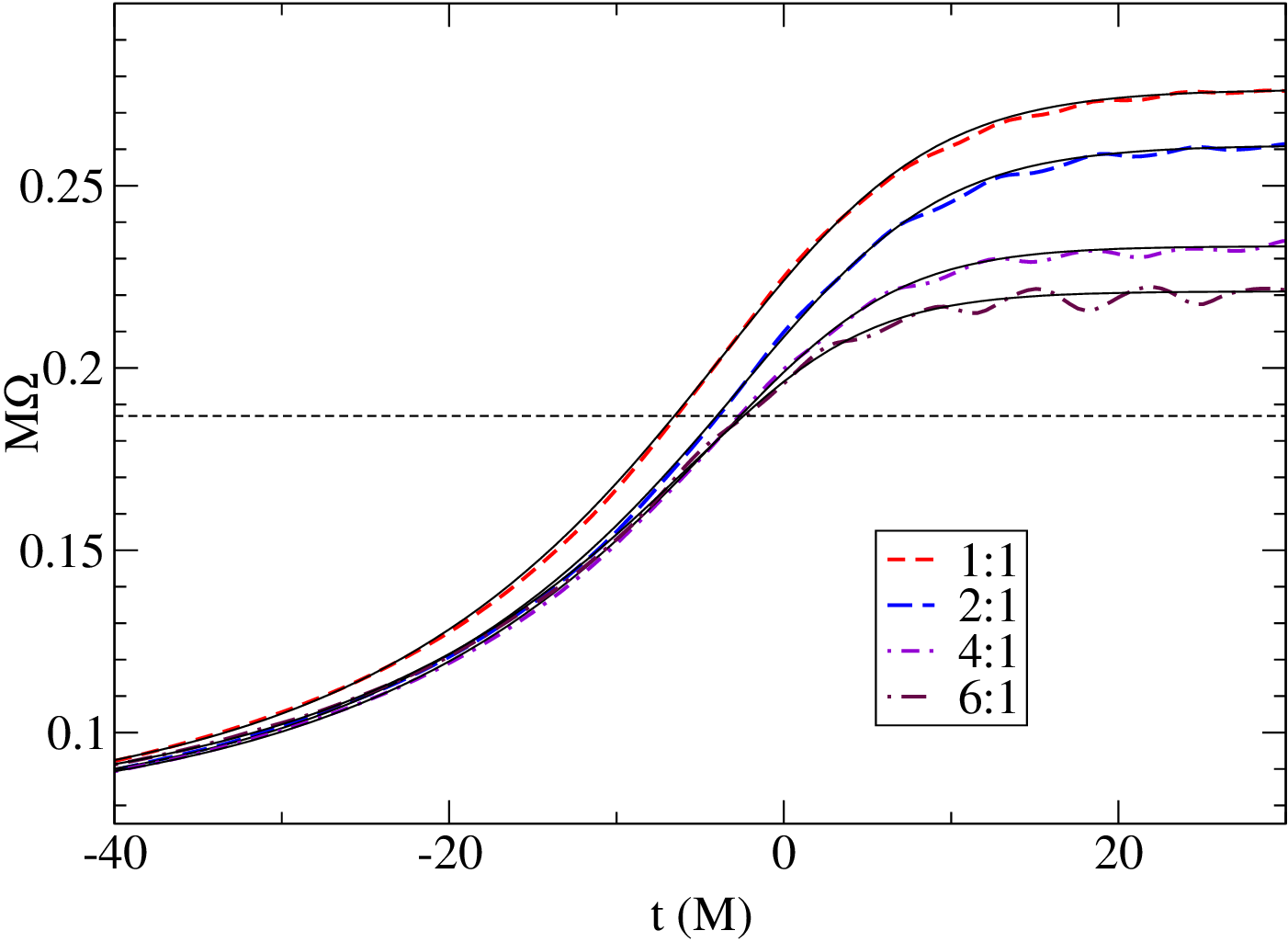}
\includegraphics*[width=3.5in]{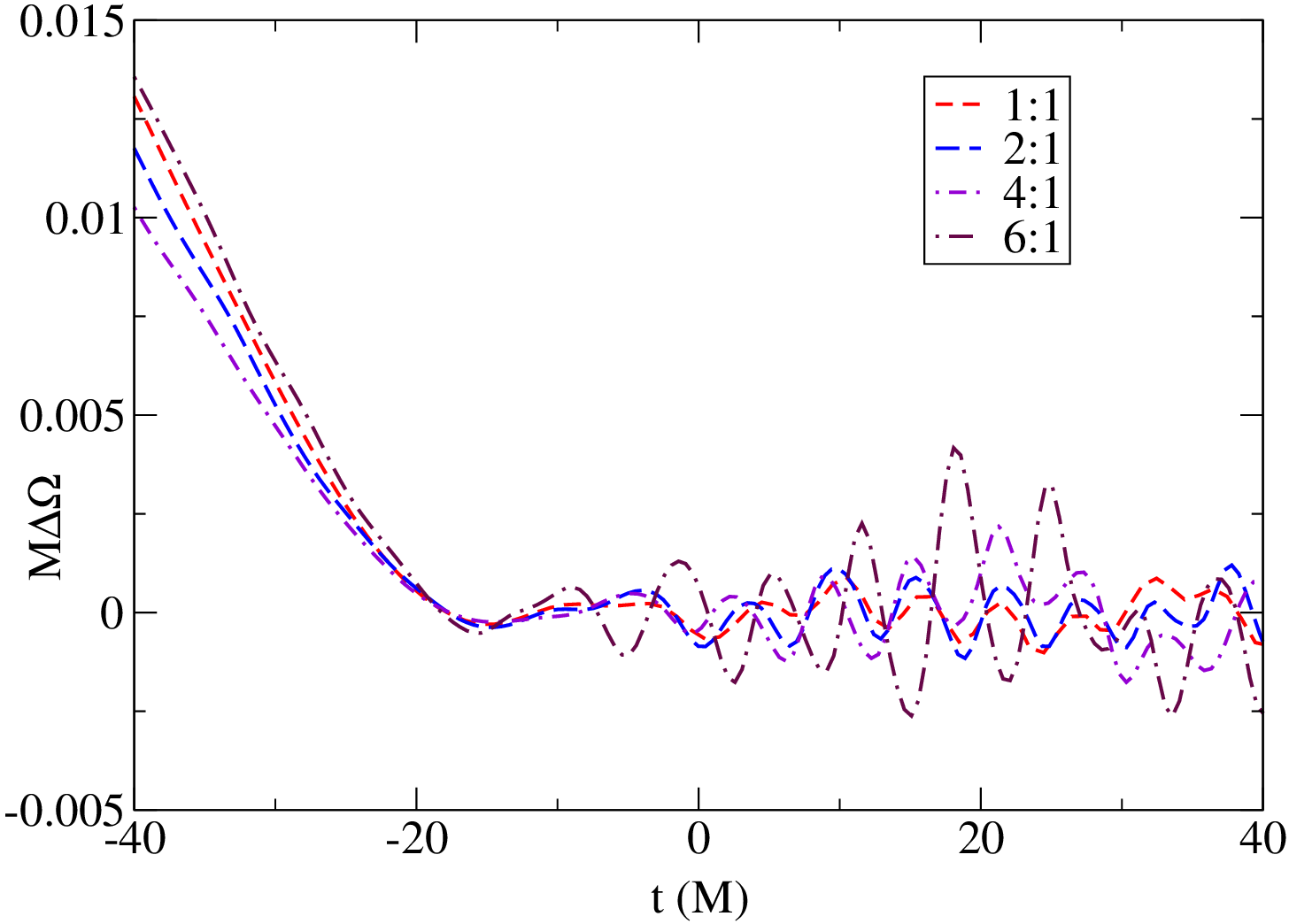}
\caption{Strain-rate rotational frequency evolution in the merger for (2,2) modes
of several mass ratios (dashed lines), with an analytic fit (solid
lines). The fit encodes a monotonically increasing frequency, which at
late times decays exponentially toward the fundamental ringdown
mode. The horizontal dashed line marks $M \Of = 0.18685$, half the
(2,2) mode QNM frequency for a nonspinning perturbed hole. The right
panel shows the residuals, which are comparable to uncertainties in
the numerical data after $20 M$ before the peak.  }
\label{fig:trans_freq_Xx}
\end{figure*}

For all other modes the frequency evolution follows a simple smoothly
evolving curve, qualitatively similar in each $(\ell,m)$ and
mass-ratio case [see Fig.~\ref{fig:trans_freq_Xx} below].  In
particular, we note that, except for small noise contributions, each
curve shows that the frequency increases monotonically, ultimately
saturating at a frequency set by the fundamental quasinormal ringdown
mode.  This monotonic frequency development is a universal
characteristic of the radiation from inspiral, through merger, and up
to ringdown.  In the PN analysis of quasicircular inspiral, this
characteristic makes it possible to describe the changing structure of
the hardening binary as a function of frequency instead of the more
coordinate-specific separation.  This allows us, for instance, to
write the waveform amplitude as a function of frequency.

Subsequently, we will assume monotonic frequency development throughout
the coalescence process.  This principle underlies our empirical curve
fitting of the frequency evolution, allowing more quantitative
analysis of the late-time phasing evolution.  In
Sec.~\ref{ssec:FreqAmp} we will further apply this idea as we study
relationships between late-time frequency and amplitude evolution.

To produce an empirical curve for describing the late-time frequency
evolution, we assume that each $(\ell,m)$ mode has a monotonically
increasing polarization frequency, which approaches the fundamental
ringdown frequency $\omQ$ at late times. The general expectation that
the frequency decays exponentially toward the ringdown frequency
suggests that we model frequency evolution based on the hyperbolic
tangent function.
 
Specifically, we will compare
frequency evolution of the strain-rate waveforms with a model of the form
$\Omega(t)=g(t)$, where
\begin{equation}
\label{eq:OmegaFitFn}
g(t) = \Oi+(\Of-\Oi) \left(\frac{1+\tanh[\ln\sqrt{\kappa}+(t-t_0)/b]}{2}\right)^\kappa.
\end{equation}
This provides a curve that first grows exponentially, with e-folding
time $b/(2\kappa)$, from some initial frequency $\Oi$, then decays
exponentially, with e-folding time $b/2$, to the final frequency
$\Of$.  The presence of the exponent $\kappa$ allows the early
exponential growth rate to differ from that at late times.  The early
part of this model is a coarse approximation to the growth in
frequency near the end of the inspiral.  This approximation must,
therefore, fail to fit the data if we look back sufficiently long
before the time of peak radiation, but, as we will show, it provides a
fair approximation of the approach to the peak. The rate at which the
frequency grows, the chirp-rate, increases to some maximum, then
decreases to zero on approach to the final frequency $\Of$. For a more
meaningful parametrization, we set
\begin{equation}
\Oi \equiv \Of - \frac{b}{2} \cval \left(1 + \frac{1}{\kappa}\right)^{1+\kappa}.
\end{equation}
With this choice, $dg/dt$ will peak with value $\cval$ at time $t_0$.
The model then depends on the five parameters $\kappa$, $b$, $\cval$,
$t_0$ and $\Of$.

\begin{figure*}
\includegraphics*[width=3.5in]{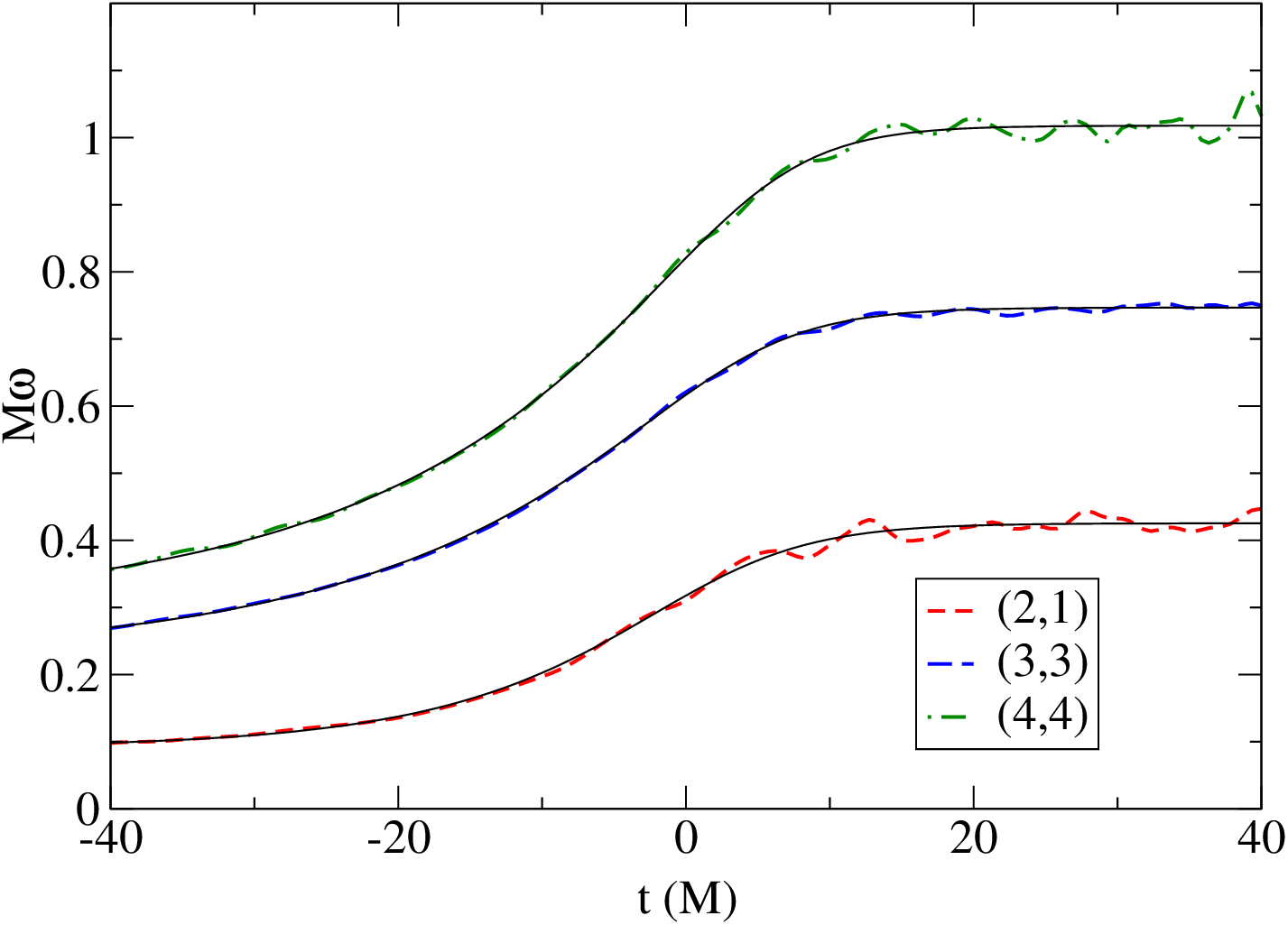}
\includegraphics*[width=3.5in]{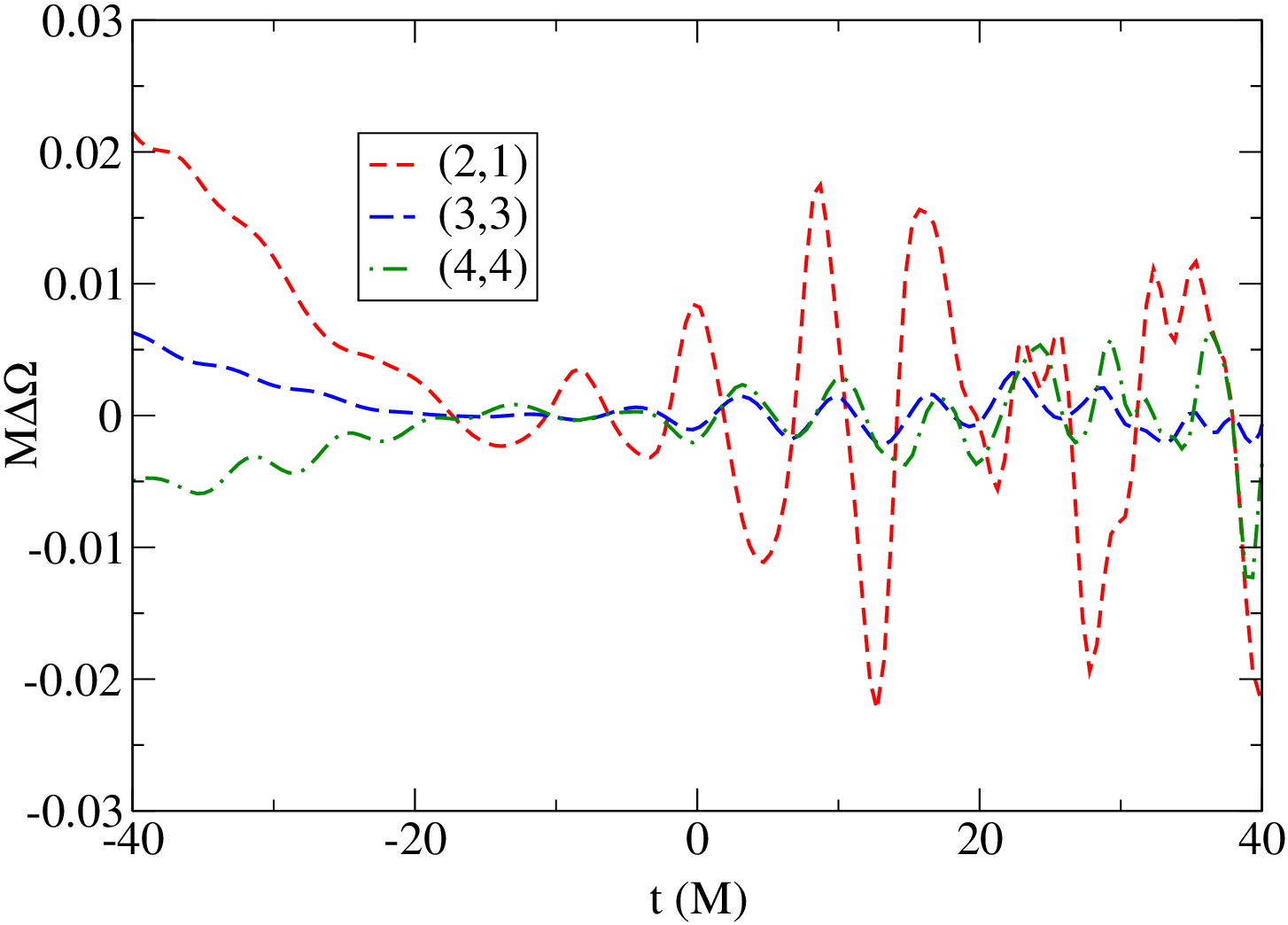}
\caption{Strain-rate frequency evolution in the 4:1 mass-ratio merger 
for several spherical harmonic modes, with an analytic fit as in
Fig.~\ref{fig:trans_freq_Xx}. The left panel shows the polarization
frequency $\omega_{\ell m}$.  The right panel shows the residuals of
the unscaled rotational frequencies $\Omega_{\ell m} = \omega_{\ell
m}/m$.  }
\label{fig:trans_freq_X4_lm}
\end{figure*}

As shown in Figs.~\ref{fig:trans_freq_Xx} and
\ref{fig:trans_freq_X4_lm}, we find that fits to the model $g(t)$
provide an excellent approximation to the numerical data for the
strain-rate rotational frequencies in most significant cases.  For the
$(2,2)$ modes of all mass ratios, and for all but one of the
significant $(\ell,m)$ modes for the representative 4:1 mass-ratio
case, we find agreement within a few percent after $t=-20 M$, with the
primary differences coming from apparent noise in the numerical
simulations, within the uncertainties in the numerical results.

Fig.~\ref{fig:trans_freq_Xx} shows the comparison for the (2,2) mode
rotational frequencies for several mass ratios. A glance at the curves
shows that the unequal-mass cases are quite similar to the equal-mass
frequency evolution, previously examined in
Refs.~\cite{Buonanno:2006ui,Hinder:2007qu}. The dominant difference
for the unequal-mass cases is that the final frequency $\Of$ decreases
with $\eta$, consistent with the decrease in the spin of the final
black hole produced. We expect $\Of$ to correspond to $\omQ/m$, where
$\omQ$ is the fundamental ($n=0$) quasinormal ringing frequency for
the specific $(\ell,m)$ mode for a black hole with the appropriate
spin. In the infinite-mass-ratio limit ($\eta\rightarrow0$), $\Of$
should correspond to half the Schwarzschild $(2,2)$ quasinormal mode
frequency $M \omQ = 0.3737$, indicated by the horizontal dashed line
in the left panel of Fig.~\ref{fig:trans_freq_Xx}.

We show a few examples of polarization frequency curves, for
subdominant modes in Fig.~\ref{fig:trans_freq_X4_lm}. The $(3,3)$ mode
is very similar to the $(2,2)$ modes shown in
Fig.~\ref{fig:trans_freq_Xx}, as are the other $\ell=m$ modes (not
shown).  The $(2,1)$ mode is of similar shape, also well approximated
by our fit.

The quantitative fit results are summarized in Table
~\ref{table:TransFreqFit}. The error bars are based on statistical fit
estimates, also incorporating the ranges of best fit results obtained
by varying the fit range starting between $t/M=-25$ and $-15$ and
ending between $t/M=20$ and $60$. The final frequencies $\Of$
approached in the fit curves in Figs.~\ref{fig:trans_freq_Xx} and
\ref{fig:trans_freq_X4_lm} were robustly determined by the fits within
a fraction of a percent.  The $\Of$ frequencies from the fits in
Fig.~\ref{fig:trans_freq_Xx} were applied in Table
~\ref{tab:KerrParams} to find final black-hole parameters consistent
with those determined by conservation of energy and angular momentum.
\begin{table*}
\caption{\label{table:TransFreqFit}
Results of fitting Eq. (\ref{eq:OmegaFitFn}). All quantities are
scaled by the final mass $\Mf = M_{\rm f,rad}$ defined in
Eq. (\ref{eqn:Mf_def}). The parameter $\Of$ can be related to the
fundamental real QNM frequency, $\omQ$, and has been used to extract
two different estimates of the final dimensionless spin $\hat{a}$ of
the hole -- see Table~\ref{tab:KerrParams}.}
\begin{tabular}{c|c|ccc|c|c|c}
\hline\hline
Mass ratio   &$(\ell,m)$&   $\kappa$ &    $b/\Mf$ &$m \cval {\Mf}^2$& $m \Of \Mf$ &$t_0/M$  & $\cval \Mf/\Of$ \\
\hline
1:1     &   (2,2) & $0.7\pm0.1$ & $13.2\pm0.3$ & $0.0112\pm0.0001$ & $0.528\pm0.001$   & $-4.6\pm0.2$ &$0.0210\pm0.0002$\\ 
\hline
2:1     &   (2,2) & $0.6\pm0.1$ & $12.3\pm0.8$ & $0.0104\pm0.0002$ & $0.5023\pm0.0006$ & $-3.6\pm0.2$ &$0.0207\pm0.0004$\\ 
\hline
4:1     &   (2,2) & $0.5\pm0.1$ & $10.5\pm0.2$ & $0.0096\pm0.0001$ & $0.4566\pm0.0001$ & $-4.8\pm0.3$ &$0.0210\pm0.0002$\\ 
        &   (2,1) & $1.0\pm0.5$ & $12\pm2$    & $0.0120\pm0.0005$ & $0.421\pm0.006$   & $-3.9\pm0.8$ &$0.0285\pm0.0012$\\ 
        &   (3,3) & $0.3\pm0.1$  & $9.7\pm0.2$  & $0.0153\pm0.0001$ & $0.730\pm0.001$   & $-3.3\pm0.1$ &$0.0210\pm0.0002$\\ 
        &   (4,4) & $0.15\pm0.1$ & $7.5\pm0.7$  & $0.0212\pm0.0003$ & $0.991\pm0.005$   & $-1\pm1$    &$0.0213\pm0.003$\\ 
\hline
6:1     &   (2,2) & $0.5\pm0.1$ & $10.4\pm0.5$ & $0.0089\pm0.0002$ & $0.4349\pm0.0004$ & $-6.5\pm0.3$ &$0.0205\pm0.004$\\ 
\hline \hline
\end{tabular}
\end{table*}

The peak in the chirp-rate $\cval$ is a particularly significant
quantity in determining the shape of curves of our general form
$g(t)$.  As shown in Table~\ref{table:TransFreqFit}, our fits
determine $\cval$ to within a few percent in all cases.  An
interesting relationship is apparent among the fits for all the
$\ell=m$ cases for all values of $\eta$ studied.  In the last column
of the table, we show the peak chirp-rate values scaled by the mass of
the final black hole $\Mf$ and the final frequency $\Of$.  In each
case where $\ell=m$ we find $\cval \Mf/\Of\sim0.02$, consistent within
the fit uncertainties.  This scaling makes some sense, since the
height of the rise in frequency through the final radiation burst is
largely determined by $\Of$, while the time scale over which this rise
occurs seems to be similar when time is scaled by $\Mf$.  In
Sec.~\ref{sec:newEOB} we will use this result to predict the phase
evolution in an analytic waveform model.

Our model for late-time frequency evolution (\ref{eq:OmegaFitFn})
describes exponential decay toward $\Of$ at an e-folding rate given by
half our fitting parameter $b$.  For all cases, the values of $b$ are
within about 30\% of $10M$.  In some cases, the fits for $b$ are
rather sensitive to the initial starting time, varying by up to 20\%
or 30\% in the (2,1) and (4,4) modes of the 4:1 case.  At this coarse
level, we note that the values for $b$ are similar to the exponential
decay rates for quasinormal ringdown mode \emph{amplitudes} listed in
Table~\ref{table:TransAmpFit}.  We will consider this relationship
further in Sec.~\ref{ssec:FreqAmp} below.

The other parameters in our fit are $\kappa$, relating to the shape of
our fit curve at early times, and $t_0$, giving the time at which the
frequency peak occurs.  The parameter $\kappa$ is not very precisely
determined; as we would expect, it depends sensitively on the starting
time of the fit interval, since the early exponential frequency growth
is only a coarse approximation of the expected behavior.  The values
for $t_0$ show that the peaks in $\dot\Omega$ generally occur roughly
$4M$ before the total energy peaks at $t_{\rm peak}$. As was the case
for the power peaks in Fig.~\ref{fig:peak_shape_lm}, the chirp-rates
of the different spherical harmonic modes peak at slightly different
times.

We have supplemented our general implicit rotating source picture with
the additional idea that the rotational frequency for each mode grows
monotonically, not only in the inspiral, but also through the merger
and ringdown. Based on this expectation we have identified an analytic
fit model for the late-time frequency evolution that precisely matches
the data for all cases but the (3,2) mode. These fits provide a
quantitative understanding of the late-time phasing yielding, in
particular, a robust result for the peak chirp rate $\cval$ for all
$\ell=m$ modes.  We will apply this information in
Sec.~\ref{sec:newEOB}.

\subsection{Late-time frequency and amplitude relationships.}
\label{ssec:FreqAmp}

The last step in our waveform analysis is to consider relationships between the
frequency evolution and the waveform amplitude.  

In the PN description of the quasicircular inspiral, the orbital
frequency not only tells us the rotational rate, but can also serve as
a label for describing the momentary state of the rotating object (in
the inspiral case this means that we can reference the state of the
system in terms of $r(\Omega)$).  The PN generalization of the
quadrupole formula, describing radiation from the rotating system,
then leads to an expression for amplitude as a function of frequency.
Our description of the gravitational radiation suggests an implicit
source rotating with monotonically increasing frequency as it
continues to ``harden'', as the system evolves smoothly into merger
and ringdown.  In this section we seek to further unify this picture
of the full coalescence process, considering an analogue of the PN
description of amplitude as a function of frequency that can describe
the radiation in the merger and ringdown.

In Sec.~\ref{ssec:amplitude} we emphasized that the radiation power
$\dot E_{\ell m}$ provides essentially the same information as the
strain-rate amplitude (\ref{eqn:Edot_modes}). If the wave frequency is
known, then the modal contribution to the total radiative angular
momentum can similarly provide information about the gravitational
wave amplitude.  Eq. (\ref{eqn:JdotAH}) gives an expression for
angular momentum flux in terms of wave amplitude and phase.  The
relation simplifies to
\begin{equation}
\label{eqn:JdotA}
\dot{J}_{\ell m} \approx \frac{1}{16\pi\Omega}(A_{\ell m})^2
\end{equation}
up to 5PN order, indicating that the radiation carries maximal angular
momentum $\dot J=\dot E/\Omega$, as we generally expect for circularly
polarized radiation.  If we know how the rotational frequency evolves,
we can derive the mode amplitudes $A_{\ell m}$ from the mode-by-mode
relationship of either energy or angular momentum with frequency,
${E}_{\ell m}(\Omega)$ or ${J}_{\ell m}(\Omega)$.

\begin{figure*}
\includegraphics*[width=3.5in, angle=0]{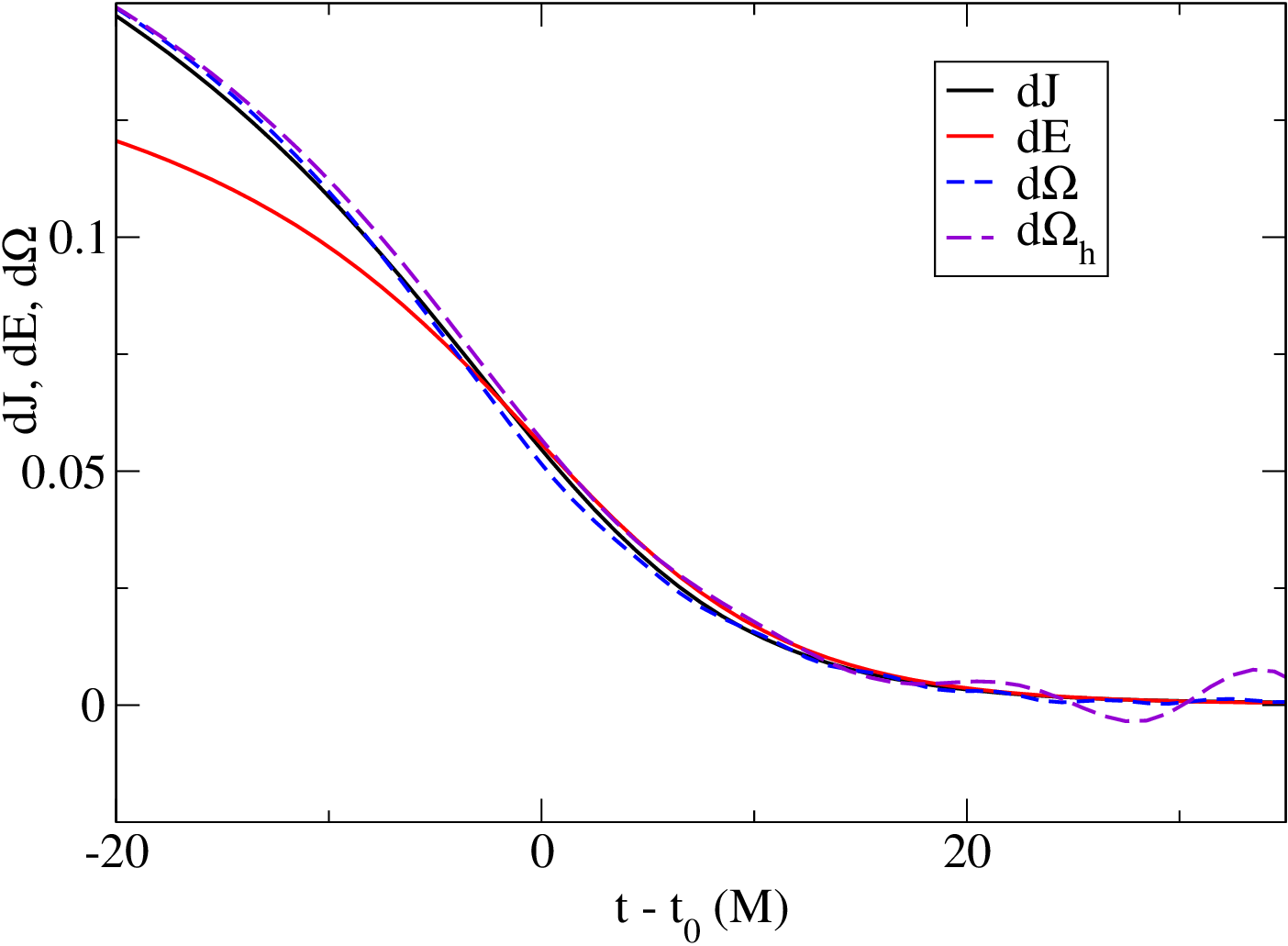}
\includegraphics*[width=3.5in, angle=0]{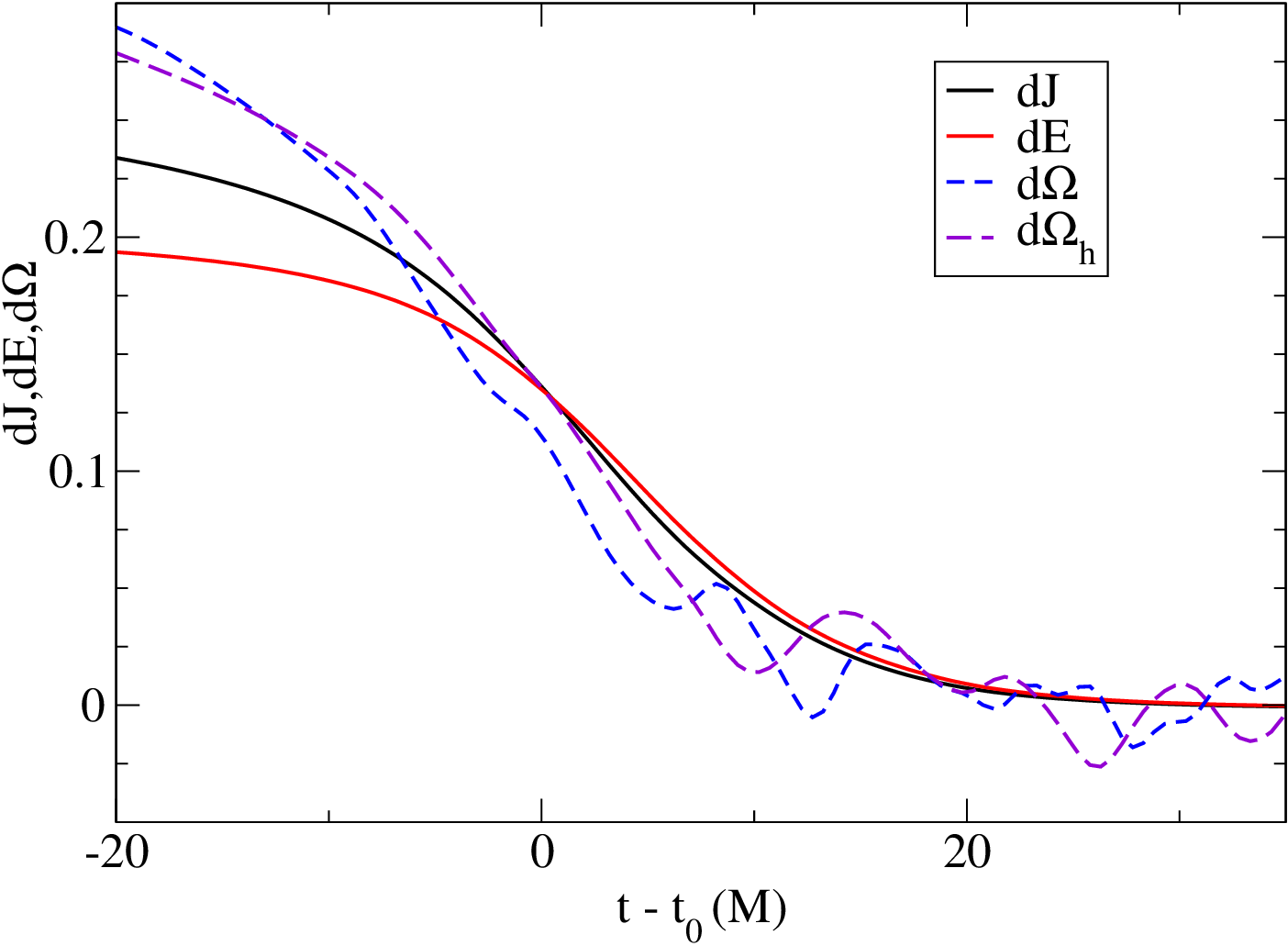}
\caption{Relationship between angular momentum and frequency evolution
approaching ringdown for the (2,2) mode of the 1:1 case (left panel)
and the (2,1) mode of the 4:1 case (right panel). We show the
differences $dE$ and $dJ$ from the final mass and angular momentum as
measured from the radiation, together with the difference from the
final ringdown rotational frequency, measured from $\dot{h}$
($d\Omega$) and from the strain $h$ ($d\Omega_h$).  We see similar
evolution for the dominant (2,2) mode of each mass ratio simulation,
and for all modes other than (3,2). In the left panel the angular
momentum and energy (in geometrical units) have been scaled up by
factors of $2.5$ and $10$, respectively, to roughly match the
frequency residuals at $t=0$. In the right panel the rescaling factors
are $760$ for angular momentum and $2000$ for energy.  The agreement
of the curves, with these rescalings, is consistent with
$dJ_{22}/d\Omega_{22}\approx0.40$ from $t=20$ through the peak
radiation. The more approximate agreement in the right panel is
consistent with $dJ_{21}/d\Omega_{21}=0.0013$, through the peak
radiation.  }
\label{fig:Jvsom}
\end{figure*}
To approach an understanding of the late-time relationships between
amplitude and frequency, we compare how the system's energy and
angular momentum approach their final values, with how the system's
frequency approaches its final value.  In Fig.~\ref{fig:Jvsom} we
examine the radiative loss of energy $dE$ and angular momentum $dJ$ as
the coalescing system approaches its final quiescent state, comparing
these with the remaining difference of gravitational-wave frequency
from its late-time limit $\Of$, as determined in
Table~\ref{table:TransFreqFit}. As well as our standard final
rotational frequency $\Omega=\dot\Phi$ defined in
(\ref{eqn:strainrate_modes}), we also show the frequency: based on the
strain, $\Omega_h=\dot\Phi_h$ as defined in (\ref{eqn:strain_modes}).
We have also rescaled the energy and angular momentum by a constant,
selected so that the value matches that of $\Omega_h$ at the time of
peak radiation power.

Fig.~\ref{fig:Jvsom} indicates a general correspondence between how
the angular momentum approaches its final state, and how the
gravitational-wave frequency approaches its final state.  Of the
several cases of spherical harmonic modes and mass ratios that we have
examined in this manner, we show two examples: the $(2,2)$ case for
the equal-mass simulation, where the evolution of angular momentum and
frequency correspond most closely (left panel), and the $(2,1)$ mode
from the 4:1 mass ratio simulation, with the weakest correspondence
(aside from the nonconforming $(3,2)$ case) (right panel).  The
correspondence is closest, holding to a better approximation over a
longer period of time, in associating $dJ$ with $\Omega_h$.  The
association with energy is slightly weaker.

Though these plots suffer significantly from small modulations in
frequency that we have not resolved numerically, the results suggest
an approximate relationship between rotational frequency and angular
momentum in particular.  If the frequency evolution is otherwise
known, then the late-time evolution of angular momentum for each mode
could be approximately described by $J(t)=J_{\rm
f}+\xi(\Omega(t)-\Of)$, where $\xi$ is a case-dependent constant that
we will not attempt to specify generically.  As $\xi=dJ/d\Omega$ we
may refer to it as the dynamical moment of inertia of the implicit
rotating source. This late-time expectation creates the possibility of
extending our PN-based understanding of angular momentum flux into the
late time waveforms, giving us the additional information we need for
a full (approximate) description of $J(\Omega)$.

This relation between frequency, angular momentum, and amplitude
provides a connection between the peaks in modal radiation power,
shown in Fig.~\ref{fig:relative_power}, and the peaks in chirp rate
given in Table~\ref{table:TransFreqFit}. Assuming $dJ/d\Omega = \xi$,
a constant, yields
\begin{eqnarray}
\dot{E} &\approx& \xi \Omega\dot{\Omega} \label{eqn:EdotOfOm}\\
\Rightarrow \ddot{E} &\approx& \xi \left( \Omega\ddot{\Omega} + \dot{\Omega}^2 \right) \label{eqn:EddotOfOm}
\end{eqnarray}
Since the $\Omega(t)$ curve is steeply increasing, we would expect the
peak in $\dot E$ to be near the peak in $\dot\Omega$ but slightly
delayed. Expanding (\ref{eqn:EddotOfOm}) linearly about $t_0$, when
$\ddot{\Omega} = 0$, we find that $\dot{E}$ reaches its peak at a time
\begin{equation}
t_{\rm peak} \approx t_0 - \frac{\dot{\Omega}_0^2}{\Omega_0 \dddot{\Omega}_0} > t_0,
\end{equation}
where $\Omega_0$ and $\dddot{\Omega}_0$ are the frequency and its
third time derivative at $t_0$. These can be evaluated from
(\ref{eq:OmegaFitFn}) and the fit parameters $\kappa$, $b$, $\cval$,
and $\Of$ in Table~\ref{table:TransFreqFit}; we find that the values
for $t_{\rm peak} - t_0$ all lie in the approximate range
$(2.16,3.17)$, in rough agreement with the direct fit for $t_0$ in
Table~\ref{table:TransFreqFit}.  

Conceptually, the monotonic evolution of frequency might lead us to
hypothesize a relationship between the structure of the implicit
rotating source and its rotational frequency, such that the changes in
the structure of this radiating rotator will be associated with finite
changes in frequency. Knowing that the frequency growth must be
limited by the quasinormal ringing frequency implies a peak in the
chirp rate.  This leads to the expectation that finite changes in
energy and angular momentum are associated with finite changes in
frequency, so that $dE/d\Omega$ and $dJ/d\Omega$ approach constants
once the evolution in frequency slows.  The peak in radiation power
might then be viewed as a consequence of the peak in chirp-rate.  More
quantitatively we find that $dJ/d\Omega$ seems to be roughly constant
even before the evolution in frequency slows down (i.e. before $t=0$).
This correspondence will be applied in the next section to provide a
model for amplitude evolution through the peak, based on information
about the frequency evolution.

If we could postulate the constancy of $dJ/d\Omega$, we might also
apply that assumption to ``explain'' some of what we have seen above.
In Sec.~\ref{ssec:TransFreq} we noted a rough agreement between the
timescale $b$ in our frequency fitting curve (\ref{eq:OmegaFitFn}) and
the quasinormal ringdown amplitude decay rates for the corresponding
quasinormal modes.  Following the discussion above, this relationship
could be derived, in the $\Omega \rightarrow \omQ / m$ limit, from the
constancy of either energy or angular momentum losses with respect to
change in frequency. For constant $\xi=dJ/d\Omega$,
Eqs. (\ref{eqn:Edot_modes}), (\ref{eqn:strainrate_modes}) and
(\ref{eqn:EdotOfOm}) imply that
\begin{equation}
A^2 = |R\dot h|^2 = 16\pi \dot{E} \approx 16 \pi \xi \Omega\dot{\Omega}.
\label{eqn:AmpFreq}
\end{equation}

In the $\Omega\rightarrow\Of$ limit, the strain-rate amplitude decays
at the rate predicted from black-hole perturbation theory, $A_{\ell
m}\rightarrow A_{0 \ell m}\exp{(-t/\tau_{\ell m})}$, where $\tau_{\ell
m}$ is the e-folding rate for the amplitude decay for the fundamental
$(\ell,m)$ quasinormal mode.  In this limit, our frequency evolution
fit model reduces to
\begin{eqnarray}
\Omega(t)
&\rightarrow& \Of - (\Of-\Oi) e^{-2(t-t_0)/b}.
\end{eqnarray}
Applying these limiting expressions for amplitude and frequency in
(\ref{eqn:AmpFreq}) in the limit $\Omega\rightarrow\Of$ yields
\begin{equation}
\label{eq:b-tau}
(A_0)^2 e^{-2t/\tau}\approx\frac{32\pi\xi\Of}{b}(\Of-\Oi) e^{-2(t-t_0)/b},
\end{equation}
where the left-hand-side derives from the amplitude, and the
right-hand-side from frequency.  Ignoring the constant coefficients,
this implies that $b=\tau$.

If we adventurously assume the constancy of $\xi$ on approaching the
ringdown, and expand the amplitude in powers of
$\epsilon=\exp{(-t/b)}=\exp{(-t/\tau)}$, the implied amplitude
frequency relation might also provide more information about the
amplitude evolution.  Since $\Omega(t)$, and consequently the
right-hand-side of (\ref{eq:b-tau}), contains only even powers of
$\epsilon$, the next term in the expansion for amplitude should be
${\cal O}(\epsilon^3)$.  This suggestion motivated our expansion
(\ref{eq:ImOmegaFitFn}) applied in fitting the late-time amplitudes in
Sec.~\ref{sec:simulations}.

\section{Variations on the EOB model}
\label{sec:newEOB}

An approach to modeling black-hole binary radiation known as the
effective-one-body (EOB) model has been presented in the literature
\cite{Buonanno:1998gg,Damour:2000we,Buonanno00a,Buonanno:2000jz,
Damour:2001tu,Buonanno:2005xu,Damour:2007cb}.  The late-time waveforms
in these models are based on a now-common description of the merger
process as an epoch of radiation from spiraling particlelike
trajectories, followed, in a sudden transition, by black-hole ringdown
dynamics with waveforms described by a superposition of quasinormal
frequencies.  Our waveform analysis provides a complementary
description of black-hole binary merger radiation that can be applied
in an alternative late-time waveform model.

A recent promising approach along these lines is the so-called
\emph{pseudo-4PN} (p4PN) EOB model \cite{Buonanno:2007pf}. This model extends
the 3PN-accurate EOB metric with a term of 4PN order with a tunable
multiplier $\lambda$. The phasing obtained from this expansion is
combined with leading-order PN strain amplitudes (the restricted
approximation) to obtain mode-by-mode waveforms valid for
inspiral. For the merger and ringdown, the phasing and amplitude are
derived simultaneously from a superposition of quasinormal modes.  For
each $(\ell,m)$ angular mode, the fundamental ringdown modes and a few
overtones are summed in proportions as required for continuity with
the late end of the radiation from the inspiral phase.  The value of
the p4PN multiplier $\lambda$ is then chosen to match the pre- and
post-merger waveform portions, optimizing the agreement with
full-numeric waveforms.

In this section, we show that it is possible to develop variations on
the p4PN EOB model that usefully encode some of the waveform phase and
amplitude relationships we have described above.  A key difference
with the new variant is our prescription for the transition from
inspiral to merger-ringdown radiation.  In contrast with
\cite{Buonanno:2007pf}, for each $(\ell,m)$ mode we consider the
entire wave train as that of a slowly varying instantaneously rigid
rotator, consistent with the dominant ``circular-polarization''
waveform pattern encoded in the radiation.  The phase evolution might
be thought of as arising from the rotation rate of the corresponding
$(\ell,m)$ source structure, which is continued through ringdown by
continuous matching to a function of the form (\ref{eq:OmegaFitFn}).
The wave amplitude will be derived directly from expectations for the
energy or angular momentum content of the radiation.

We present two specific models of this nature. Model 1 is based on
exactly the same EOB-based prescription for inspiral-plunge
trajectories in \cite{Buonanno:2007pf}, while Model 2 shows the effect
of a slight variation in the underlying EOB model.  In both variants,
as in \cite{Buonanno:2007pf}, we derive the waveform phasing directly
from the EOB trajectories (with $\lambda=60$ for the strength of the
p4PN term) up until some matching time, which we take simply as the
time at which the (2,2) wave frequency is half the frequency of the
fundamental (2,2) ringdown mode.  After this point we use our fit
model (\ref{eq:OmegaFitFn}) to describe the subsequent phase
evolution.

Recall that this model depends on several parameters: $\kappa$, $b$,
$\cval$, $t_0$, and $\Of$.  The results of our analysis in Sec.
\ref{ssec:FreqAmp} guide us in producing a fully specified model for
these parameters.  We take $\Of = \omQ/m$ from the fundamental
ringdown frequency $\omQ$ of the radiation.  For the time-constant for
frequency decay $b$ we use the fundamental quasinormal mode amplitude
decay time constant.  While this is not clearly implied by our fits in
the last section, it will lead to the correct amplitude fall-off, as
specified below.  For the strongest $\ell=m$ modes, our fits indicate
$\cval=0.021 \omQ/m \Mf$.  Lacking any better model we simply increase
this by a factor $4/3$ when $\ell\neq m$, roughly consistent with the
higher value of $\cval$ found for the (2,1) mode.  For these models we
derive the quasinormal modes using the fit for the final black hole
mass and spin described in \cite{Buonanno:2007pf}.  The remaining
parameters $t_0$ and $\kappa$ are chosen to provide continuity, up to
the second time-derivative of phase, with the direct EOB-based phasing
at the matching time.

Our prescription for the wave amplitudes differs from the restricted
amplitude description applied in \cite{Buonanno:2007pf}.  As we showed
in Fig.~\ref{fig:relative_power}, we can improve on the restricted
amplitude approximation by using the leading-order PN expressions for
waveform $(\ell,m)$ mode amplitudes only to fix the
\emph{partitioning} of radiation power into angular modes.  We set the
total power independently, from the full-order EOB model description
of the radiation power; the resulting waveforms are then energetically
consistent with the EOB description of the dynamics, and also show
better agreement with the numerical results (note that in this model
there is no radiation in the nonrotational $m=0$ modes).  After
reaching the matching frequency, we continue the amplitude evolution
based on the assumption, suggested in Sec.~\ref{ssec:FreqAmp}, that
the amplitude is roughly constant through the radiation peak.

In our new model, we set the late-time amplitude by asserting the
approximate relationship (\ref{eqn:AmpFreq}), written this time in
terms of the polarization frequency $\omega = m \Omega$:
\begin{equation}
A_{\ell m}^2 \approx 16\pi \xi_{\ell m} \frac{\omega_{\ell m}\dot{\omega}_{\ell m}}{m^2},
\end{equation}
setting the value of $\xi_{\ell m}$ for amplitude continuity at the
match-frequency.  With this model for the amplitude, the peak in the
gravitational-wave amplitude is a direct consequence of the peak in
the time-derivative of gravitational frequency, fixed by $\cval$ in
(\ref{eq:OmegaFitFn}). The exponential decay in amplitude also follows
directly from the exponential approach of the wave frequency to
$\omQ$.
\begin{figure*}
\includegraphics*[width=3.5in, angle=0]{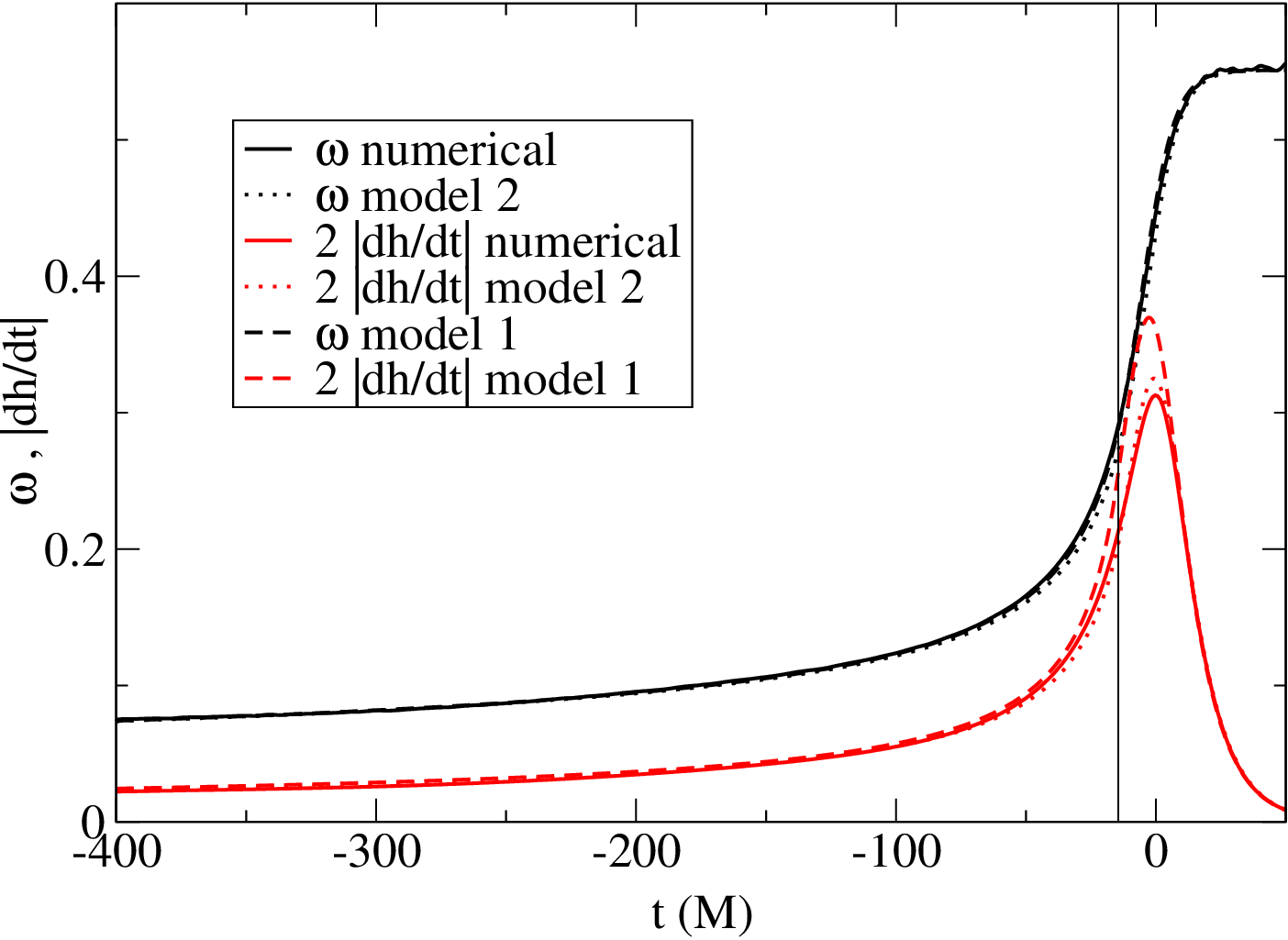}
\includegraphics*[width=3.5in, angle=0]{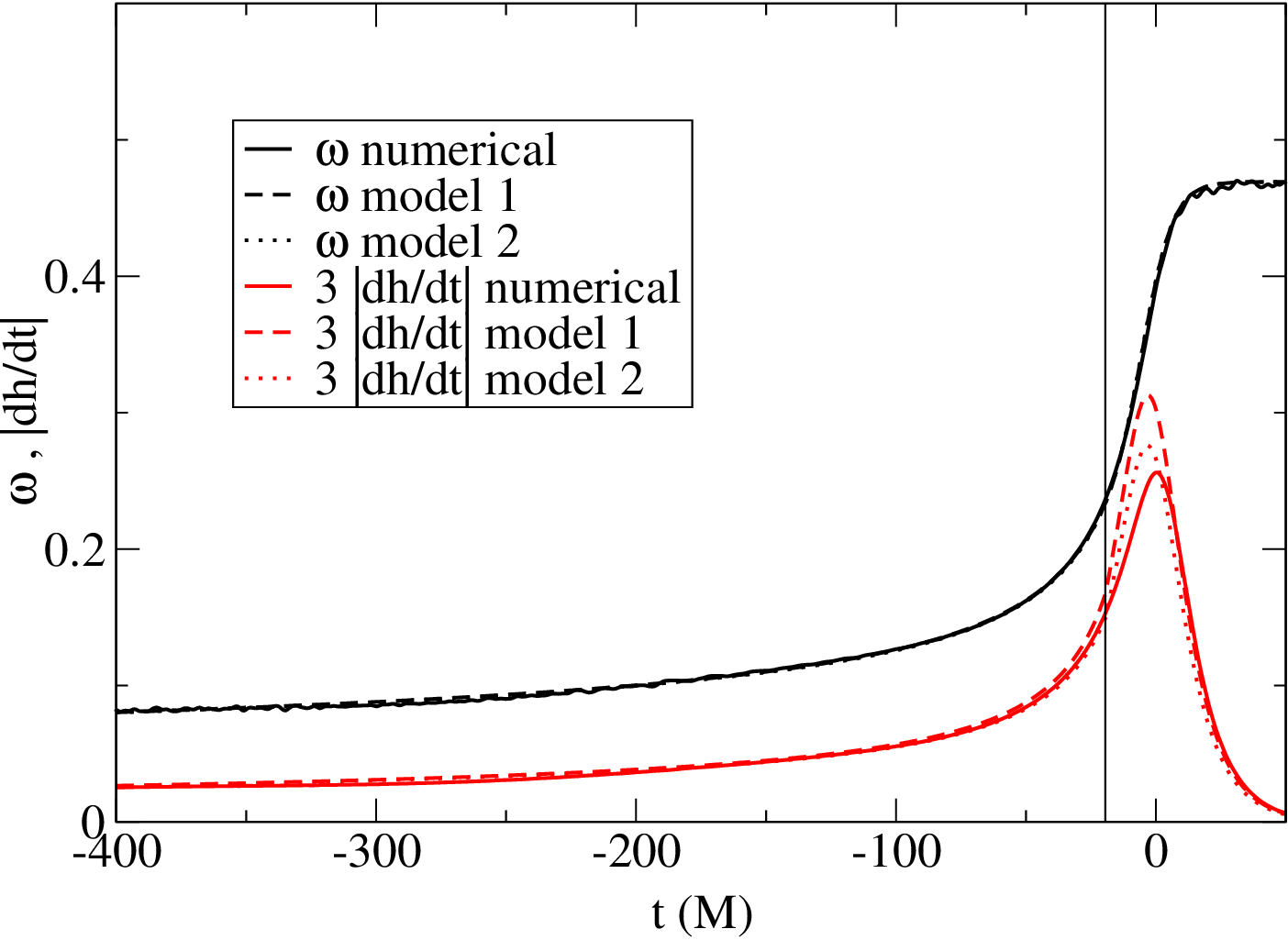}
\caption{Comparisons of new EOB-based models with numerical results 
for the (2,2) waveform frequency and amplitude for the 1:1 (left) and
4:1 (right) mass ratio cases.  The models are variations on the p4PN
EOB waveform model with a flux-based determination of the wave
amplitudes and an alternative, based on Eq.~\ref{eq:OmegaFitFn}, to
quasinormal mode summing for continuing the waveforms through
ringdown.  The vertical bar indicates our matching frequency where we
transition from direct EOB phasing and flux.  Models 1 and 2
correspond to different versions of the radiative flux, which
primarily affects the consequent wave amplitudes; see the text for
more details.  }
\label{fig:newEOB22}
\end{figure*}
\begin{figure*}
\includegraphics*[width=3.5in, angle=0]{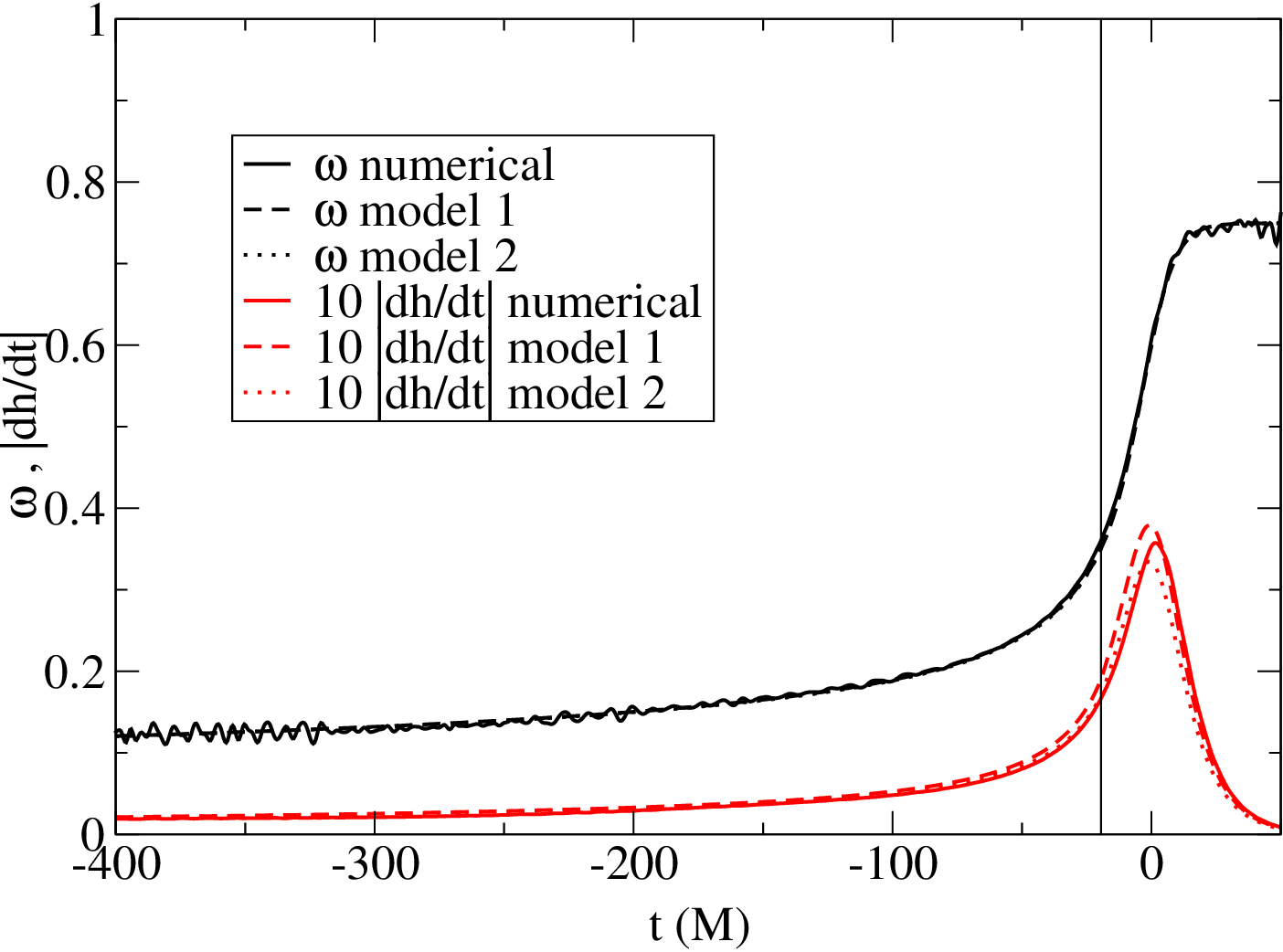}
\includegraphics*[width=3.5in, angle=0]{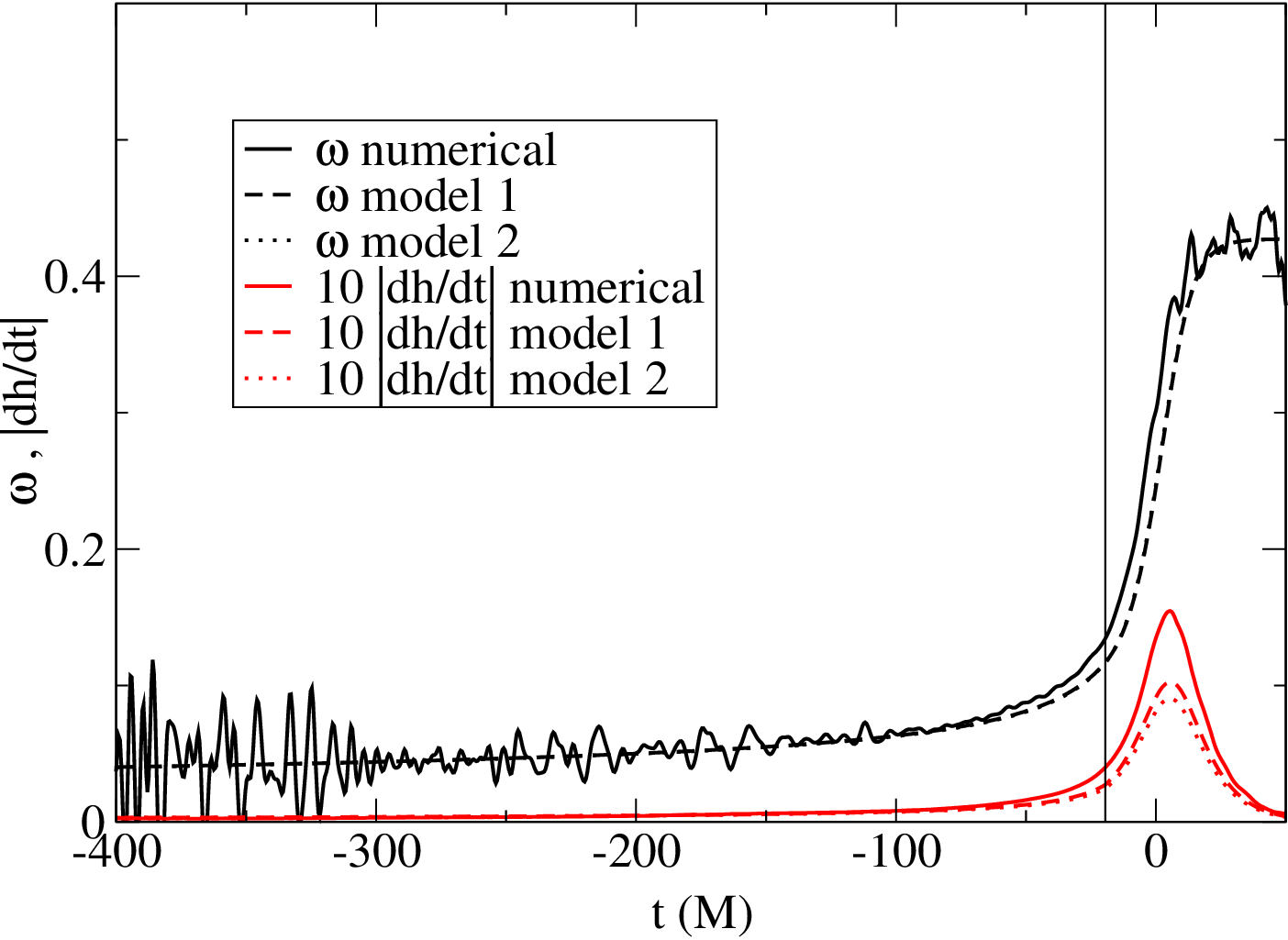}
\caption{Comparisons of new EOB-based models with numerical results for the 4:1 case,
for the (3,3) (left)  and (2,1) (right) waveform modes.
The plots show frequency and amplitude, labeled as in Fig.~\ref{fig:newEOB22} 
}
\label{fig:newEOBsubdom}
\end{figure*}

We compare the frequency and amplitude of the modeled waveform $(2,2)$
component with the corresponding numerical result for the 1:1 and 4:1
mass ratio cases in Fig.~\ref{fig:newEOB22}.  A similar comparison is
shown for some of the subdominant modes in the 4:1 case, in
Fig.~\ref{fig:newEOBsubdom}.  The model we have so far described,
based on the p4PN EOB trajectories is labeled Model 1 in the figures.
The matching frequency is indicated as a vertical line in each plot.
The frequency curves indicate very good phasing agreement for all
cases except the $(2,1)$ mode of the 4:1 mass ratio case.  In that
case the sharp rise in frequency occurs a few $M$ too late, several
times worse than the agreement shown for the other modes.  The
difference is a consequence of the slightly higher frequency early on,
as the numerical $(2,1)$ frequency begins to grow already before the
matching point as compared with the rotational frequency consistent
with the numerical $\ell=m$ modes.

The generally good amplitude agreement shown before merger represents an 
improvement over the simple restricted amplitude model employed in 
\cite{Buonanno:2007pf}.  For the $\ell=m$ modes, Model 1 overestimates the
amplitudes just before and consequently after matching.  This
indicates that the PN-based flux expression applied to the p4PN EOB
model, based on \cite{Buonanno:2002ft} and \cite{Damour:1997ub},
overestimates the flux at high frequencies.  This is perhaps not
surprising, since the flux formula is formulated with a pole at finite
frequency.  Physically we expect the flux to decrease at late times,
when the frequency approaches the quasinormal ringdown frequency.

To correct for this, we show also results for a second variation,
Model 2, in which we have introduced a zero in the flux function at
ringdown frequency.  Following the notation of
\cite{Buonanno:2002ft}, we modify their Eq. (45) for the flux to read
\begin{equation}
{\cal F}_{P_N}=\frac{32}{5}\eta^2v^{10}\frac{1-\omega/\omQ}{1-v/v_{\rm pole}}f_{P_N}(v;\eta),
\end{equation}
and then respecify the coefficients $c_1-c_7$ in their Eq. (50) to
again provide consistency to 3.5PN order with the Taylor series
expansion for the flux.  Note that the flux now depends on $\ell$ and
$m$ via the quasinormal mode frequency $\omQ$ and
$v=(M\Omega)^{1/3}=(M\omega/m)^{1/3}$.  The modified flux function
anticipates that the radiation will cut off at the ringdown frequency.
We find that the new EOB model provides a very good approximation to
the original p4PN EOB phasing (with $\lambda=60$) if we choose
$\lambda=27$ for the new version with modified flux.  The figures show
that, for the $\ell=m$ waveform modes, the amplitudes based on Model 2
with the modified flux show better agreement with the numerical
results leading up to the matching point, and also at the peak, while
the frequency evolution is nearly identical. For the (2,1) case, Model
2 suffers the same problems as Model 1.

We have applied the observations made in previous sections to
successfully predict late-time waveforms.  This provides an
alternative description of the transition to ringdown, distinct from
the widely-applied approach of summing quasinormal modes
\cite{Buonanno:2006ui,Damour:2007vq,Berti:2007fi}.  In
Ref.~\cite{Buonanno:2007pf}, we applied it to nonspinning mergers.

This raises the question: how is it that each of these quite distinct
approaches to approximately describing the late-time radiation can be 
simultaneously effective?  We can explore this question
by considering the polarization frequency evolution in a waveform
constructed from a sum of quasinormal modes (suppressing $\ell$ and $m$ labels)
\begin{equation}
s(t)=\sum_{n=0}^\infty A_{n} e^{-\sigma_{n}t+i\varphi_{n}(t)},
\end{equation}
with $\varphi_{n}(t)=\varphi_{n}(0)+\omega_{n} t$ and where
$\omega_{n}$ and $\sigma_{n}=1/\tau_{n}$ correspond to the $n^{\rm
th}$ quasinormal overtone mode.  Next we restrict to the first two
terms and to linear order in $\epsilon(t) = \exp[-(\sigma_1-\sigma_0)
t]$, which vanishes at late times. This yields
\begin{eqnarray}
s(t)&\approx& A_{0} e^{-\sigma_{0}t+i\varphi_{0}(t)}
\left( 1+\epsilon e^{i(\varphi_1(t)-\varphi_0(t))}\right)\\
    &\approx&A_0e^{-\sigma_{0} t+\epsilon \cos{(\Delta\varphi)} +i\left[\varphi_{0}+\epsilon \sin{(\Delta\varphi)}\right]},
\label{eq:swave}
\end{eqnarray}
where $\Delta\sigma=\sigma_1-\sigma_0$ and
$\Delta\varphi(t)\equiv\varphi_1(t)-\varphi_0(t)$. Taking the
derivative of the expression in square brackets gives the polarization
frequency of $s(t)$
\begin{equation}
\omega_s(t)\approx\omega_0- \epsilon
\left[\Delta\sigma \sin{(\Delta\varphi)} - \Delta\omega \cos{(\Delta\varphi)}\right],
\end{equation}
where $\Delta\omega \equiv \omega_1 - \omega_0$.  Note that the
expression in brackets is periodic with period $2\pi/\Delta\omega$,
where $\Delta\omega$ is the difference between the fundamental
quasinormal ringing frequency and its first overtone [see
\cite{Berti:2005ys} for a table of these overtones].

For the waveform modes we consider here $M\Delta\omega$ is quite
small, generally 1 or 2\%, which means that the period of the
expression in brackets is $\gtrsim100M$.  We rewrite $\omega_s$, only
keeping terms linear in $\Delta\omega$, as
\begin{equation}
\omega_s(t)\approx\omega_0 - \epsilon\Delta\sigma \sin{(\Delta\varphi)}
\left[1 - \frac{\Delta\omega}{\Delta\sigma} \cot{(\Delta\varphi)}\right].
\end{equation}
Considering the second term, we note that $M\Delta\sigma$ is generally
just under $0.2$ for the cases we've studied so that
$(\Delta\omega/\Delta\sigma)\lesssim0.1$.  The other factor,
$\cot{(\Delta\varphi)}$, is not predictable without knowledge of the
initial conditions, but most values of $\Delta\varphi$ leave the
second term somewhat less than one.

If these conditions hold near the onset of ringdown, we might neglect
the time dependence in the second term.  The late-time frequency
evolution is approximated by a exponential decay with a time constant
of $1/\Delta\sigma$.  Working under these assumptions, and comparing
with the late-time frequency evolution model in this paper, yields the
association $2b=1/\Delta\sigma$, where $b$ is the fitting parameter in
our late time frequency evolution model. Again, looking at the
quasinormal mode values, we comment that $\sigma_1/\sigma_0\sim 3$ so
that $1/(2\Delta\sigma)\sim b \sim\tau_0$, the approximate
relationship noted in Sec.\ref{ssec:TransFreq}, which was also
approximately derived from the assumption that $dJ/d\omega$ is
constant at late times.  Consistently, applying the relation
$\sigma_1/\sigma_0\sim 3$ in Eq.~(\ref{eq:swave}) also give an
expression for late-time amplitude consistent with
Eq.~(\ref{eq:ImOmegaFitFn}).

\section{Discussion}
\label{sec:discussion}

With hope of reaching out to a wide range of researchers interested in
gravitational radiation from black-hole binary mergers, we have
provided a descriptive walk-through of many of the general features of
late-time waveforms from generic mergers of nonspinning binary black
hole systems, based on a series of numerical simulations covering
systems from equal-mass up to mass ratio 6:1.  In this basic waveform
description we have examined waveform phase and amplitude, comparing
results among different mass ratios, as well as among the different
spin-weighted spherical harmonic $(\ell,m)$ component modes.

In our presentation, we have attempted to describe the radiation in
the simplest physical terms, pointing out traits in the waveforms that
are similar through the inspiral, merger and ringdown stages.
Throughout the coalescence, we find simple waveforms in each
$(\ell,m)$ mode, each exhibiting strong circular polarization and
monotonically increasing polarization frequency.

In our amplitude comparisons, we find that the leading-order
PN-prediction for energy-partitioning provides a good estimate of the
amplitude until late in the merger for $\ell=m$ modes.  In
astrophysical units, our fit for the peak-power in the dominant
$(2,2)$ mode is
\begin{equation}
\dot{E}_{2,2} (\eta) = \left( 1.60 \eta^2 + 19.70 \eta^4 \right) \times 10^{57} {\rm erg \, s^{-1}}.
\end{equation}
Scaling the amplitudes by the peak values yields a very similar shape
through the peak for the $(2,2)$ mode amplitudes for all mass ratios
we have studied. For a particular mass-ratio case, the peak-widths
remain similar, though there is some variation in peak-time.

For each mass ratio, the phase (and frequency) of the different
$(\ell,m)$ components are strongly related.  While this should be
expected for the early inspiral, where the waveform phase is directly
connected to the orbital phase, we show that, for $\ell=m$ modes, the
same relationship holds through the merger and into the inspiral.  We
compare the phasing among simulations with different mass ratios in
two ways, with time scaled by chirp-mass, as is appropriate in the
early-time limit, and then with time scaled by total system mass
$M$. With the latter scaling, the dominant $(2,2)$ waveforms are
similar in phase (and $\eta$-scaled amplitude) for the last $\sim
200M$.
   
In the near-peak waveform comparisons the $(3,2)$ mode does not
exhibit the same simple behavior as the other modes.  This mode is
easily subject to coupling with the much stronger $(2,2)$ mode.  It
has been seen in \cite{Buonanno:2006ui,Schnittman:2007ij} that the
(3,2) mode demonstrates significant mode mixing with the dominant
(2,2) mode -- the QNM ringdown part of the (3,2) waveform contains the
fundamental frequencies of both the (3,2) and the (2,2) modes. We
speculate that this mixing might be partly due to the use of
coordinate extraction spheres that are systematically warped from the
areal-radius spheres appropriate for correct radiation extraction. A
more refined choice of extraction spheres and perhaps a better tuned
decomposition basis (e.g. spheroidal harmonics) would make it possible
to represent the $(3,2)$ waveform content in a manner which is like
that seen with the other modes.  Thereby we expect that a similar
simple physical representation of the waveform content can be extended
to all $m\neq0$ modes.

We suggest a simple conceptual interpretation that applies through the
full coalescence.  We think of the radiation as being generated by an
\emph{implicit rotating source}, with each $(\ell,m)$ mode generated
separately by the $(\ell,m)$ moment of some implicit source (which we
understand here only in the context of the radiation).  The nearly
fixed relationship among the $(\ell,m)$ phase moments is interpreted
to indicate that the implicit source maintains some structural
integrity throughout the coalescence, without shearing among the
various modal components.  For the $\ell=m$ modes, this rigidity is
maintained through the merger and into the ringdown, a relationship
made possible by the approximate equality for each of the $\ell=m$
quasinormal modes $\omQ/m\sim\Omega_{\rm LR}$, where $\Omega_{\rm LR}$
is the orbital frequency of unstable circular prograde graviton (or
photon) orbits.

The following physical picture may underlie these relationships.  For
well-separated black holes, the fields that embody the implicit source
object evidenced in the radiation may be tied directly to the
pointlike centers of the orbiting black holes.  The source rotation
frequency is the orbital frequency of the timelike trajectories
traced out by the black holes.  As the binaries spiral together, the
pair can continuously be viewed as a shrinking, distributed
dumbbell-like rotator.  Eventually, most of this dumbbell shrinks
inside the light-ring, which roughly coincides with the potential
barrier in the wave mechanics of gravitational perturbation theory.
From inside, little radiation can escape to a distant observer, and
the timelike motion of the black hole centers disconnects from the
radiation.  At late times, the effective radiation source becomes a
gravitational disturbance orbiting the forming black hole at the light
ring.  This is a seamless transition, with nearly consistent
rotational phasing among all $\ell=m$ modes throughout the process.
For the $\ell\neq m$ modes the associated quasinormal-ringing dynamics
are somewhat distinct, and the phasing and amplitude relationships
begin to peel away from the main $\ell=m$ trend through the merger
process.
 
For the late-time portions of the waveforms, including the approach to
the peak and the ringdown, we have introduced a quantitative fitting
model based on a monotonically increasing polarization frequency for
each mode, which decays exponentially toward the expected fundamental
quasinormal ringdown frequency $\omQ$ at late times.  These fits
provide an excellent match for the frequency evolution beginning
$\sim20M$ before the peak, and allow precise estimates of $\omQ$, as
well as the peak rate of change in frequency $\dot\omega_0$.  Scaling
the latter quantity by the final black hole mass $\Mf$ and $\omQ$, we
find $\dot\omega_0 \Mf/\omQ\approx0.021$ for all $\ell=m$ waveforms we
have looked at, including the $(2,2)$ modes of each mass-ratio, and
modes up to $(4,4)$ for the 4:1 mass-ratio case.

Conceptually, the monotonicity of the frequency evolution suggests
that, as is the case for inspiralling systems, the frequency can be
taken to label the state of the adiabatically changing implicit
rotating source that we interpret as the source of the radiation.  If
we suppose that changes in the source structure are tied to loss of
angular momentum, then we would expect that finite changes in
frequency would be associated with finite angular momentum loss, so
that $dJ(\omega)/d\omega$ has a finite, nonzero value even at late
times.  Since $\omega$ approaches a nonzero constant at late times,
we would likewise expect $dJ/d\omega$ to approach a constant value.
In Sec.~\ref{ssec:FreqAmp}, we show that the late-time evolution of
$J$ and $\omega$ are approximately consistent, mode-by-mode, with
constant $dJ/d\omega$ beginning about $\sim 20M$ before merger.

Such a relation between frequency and angular momentum also implies a
connection between frequency and amplitude.  The moment of peak
amplitude is expected to be near the peak in $\dot\omega(t)$.  At very
late times, constant $dJ/d\omega$ implies a connection between the
rates at which the frequency and amplitude approach their quasinormal
late time state, namely that our fitting parameter $b\approx\tau$,
where $\tau$ is the damping time of the quasinormal amplitude decay
[see Eq.~\ref{eq:OmegaFitFn}].

The simple relationships between the waveform modes and simple
dependence on mass-ratio make it possible to specify much of the
late-time waveform information developed in our numerical simulations
in terms of just a few quantities.  This information can then be
combined with information from the PN approximation about the inspiral
trajectories to provide analytical models for full-coalescence
waveforms.  In Sec.~\ref{sec:newEOB} we have applied the PN-consistent
EOB-based p4PN trajectory model presented in \cite{Buonanno:2007pf}
together with assumptions asserting several of the approximate
waveform features observed in Sec.~\ref{sec:description}.  As in
\cite{Buonanno:2007pf}, early waveform phasing is derived from the
p4PN EOB trajectories, with waveform amplitudes based on our PN-based
power partitioning, together with the PN flux model. As the waveform
approaches the anticipated peak, we match to a waveform phasing model
based on our fit model, with parameters specified according to the
approximate relationships identified in Sec.~\ref{sec:description},
and with amplitudes derived from $dJ/d\omega$=constant.  The fits show
excellent phase agreement with the numerical simulation results for
the most significant $\ell=m$ modes.  A variation in the PN flux model
that enforces that the flux for each $(\ell,m)$ mode vanish as
$\omega\rightarrow\omQ$ gives better late-time amplitude agreement.

Lastly, we observe that our description of the late-time phasing and
amplitudes provides a picture complementary to another approach
applied in several previous studies
\cite{Buonanno:2007pf,Damour:2007vq,Berti:2007fi,Buonanno:2006ui},
which successfully treat the late-time waveforms as a sum of
quasinormal fundamental and overtone modes for each $(\ell,m)$
waveform component.  This is motivated by the expectation that
waveforms from generic initially compact distortions of the forming
black hole will quickly reduce to a sum of these quasinormal harmonics
\cite{Teukolsky:1974yv}.  In \cite{Buonanno:2007pf}, we have shown
that in comparisons with some of the runs presented here, for mass
ratios up to 2:1, this assumption can lead to a predictive waveform
model with similar accuracy to our alternative model presented in
Sec.~\ref{sec:newEOB}.  As a link between the two approaches, we have
shown that at late times the combination of the fundamental and first
QNM overtones for a particular mode may, under reasonable
circumstances, mimic the amplitude decay properties of our model.

This work suggests several directions for further study.  In the
immediate future, we plan to assess the fidelity of the available
nonspinning waveforms and models, and the impact of mass ratio on the
overall detectability of the merger signal.  We also plan to apply our
implicit-rotating-source description as a baseline in analyzing future
higher-precision numerical simulations.  This might provide insight
into understanding finer features of the merger physics, some of which
could violate our simplified description.  Further understanding the
anomalous $(3,2)$ mode waveforms will be a first step in this
direction.  We must also investigate whether our description of the
merger radiation applies also to spinning black holes. It is plausible
that even precessing systems might be analyzed in this way using a
spherical harmonic basis that tracks the orbital axis
\cite{Gualtieri:2008ux}.  This may make it possible to extend the
analytic EOB-based waveform model presented here to include spin
effects.  Including spins in such analytic models will be necessary
for observational data analysis applications.

\acknowledgments
We thank Emmanuele Berti and Alessandra Buonanno for interesting
discussions, and Luciano Rezzolla for useful comments.
This work was supported in part by NASA grant
05-BEFS-05-0044 and 06-BEFS06-19. The simulations were carried out
using Project Columbia at the NASA Advanced Supercomputing Division
(Ames Research Center) and at the NASA Center for Computational
Sciences (Goddard Space Flight Center). B.J.K.  was supported by the
NASA Postdoctoral Program at the Oak Ridge Associated Universities.

\begin{appendix}
\section{Radiation Extraction}
\label{appendix:Radiation}

We extracted radiation using the outgoing Weyl scalar $\psi_4$,
defined as in \cite{Baker:2001sf}, calculated with a symmetric
tetrad. This is related to the complex gravitational-wave-strain $h
(t,\vec{x})$ via
\begin{eqnarray}
\psi_4 (t,\vec{x}) &=& - \ddot{h}_+(t,\vec{x}) + i \ddot{h}_{\times}(t,\vec{x}).
\end{eqnarray}
$\psi_4$, the strain $h$, and its time-derivative $\dot{h}$ (which we
call the \emph{strain rate})
are all functions of time $t$, extraction radius $\Rext$,
and polar angles $\theta$, $\phi$. As is customary in
numerical relativity, we decompose the radiation into
spin-(-2)-weighted spherical harmonic components:
\begin{eqnarray}
\psi_4 &=& \sum_{\ell=2}^{\infty} \sum_{m=-\ell}^{\ell} C_{\ell m}(t,R) \,_{-2}Y_{\ell}^m(\theta,\phi). \label{eqn:psi4lm_def}\\
\dot{h} &=& \sum_{\ell=2}^{\infty} \sum_{m=-\ell}^{\ell} \dot{h}_{\ell m}(t,R) \,_{-2}Y_{\ell}^m(\theta,\phi), \label{eqn:hdotlm_def}\\
h &=& \sum_{\ell=2}^{\infty} \sum_{m=-\ell}^{\ell} h_{\ell m}(t,R)
\,_{-2}Y_{\ell}^m(\theta,\phi). \label{eqn:hlm_def}
\end{eqnarray}
With this, two time integrations of our measured quantity, $\psi_4$,
yields the more familiar gravitational-wave strain $h$.

To extract the radiation from a simulation, we define a series of
coordinate spheres of different radii $\Rext$; here we use extraction
spheres having radii between $\Rext=40M$ and $\Rext=100M$.  We
extracted the radiation in modes by integrating $\psi_4$ against
different $\,_{-2}Y_{\ell}^m(\theta,\phi)$ over these coordinate
spheres, using fourth-order interpolation onto each sphere followed by
Newton-Cotes angular integration.

The gravitational waves produced by the binary carry both energy and
angular momentum.  Overall, the rate of energy emission is given by an
angular integral of the squared strain rate $|\dot{h}|^2$ over a
coordinate sphere [see Eq. (5.1) of \cite{Baker:2002qf}]:
\begin{equation}
\label{eqn:Edot_total}
\frac{dE}{dt} = \lim_{R\rightarrow\infty}\frac{R^2}{16 \pi}\oint d\Omega |\dot{h}|^2.
\end{equation}
Then using Eq. (\ref{eqn:hdotlm_def}), we can express the total energy
flux (\ref{eqn:Edot_total}) as a sum over modes:
\begin{eqnarray}
\dot{E}_{\ell m} &\equiv& \left( \frac{dE}{dt} \right)_{\ell m} = \lim_{R\rightarrow\infty} \frac{1}{16 \pi} |R \dot{h}_{\ell m}|^2, \label{eqn:Edot_modes} \\
&=& \frac{(A_{\ell m})^2}{16\pi} \label{eq:energy-amplitude},
\end{eqnarray}
where we have used the strain-rate decomposition
(\ref{eqn:strainrate_modes}) and taken the limit $R\rightarrow\infty$
to go from (\ref{eqn:Edot_modes}) to (\ref{eq:energy-amplitude}).

Similarly, the rate of radiation of the $z$-component of angular
momentum can be expressed as a sum over modes \cite{Lousto:2007mh}:
\begin{equation}
\label{eqn:Jdot_modes}
\dot{J}_{\ell m} \equiv \left( \frac{dJ^z}{dt} \right)_{\ell m} = \lim_{R\rightarrow\infty} \frac{m}{16 \pi} R^2 {\rm Im}(\dot{h}_{\ell m} h^*_{\ell m}).
\end{equation}

Substituting the definitions (\ref{eqn:strainrate_modes}) and
(\ref{eqn:strain_modes}) into expression (\ref{eqn:Jdot_modes}) for
the angular momentum, and taking the limit $R\rightarrow\infty$ yields
\begin{equation}
\label{eqn:JdotAH}
\dot{J}_{\ell m} = \frac{|m|}{16 \pi} A_{\ell m} H_{\ell m} \cos(m(\Phi_{\ell m}-\Phi^{(h)}_{\ell m})).
\end{equation}

\section{Convergence}
\label{appendix:Convergence}

We carried out three runs of the 4:1 mass ratio model at different
resolutions to study the convergence properties of our simulations.
For these cases, the mesh spacing of the finest grids (the ones
including the smaller puncture) was taken to be $h_f = 3M/160$ (low
resolution), $h_f = M/64$ (medium resolution), and $h_f = 3M/224$
(high resolution).  To facilitate comparisons among these cases, the
overall grid structure of the runs was kept the same.  In this
Appendix, we discuss the convergence properties of the constraints and
gravitational waveforms.

In comparing our medium and high resolutions for the 4:1 mass ratio
case, the Hamiltonian constraint was found to be manifestly
fourth-order-convergent in the dynamical strong-field, where the black
holes move, evidently dominated by the expected fourth-order error
from refinement interfaces.  The convergence falls off to an apparent
rate closer to first order in the coarsest regions. This seems to
result from stronger dissipation of high frequency noise. Though less
noise is generated in the higher resolution simulations, a greater
portion of it survives propagation into the distant coarser regions.
As shown below, this does not appear to affect the waveforms, which
are well-resolved in the wave zone.  The momentum constraint appeared
to be at least second-order-convergent in the dynamical strong-field
region, but also fell off to an apparent rate closer to first order in
the coarsest regions -- from the wave-extraction region outwards.

In our waveform mode analysis, we have typically time- and
phase-shifted the data so that the amplitude peak fell at time $t=0$
\cite{Baker:2006yw}. In Fig.~\ref{fig:hdot22_44}, the real part of the (2,2) and (4,4)
strain-rate harmonics is shown for all three resolutions. The
agreement between the resolutions is then seen to be excellent. In
Figs.~\ref{fig:amperrorMAQ} and \ref{fig:phaseerrorMAQ}, we showed the
errors to be expected from the maquillaged strain-rate waveform data.

\begin{figure}
\includegraphics*[width=3.5in]{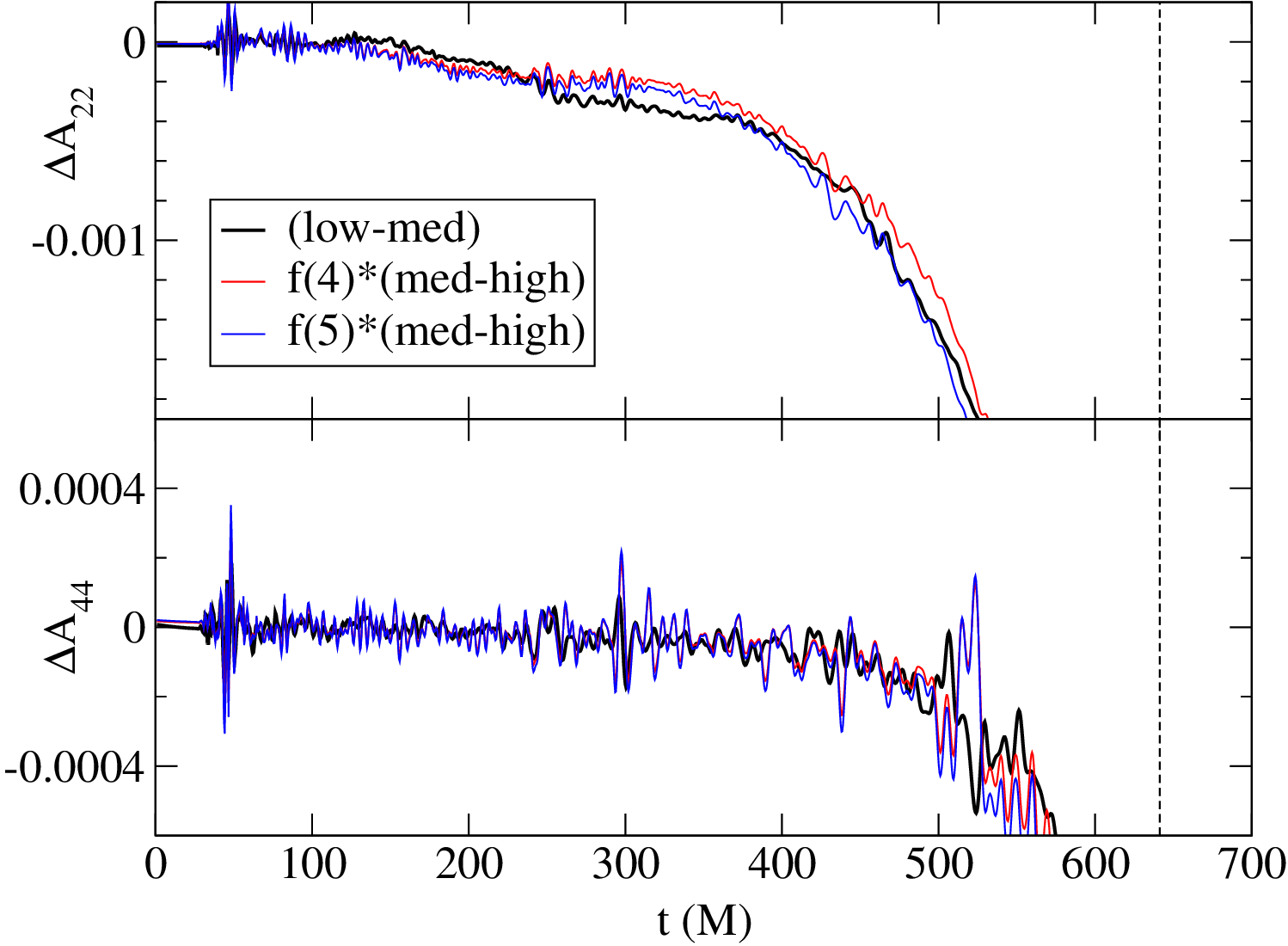}
\caption{Convergence of strain-rate amplitude for (2,2) (upper panel)
and (4,4) (lower panel) modes. The medium-high differences are scaled
for both fourth- and fifth-order convergence. Three-level convergence
is lost around $150 M$ before merger (vertical dashed line).}
\label{fig:amp_convRAW}
\end{figure}

\begin{figure}
\includegraphics*[width=3.5in]{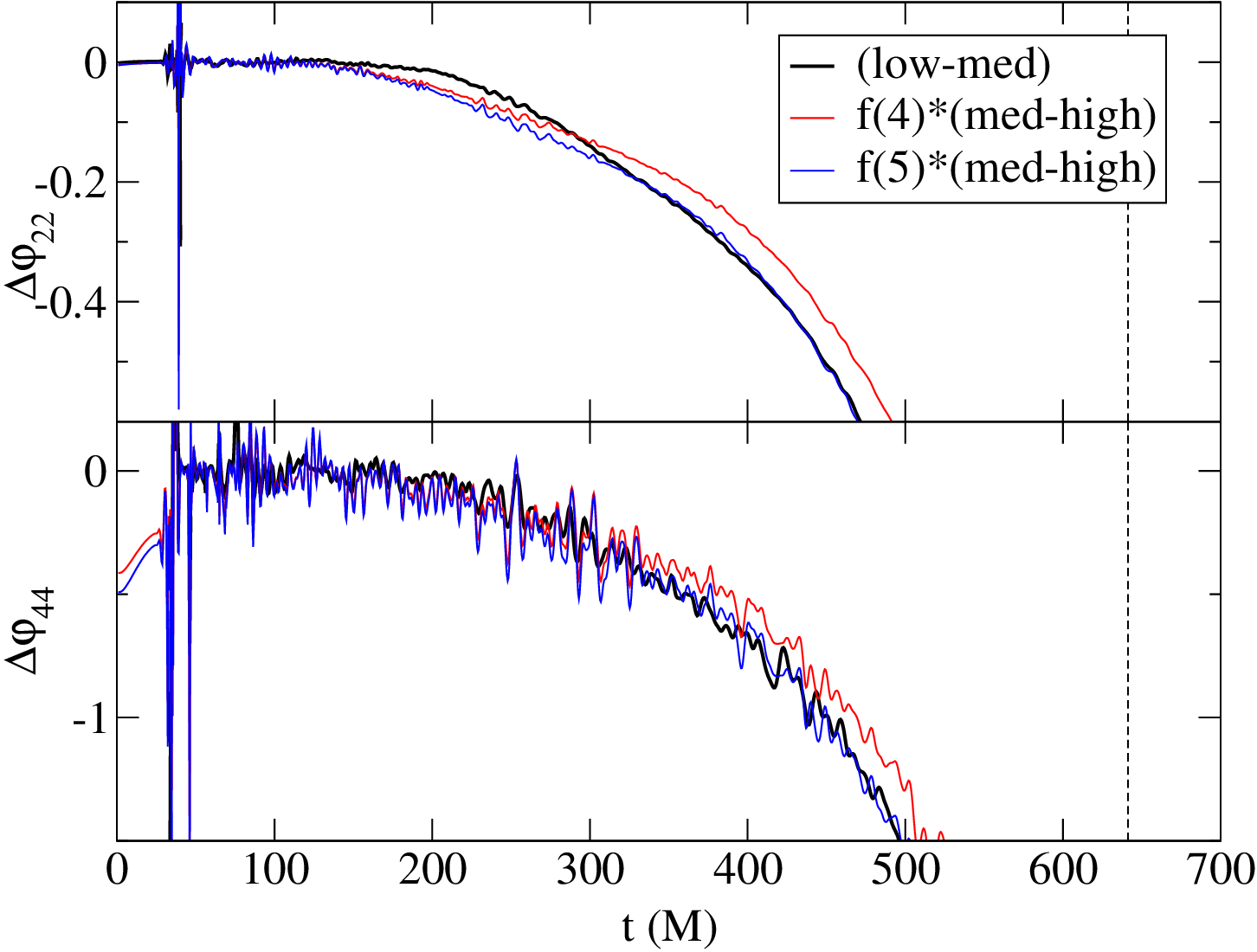}
\caption{Convergence of strain-rate phase for (2,2) (upper panel)
and (4,4) (lower panel) modes. The medium-high differences are scaled
for both fourth- and fifth-order convergence. Three-level convergence
is lost around $150 M$ before merger (vertical dashed line).}
\label{fig:phase_convRAW}
\end{figure}

However, the presence of considerable eccentricity in the binary makes
it difficult to compare such time-shifted waveforms between
resolutions and establish an unambiguous order of convergence. In
In Figs. \ref{fig:amp_convRAW} and \ref{fig:phase_convRAW} we show the
amplitude and phase errors, respectively, for our three resolutions
\emph{without} time-shifting -- that is, we plot the data
starting from the initial time in each case, scaling the medium - high
differences for fourth and fifth-order convergence. It is clear that
we observe convergence between fourth and fifth order throughout the
evolution until around $150 M$ before merger, when the difference in
merger times among the runs becomes important. Additionally, it is
possible that at this higher-frequency stage of merger, the
lowest-resolution data is no longer in the convergence regime. We do
note, however, that the rate of growth of amplitude difference between
the medium and high-resolution runs is comparable here to that
observed closer to the merger for time-shifted data [see
Fig. \ref{fig:amperrorMAQ}]. From this we deduce that the medium and
high-resolution runs are still in the convergence regime, with errors
consistent with fourth-order convergence.

\section{End states}
\label{appendix:EndState}

In this Appendix we discuss the state of the final black hole formed
in our simulations.  This can be measured by several independent
means, which we compare, finding agreement to within $0.4 \%$ for
$\Mf$ and $2.1\%$ for $\hat{a}$.  Table~\ref{table:RunResults}, we
present ``coarse'' results for the simulations.  The total radiated
energy $\Delta E_{\rm rad}$ is obtained by integrating
Eq. (\ref{eqn:Edot_total}), and the total radiated angular momentum
$\Delta J_{\rm rad}$ is obtained by summing over mode contributions
(\ref{eqn:Jdot_modes}).  The final mass and angular momentum of the
post-merger black hole are calculated using
\begin{eqnarray}
M_{\rm f,rad} &\equiv& \MADM - \Delta E_{\rm rad}, \label{eqn:Mf_def}\\
J_{\rm f,rad} &\equiv& J_0 - \Delta J_{\rm rad}, \label{eqn:Jf_def}
\end{eqnarray}
where, by the symmetries of the current simulations, we only deal with
the $z$ component of angular momentum.  For the 1:1 simulations, we
only had the leading-order $(2,\pm2)$ radiation modes available, so
our $\Delta J_{\rm rad}$ estimate is significantly truncated, by as
much as 11\% (a conservative error estimate based on the effect of
similarly truncating mode contributions past $\ell = 2$ for the 4:1
case); we have marked this and derived values. We also quote the
measured value of $t_{\rm peak}$, the time at which $\dot{E}$ reaches
its peak. All waveform and derived plots in this paper have been
time-shifted by subtracting this time, as an approximate marker of the
time of merger, unless otherwise indicated.

Table~\ref{table:RunResults} also contains data about the common
apparent horizon (CAH) of the merged binary: the time $t_{\rm CAH}$ at
which the CAH was first detected, the CAH's irreducible mass $M_{\rm
irr, CAH}$, and its full (horizon) mass, obtained from the
Christodoulou \cite{Christodoulou:1970wf} formula:
\begin{equation}
M_{\rm f,CAH}^2 = M_{\rm irr}^2 + \frac{J^2}{4 M_{\rm irr}^2},
\end{equation}
where we use $J = J_{\rm f,rad}$ for the final hole's angular
momentum.

\begin{table*}
\caption{\label{table:RunResults}
Results of radiation and apparent-horizon analysis of merger. The
total energy flux $\dot{E}$ (\ref{eqn:Edot_total}) reaches its peak at
time $t_{\rm peak}/M$.  $\Delta E_{\rm rad}$ and $\Delta J_{\rm rad}$
are the total energy and ($z$) angular momentum radiated during the
simulation (the latter calculated as a sum over modes
(\ref{eqn:Jdot_modes})), resulting in final mass $M_{\rm f,rad}$ and
angular momentum $J_{\rm f,rad}$ (For the 1:1 case, only the $\ell =
2$ modes were available, so the emitted $\Delta J_{\rm rad}$ will
underestimate, and $J_{\rm f,rad}$ will overestimate, the physical
results.).  $M_{\rm irr, CAH}$ is the irreducible mass of the common
apparent horizon, first detected at simulation time $t_{\rm CAH}$.
$M_{\rm f,CAH}$ is the mass deduced from this and $J_{\rm f,rad}$
using the Christodoulou formula \cite{Christodoulou:1970wf}.}
\begin{tabular}{cc||c|rcc|cc||rcc}
\hline\hline
Mass  & $h_f$   & $\Rext/M$ & $t_{\rm peak}/M$ & $\Delta E_{\rm rad} (\times 10^{-2})/M$ & $M_{\rm f,rad}/M$ & $\Delta J_{\rm rad} (\times 10^{-2})/M^2$ & $J_{\rm f,rad}/M^2$ & $t_{\rm CAH}/M$ & $M_{\rm irr, CAH}/M$ & $M_{\rm f,CAH}/M$ \\
ratio & & & & & & & & &\\
\hline
1:1     & $M/32$  &  60      & 1303.7 & 3.5934 & 0.9548 & 33.79 & 0.6468 & 1196 & $\cdots$ & $\cdots$ \\
\hline
2:1     &$3M/160$ &  45      &  627.7 & 2.8306 & 0.9606 & 23.34 & 0.5937 &  581 & 0.9063 & 0.9637 \\
\hline
4:1     &$3M/224$ &  45      &  641.9 & 1.4327 & 0.9786 & 12.78 & 0.4615 &  588 & 0.9489 & 0.9796 \\
        & $M/64$  &  45      &  652.6 & 1.4262 & 0.9786 & 12.99 & 0.4594 &  599 & 0.9489 & 0.9793 \\
        &$3M/160$ &  45      &  677.9 & 1.4102 & 0.9789 & 12.90 & 0.4603 &  610 & 0.9492 & 0.9797 \\
\hline
6:1     &$M/64$   &  45      &  564.9 & 0.9212 & 0.9850 & 07.83 & 0.3666 &  513 & 0.9667 & 0.9851 \\
\hline \hline
\end{tabular}
\end{table*}

We can use the data from Table~\ref{table:RunResults} to estimate the
mass and dimensionless spin of the end-state Kerr black hole through
different methods. Given the final mass estimates $M_{\rm f,rad}$ and
$M_{\rm f,CAH}$, we can calculate $\hat{a} \equiv J_{\rm f}/\Mf^2$.

Another means of characterizing the final black hole comes from
studying the characteristics of the radiation after the peak. This is
expected to be a sum of the hole's quasinormal modes (QNMs). We can
determine the imaginary part of the mode frequency by fitting the
waveform amplitude $A_{l,m}$ from (\ref{eqn:hdotlm_polar}) post-peak
to a decaying exponential. The damping time in this fit corresponds to
the imaginary frequency. Specifically, we fit to a functional form:
\begin{equation}
\label{eq:ImOmegaFitFn}
A(t) = A_0 e^{-t/\tau_{\rm QNM}} \left( 1 - C e^{-2 t/\tau_{\rm QNM}} \right).
\end{equation}
The presence of the additional damping term, parametrized by $C$,
allows more freedom for nonlinear decay early in the ringdown. Note
that this factor of 3 between primary and secondary damping times
approximately mirrors the difference in damping times one sees between
the first two QNMs of a Kerr hole \cite{Leaver85}. The results of this
fit are given in Table~\ref{table:TransAmpFit}. The value of
$\tau_{\rm QNM}$, together with an estimate of the real QNM frequency
$\omQ$, uniquely determines the mass and dimensionless spin of the
Kerr hole. We combine our present fit for $\tau_{\rm QNM}$ with the
real frequency $\omQ$, as presented in Table~\ref{table:TransFreqFit},
to obtain ($M_{\rm QNM}$,$\hat{a}_{\rm QNM}$).
\begin{table}
\caption{\label{table:TransAmpFit}
Fit results for $\tau_{\rm QNM}$ from
Eq. (\ref{eq:ImOmegaFitFn}). $\tau_{\rm QNM}$ is used for one of the
determinations of final Kerr parameters in Table
\ref{tab:KerrParams}.}
\begin{tabular}{c|c|cc}
\hline \hline
Mass ratio   &$(\ell,m)$&   $A_0 ( \times 10^{-2})$ & $\tau_{\rm QNM}/M$\\
\hline\hline
1:1     &   (2,2) & $31.3  \pm 0.1$  & $11.68 \pm 0.01$\\
\hline
2:1     &   (2,2) & $26.80 \pm 0.06$ & $11.57 \pm 0.01$\\
\hline
4:1     &   (2,2) & $17.23 \pm 0.09$ & $11.37 \pm 0.01$\\
        &   (2,1) &  $4.1  \pm 0.1$  & $11.42 \pm 0.11$\\
        &   (3,3) &  $8.2  \pm 0.2$  & $10.96 \pm 0.06$\\
        &   (4,4) &  $2.85 \pm 0.05$ & $10.88 \pm 0.06$\\
\hline
6:1     &   (2,2) & $12.77 \pm 0.08$ & $11.35 \pm 0.02$\\
\hline \hline
\end{tabular}
\end{table}

If we already have a robust determination of the Kerr mass, then we
only need one component of the QNM frequency to lock down
$\hat{a}$. In Table~\ref{table:TransFreqFit}, we use $\omega_{\rm
QNM}$ to obtain $\hat{a}_{\rm QNM, re}$, assuming $M_{\rm QNM} =
M_{\rm f,rad}$.

We have gathered all the estimates of $\Mf$ and/or $\hat{a}$ discussed
above in Table~\ref{tab:KerrParams}.
\begin{table*}
  \caption{\label{tab:KerrParams} The parameters of the post-merger
  Kerr black hole, estimated by different methods.  $M_{\rm f,rad}$
  and $\hat{a}_{\rm rad}$ are based on energy and angular-momentum
  balance in the gravitational radiation over the whole
  evolution. $M_{\rm f,CAH}$ is the mass of the common apparent
  horizon, as determined from the horizon's irreducible mass [see
  Table~\ref{table:RunResults}]. $M_{\rm f,QNM}$ and $\hat{a}_{\rm
  QNM}$ are the mass and spin determined from the real frequency
  $\omQ$ and damping time of the post-merger radiation [see Tables
  \ref{table:TransAmpFit} and \ref{table:TransFreqFit}], while
  $\hat{a}_{\rm QNM, re}$ was determined from the real frequency
  $\omQ$ only, assuming the final mass $M_{\rm f,rad}$.} 
 \begin{tabular}{c|cc|c|cc|c}
    \hline\hline
    Mass ratio & $M_{\rm f,rad}/M$ & $\hat{a}_{\rm rad}$ & $M_{\rm f,CAH}/M$ & $M_{\rm f,QNM}/M$ & $\hat{a}_{\rm QNM}$ & $\hat{a}_{\rm QNM, re}$ \\
    \hline
    1:1 & $0.9548$ & 0.7095          & $\cdots$ & $0.951 \pm 0.002$ & $0.684 \pm 0.002$ & $0.689 \pm 0.002$ \\
    2:1 & $0.9606$ & 0.6434          & 0.9637 & $0.961 \pm 0.002$ & $0.624 \pm 0.001$ & $0.624 \pm 0.002$ \\
    4:1 & $0.9786$ & 0.4819          & 0.9796 & $0.979 \pm 0.001$ & $0.472 \pm 0.002$ & $0.4710 \pm 0.0004$ \\
    6:1 & $0.9850$ & 0.3779          & 0.9851 & $0.989 \pm 0.002$ & $0.383 \pm 0.004$ & $0.377 \pm 0.002$ \\
    \hline\hline
  \end{tabular}
\end{table*}
It is instructive to compare these values with recent predictions for
the final mass and spin from analytic and numerical methods. Working
with a subset of the data supplied here, and also invoking the
test-mass limit, \cite{Buonanno:2007pf} suggest the one-parameter fits:
\begin{eqnarray}
\Mf/M &=& 1 - (1 - \sqrt{8/9}) \eta - (0.498 \pm 0.027) \eta^2, \label{eqn:nreob_Mfit}\\
\hat{a} &=& \sqrt{12} \eta - (2.900 \pm 0.065) \eta^2. \label{eqn:nreob_afit}
\end{eqnarray}
The numerical data from Table~\ref{tab:KerrParams} also fits well the
wholly numerical formula (3.17a) from \cite{Berti:2007fi}, as well as
the parametrized formula of \cite{Rezzolla:2007rd}.

We note that when discussing the mass $\Mf$ of the end-state Kerr hole
in Secs. \ref{ssec:TransFreq} and later, we have consistently taken
the radiation-derived value, $\Mf \equiv M_{\rm f,rad}$, as results
from \cite{Baker:2006kr} indicate this has errors up to a factor of
3 less than QNM-based mass. Since a far larger proportion of the
initial angular momentum than mass is emitted during the inspiral and
merger, there is greater uncertainty in the radiation-derived final
spin than the final mass. For this reason, our preferred measure of
the final dimensionless spin is the QNM-derived value $\hat{a}_{\rm
QNM, re}$.
\end{appendix}

%

\end{document}